\documentclass[aps,pra,amsmath,twocolumn,amssymb,superscriptaddress]{revtex4-1}

\usepackage{amssymb,amsfonts,amsmath}

\usepackage{color}
\usepackage[apple mac]{inputenc}
\usepackage[english]{babel}
\usepackage{times}
\usepackage{latexsym}
\usepackage{fancyhdr}
\usepackage{verbatim}
\usepackage{graphicx}
\usepackage{epsf}
\usepackage{subfigure}
\usepackage{bm}
\usepackage{epstopdf}\usepackage[
top    = 1.5cm,
bottom = 1.50cm,
left   = 1.5cm,
right  = 1.0cm]{geometry}

\definecolor{Nathanblue}{rgb}{0.,0.24,0.51}
\newcommand{\blue}{\color{Nathanblue}}
\definecolor{orange}{rgb}{0.96,0.24,0.00}

\def\be{\begin{equation}}
\def\ee{\end{equation}}
\def\bs#1{\boldsymbol{#1}}

\begin{document}
 
%\title{{\blue Gauge fields in periodically-driven quantum systems: \\ Effective Hamiltonians and the use of the Trotter expansion}}

\title{{\blue Periodically-driven quantum systems: \\ Effective Hamiltonians and engineered gauge fields}}

\author{N. Goldman}
\email[]{nathan.goldman@lkb.ens.fr}
\affiliation{Coll\`ege de France, 11, place Marcelin Berthelot, 75005 Paris, France}
\affiliation{Laboratoire Kastler Brossel, CNRS, UPMC, ENS, 24 rue Lhomond, 75005, Paris, France}

\author{J. Dalibard}
\email[]{jean.dalibard@lkb.ens.fr}
\affiliation{Coll\`ege de France, 11, place Marcelin Berthelot, 75005 Paris, France}
\affiliation{Laboratoire Kastler Brossel, CNRS, UPMC, ENS, 24 rue Lhomond, 75005, Paris, France}

\date{\today}

\begin{abstract}
Driving a quantum system periodically in time can profoundly alter its long-time dynamics and trigger topological order. Such schemes are particularly promising for generating non-trivial energy bands and gauge structures in quantum-matter systems. Here, we develop a general formalism that captures the essential features ruling the dynamics: the effective Hamiltonian, but also the effects related to the initial phase of the modulation and the micro-motion. This framework allows for the identification of driving schemes, based on  general $N$-step modulations, which lead to configurations relevant for quantum simulation. In particular, we explore methods to generate synthetic spin-orbit couplings and magnetic fields in cold-atom setups.
\end{abstract}

\maketitle

%\tableofcontents

%%%%%%%%%%%%%%%%%%%%%%%%%%%%%%%%%%%%%%%%%%%%%%

%\section{Introduction}

\section{Introduction}

%\subsection{General introduction}

Realizing novel states of matter using controllable quantum systems constitutes a common interest, which connects various fields of condensed-matter physics. Two main routes are currently investigated to reach this goal. The first method consists in fabricating materials \cite{topological,superconductivity}, or artificial  materials  \cite{artificial_graphene0,artificial_graphene,artificial_graphene2,photonic1,photonic2,photonic3,photonic4,Tarruell:2012}, which present \emph{intrinsic} effects that potentially give rise to interesting phases of matter. For instance, this is the case for topological insulating materials, which present large intrinsic spin-orbit couplings \cite{topological}. The second method, which is now commonly considered in the field of quantum simulation, consists in driving a system using external fields \cite{Dalibard:2011,Goldman:2014Review,Zhai:Review} or mechanical deformations \cite{Guinea,artificial_graphene} to generate \emph{synthetic}, or effective, gauge structures. Formally, these driven-induced gauge fields enter an effective Hamiltonian, which captures the  essential characteristics of the modulated system. This strategy exploits the fact that  modulation schemes can be tailored in such a way that effective Hamiltonians reproduce the Hamiltonians of interesting  static systems. Furthermore, the versatility of driving schemes might enable one to explore situations that remain unreachable in static fabricated systems.   

In this context, several works have proposed methods to engineer effective magnetic fields or spin-orbit coupling based on driven cold-atom or ion-trap systems \cite{Sorensen,Ueda,Anderson,Creffield:2011,Hemmerich:2010,Eckardt:2010,Lim:2008,Struck:2012,Hauke:2012,Bermudez:2012,Bermudez:2011,Kolovsky:2011,Baur:2014,Creffield:2014,Zheng:2014}. Recently, these proposals led to the realization of the Hofstadter model [i.e. a lattice system penetrated by a uniform magnetic field \cite{Hofstadter:1976}], using ``shaken" optical lattices \cite{Aidelsburger:2011,Aidelsburger:2013,Miyake:2013}, and to frustrated magnetism using triangular optical lattices \cite{Struck:2011,Struck:2013}. Besides, artificial magnetic fields have been created in strained graphene \cite{Guinea}. Another field of research focuses on the possibility to create ``Floquet topological insulating states" by subjecting \emph{trivial} insulators or semi-metals [e.g. semiconductors or graphene] to external electromagnetic radiation \cite{Floquet_top1, Floquet_top2,Floquet_top3,Floquet_top4,Floquet_top5,Floquet_top6,Floquet_top7,Floquet_top8}. This strategy has been generalized for superfluids, where ``Floquet Majorana fermions" could be created by driving superconducting systems \cite{Floquet_Majorana_1,Floquet_Majorana_2,Floquet_Majorana_3,Floquet_Majorana_3_bis,Floquet_Majorana_4}.  The topological invariants [i.e. winding numbers] and the edge-state structures proper to driven systems were analyzed in Refs. \cite{Kitagawa,Morell:2012,Rudner2013,Leon:2013,Lababidi:2014,Perez:2014,Reichl:2014}. Finally, we note that time-periodic modulations can also be simulated in photonics crystals, where  time is replaced by a spatial direction \cite{photonic1}; such photonics systems have been fabricated recently \cite{photonic3} with a view to observing the anomalous quantum Hall effect \cite{Haldane:1988},  through the imaging of the related topological edge states. \\

In this work, we develop and explore  a general framework that describes periodically-driven quantum systems, and which  generalizes the formalism introduced by Rahav \emph{et al.} in Ref. \cite{Rahav}. In contrast with the standard Floquet analysis \cite{Maricq,Grozdanov:1988} or the effective-Hamiltonian method presented by Avan \emph{et al.} in Ref. \cite{Avan:1976}, the present method clearly isolates and identifies the three main characteristics of modulated systems: (1) the effective Hamiltonian underlying the long-time dynamics; (2) the micro-motion; and (3) the effects associated with the initial phase of the modulation. These distinct effects will be largely illustrated in this work, based on different examples relevant for the quantum simulation of gauge structures, e.g. magnetic fields and spin-orbit couplings. Moreover, this work provides general formulas and methods, which can be easily exploited to identify wide families of promising driving schemes. \\

Before presenting the outline of the paper [Section \ref{section:outline}], we briefly summarize some important notions related to driven quantum systems, based on basic illustrative examples.

\subsection{Effective Hamiltonians and the micro-motion: \\ Two simple illustrations}

We start the discussion by presenting two very simple situations, which illustrate in a minimal manner the basic notions and effects encountered in the following of this work. 

\subsubsection{The Paul trap}\label{section-paul_trap}

This first illustrative and basic example consists in a particle moving in a modulated harmonic trap \cite{foot_book}. The Hamiltonian is taken in the form
\be
\hat H(t)=\hat H_0 + \hat V  \cos (\omega t)= \frac{\hat p^2}{2m} + \frac{1}{2}m \omega_0^2 \hat x^2 \cos (\omega t),\label{paul_operators}
\ee
where $\omega=2 \pi /T$ [resp. $\omega_0$] denotes the modulation [resp. harmonic trap] frequency. The evolution operator after one period of the modulation is evaluated in  Appendix \ref{appendix:paul_trap}, and it reads
\begin{align}
&\hat U(T) \!=\! \exp \left (-i T \hat H_{\text{eff}}  \right ), \quad \hat H_{\text{eff}} \approx \hat H_0 + \frac{1}{2} m \Omega^2 \hat x^2, \label{paul_result_main_text}
\end{align}
expressing the fact that the particle effectively moves in a harmonic trap with frequency $\Omega=  \omega_0^2 / \sqrt{2} \omega$. An additional insight is provided by a classical treatment, in which one partitions the motion $x(t)=\bar x(t) + \xi (t)$ into a slow and a fast (micro-motion) component. As shown in Appendix \ref{appendix:paul_trap}, this analysis shows that the effective harmonic potential with frequency $\Omega$ that rules the slow motion $\bar x(t)$ is equal to the average kinetic energy associated with the \emph{micro-motion}:
\be
\frac{1}{2}m \Omega^2 \bar x^2 = \frac{1}{2}m \langle \dot{\xi}^2 \rangle,
\ee
where $\langle . \rangle$ denotes the average over one period. This classical result  illustrates the important role played by the micro-motion in modulated systems. 

\subsubsection{The modulated optical lattice}\label{section:modulated_lattice}

As a second example, we consider a modulated 1D lattice, treated in the single-band tight-binding approximation \cite{Eckardt:2005,Lignier:2007,Eckardt:2007,Eckardt:2009,Bessel_renormalize}. The Hamiltonian is taken in the form [see Appendix \ref{appendix:bessel_one}]
\be
\hat H (t)=\hat H_0 + \kappa   \cos (\omega t) \hat V, \quad \hat V =  \sum_j j \hat a_{j}^{\dagger} \hat a_j =  \hat x, \label{modulated_lattice_operators}
\ee 
where $\hat H_0$ describes the nearest-neighbour hopping on the lattice, and where the operator $\hat a_j^{\dagger} $ creates a particle at lattice site $x=j a$, and $a$ is the lattice spacing. The effective Hamiltonian describing the slow motion of a particle moving on the modulated lattice can be derived exactly [see Appendix \ref{appendix:bessel_one}], yielding the well-known renormalization of the hopping rate by a Bessel function of the first kind 
\be
\hat H_{\text{eff}}= \mathcal{J}_0 (\kappa / \omega) \hat H_0  \approx \left ( 1 - \frac{\kappa^2}{4 \omega^2} + \dots \right ) \hat H_0 .\label{result_bessel_main_text}
\ee
This effect has been observed experimentally with cold atoms in optical lattices \cite{Lignier:2007,Eckardt:2009}. 

The micro-motion also plays an important role in this second example,  where it is associated with large oscillations in quasi-momentum space. Indeed, within the single-band approximation, a significant modification of the tunneling rate is found when the micro-motion oscillation is comparable to the width of the Brillouin zone.\\

We point out that the Paul trap and the modulated lattice share similar structures [see also Appendices \ref{appendix:paul_trap} and \ref{appendix:bessel_one}]: both systems are driven by a modulation of the form $\hat H_0 + \hat V \cos (\omega t)$, and their effective Hamiltonians both contain a non-trivial term which is \emph{second order} in the period $T$. In the present case of the modulated lattice, the term $\sim (\kappa/\omega)^2$ is the first non-trivial term of an infinite series, which can be truncated  for $\kappa /\omega <1$, see Eq. \eqref{result_bessel_main_text}. 

\subsection{The two-step modulation and the ambiguity inherent to the Trotter approach}\label{section:Trotter_ambiguity}

Motivated by the two simple examples described above, we consider a general quantum system described by a static Hamiltonian $\hat H_0$, which is periodically driven by a repeated  two-step sequence of the form
\begin{equation}
\gamma: \quad \{ \hat H_0+ \hat  V, \hat H_0- \hat V \},
\label{trivial_sequence}
\end{equation} 
where $\hat  V$ is some operator. For simplicity, we suppose that the duration of each step is $T/2$, where $T=2 \pi / \omega$ is the period of the driving sequence $\gamma$. Thus, the  square-wave sequence $\gamma$ is qualitatively equivalent to the smooth driving $\hat H_0 + \hat V \cos (\omega t)$ encountered in the two examples discussed above. \\

In the following of this work, the energy $\hbar \omega$ will be considered to be very large compared to all the energies present in the problem, justifying a perturbative treatment in $(1/ \omega)$. The small dimensionless quantity associated with this expansion, $\Omega_{\text{eff}}/\omega$, will be made explicit for the various physical problems encountered in the following Sections. Typically, $\Omega_{\text{eff}}$ will be identified with the cyclotron frequency in the case of synthetic magnetism (see Section \ref{section_four}), or with the spin-orbit coupling strength (see Section \ref{section_five}); see also Section \ref{section:convergence}. \\

Starting in an initial state $\vert \psi_0 \rangle$ at time $t_i=0$, the state at time $t=NT$ ($N \in \mathbb{N}$) is obtained through the evolution operator
\begin{align}
%\vert \psi (t=NT) \rangle &= 
\hat  U(t\!=\!NT) \vert \psi_0 \rangle &\!=\! \left ( e^{- i T (\hat  H_0- \hat V)/2} e^{- i T  (\hat  H_0+ \hat  V)/2}  \right )^N \! \vert \psi_0 \rangle \label{long-time} \\
&= e^{-i NT \hat  H_{\text{eff}}^{\mathcal{T}}} \vert \psi_0 \rangle,\notag
\end{align}
where we introduced a \emph{time-independent effective Hamiltonian}, 
\be
\hat  U(T)=  e^{- i T \hat  H_{\text{eff}}^{\mathcal{T}}} = e^{- i T  (\hat  H_0- \hat V)/2} e^{- i T  (\hat H_0+\hat V)/2}.
\ee
The product of two exponentials can be simplified  through the Baker-Campbell-Hausdorff (BCH) formula, hereafter referred to as the Trotter expansion, 
\be
e^{X} e^{Y}= \exp \left ( X\!+\!Y \!+\!\frac{1}{2} [X,Y] \!+\! \frac{1}{12} [X-Y,[X,Y]] \dots  \right ) , \label{BCH}
\ee
yielding a simple expression for the effective ``Trotter" Hamiltonian
\begin{equation}
\hat  H_{\text{eff}}^{\mathcal{T}}= \hat H_0 - i \frac{T}{4 }[ \hat H_0, \hat V] + \mathcal{O} (T^2). 
\label{Trotter1}
\end{equation}
Importantly, the sign in front of the first order term depends on the starting pulse ($\hat H_0+ \hat V$ or $\hat H_0- \hat V$) of the driving sequence \eqref{trivial_sequence}, or equivalently, on the definition of the starting time $t_i$: indeed, shifting the starting time $t_i \rightarrow t_i+(T/2)$, leads to the opposite term $+ i (T/4 )[\hat H_0,\hat V] $. Thus, the first-order term arising from the Trotter expansion  is sensitive to the initial phase $\omega t_i$ of the driving. We emphasize that the sign ambiguity is different from the phase of the micro-motion sampling, which will be illustrated later in this work. Furthermore, we note that the first-order term in Eq. \eqref{Trotter1} can be eliminated  by a unitary transformation
\be
\hat  H_{\text{eff}}^{\mathcal{T}}=\hat  S^{\dagger} \hat  H_0 \hat S + \mathcal{O} (T^2), \quad \hat S= e^{- i T \hat  V/4},
\label{unitary}
\ee 
indicating its trivial role in the effective Hamiltonian. Indeed, the long-time behavior of the system described by Eq. \eqref{long-time} can be expressed as
\begin{align} 
%\vert \psi (t=NT) \rangle = 
\hat U(t=NT) \vert \psi_0 \rangle &= e^{- i NT  \hat H_{\text{eff}}^{\mathcal{T}}} \vert \psi_0 \rangle \notag \\
&= \hat S^{\dagger} \left ( e^{- i N T \hat H_0} \right ) \hat S \vert \psi_0 \rangle + \mathcal{O} (T^2),\label{decomposition}
\end{align}
which indicates that the system first undergoes an initial kick $\hat S=e^{- \frac{i T}{4 } V}$ [when the sequence $\gamma$ is applied in this order], then evolves ``freely" for a long time $t=NT$, and finally undergoes a final sudden kick (i.e. a micro-motion). From Eq. \eqref{unitary}, we conclude that the first-order term in Eq. \eqref{Trotter1} cannot be exploited to modify the band structure of $\hat H_0$, or equivalently, to generate non-trivial gauge structures (e.g. effective magnetic fields or spin-orbit couplings). This observation is in agreement with the effective Hamiltonians obtained for the Paul trap \eqref{paul_result_main_text} and the modulated lattice \eqref{result_bessel_main_text}, where the first non-trivial terms were found to be \emph{second-order} in the period $T$, see Sections \ref{section-paul_trap}-\ref{section:modulated_lattice}.

\subsection{Outline of the paper}\label{section:outline}
The following of the text is structured as follows:
\begin{itemize}
\item {\blue {\bf Section \ref{section-two}}} presents the general formalism used to treat time-dependent Hamiltonians. The method is then applied to the simple two-step modulation introduced in Eq. \eqref{trivial_sequence}.
\item {\blue {\bf Section \ref{section:abrupt}}} illustrates the impact of the initial phase of the modulation on long-time dynamics, based on a simple example. This Section highlights the importance of the ``kick" operator $\hat K (t)$ introduced in Section \ref{section-two}.
\item {\blue {\bf Section \ref{section-three}}} derives useful formulas for the effective Hamiltonian and kick operators in the general case of $N$-step modulations ($N \in \mathbb{Z}$).
\item {\blue {\bf Section \ref{section:4phase}}} explores two specific classes of modulations, characterized by $N\!=\!4$ different steps. 
\item {\blue {\bf Section \ref{section_four}}} applies the latter results to a modulation generating an effective magnetic field in two-dimensional systems. This sequence is explored both in the absence and in the presence of a lattice.
\item {\blue {\bf Section \ref{section_five}}} proposes and explores several driving sequences realizing effective spin-orbit couplings in two-dimensional spin-1/2 systems. These sequences are also analyzed in the absence and in the presence of a lattice.
\item {\blue {\bf Section \ref{section:discussion}}} is dedicated to general discussions and conclusions. This final part analyses the convergence of the perturbative approach introduced in Section \ref{section-two}. It also briefly discusses the possibility to launch the modulation adiabatically. Finally, we present concluding remarks and outlooks.
\end{itemize}

\section{The notion of effective Hamiltonians: \\ using a reliable approach}
\label{section-two}

\subsection{The formalism}\label{section:formalism}

Having identified the subtleties proper to the analysis based on the BCH-Trotter formula  in Section \ref{section:Trotter_ambiguity}, we now consider an alternative approach inspired by Ref. \cite{Rahav}. Let us first rephrase the general problem based on our previous analysis. We act on an initial state $\vert \psi_0\rangle$ with a time-periodic Hamiltonian 
\begin{align}
&\hat H(t)=\hat H_0 + \hat V(t) , \label{ham_series}\\
&\hat V(t)=  \sum_{j=1}^{\infty} \hat V^{(j)} e^{i j \omega t} + \hat V^{(-j)} e^{-i j \omega t}, \label{ham_series_tri}
\end{align}
between times $t_i$ and $t_f$, the period of the driving being $T=2 \pi / \omega$. In Eq. \eqref{ham_series_tri}, we explicitly Fourier expand the time-dependent potential, to take higher harmonics into account. There are three distinct notions:
\begin{enumerate}
 \item The initial phase of the Hamiltonian at time $t_i$ (i.e. $\omega t_i$ mod $2\pi$): the way the driving starts, namely $\hat V(t_i)$, may have an important impact on the dynamics;
\item The evolution of the system between the interval $\Delta t=t_f-t_i$, which can be arbitrary long, and during which the Hamiltonian $\hat H(t)$ is applied;
\item The final phase of the Hamiltonian at time $t_f$ (i.e. $\omega t_f$ modulo $2\pi$): this final step describes the micro-motion.
\end{enumerate}
These concepts were illustrated in Eq. \eqref{decomposition}, for the simple two-step sequence \eqref{trivial_sequence}  presented in Section \ref{section:Trotter_ambiguity}. In order to separate these three effects in a clear manner, we generalize the approach of Ref. \cite{Rahav} and re-express the evolution operator as
\begin{align}
\hat U(t_i \to t_f )&= \hat{\mathcal{U}}^{\dagger} (t_f) e^{-i (t_f-t_i)\hat H_{\rm eff}} \hat{\mathcal{U}} (t_i),\notag \\
&= e^{-i\hat K(t_f)}e^{- i (t_f-t_i)\hat H_{\rm eff}}e^{i\hat K(t_i)}
\label{rahav}
\end{align}
where we impose that: 
\begin{itemize}
\item $\hat H_{\rm eff}$ is a time-independent operator;
\item $\hat K(t)$ is a time-periodic operator, $\hat K(t+T)= \hat K(t)$, with zero average over one period;
\item $\hat H_{\rm eff}$ does not depend on the starting time $t_i$, which can be realized by transferring all undesired terms into the ``kick" operator $\hat K(t_i)$. Similarly, $\hat H_{\rm eff}$ does not depend on the final time $t_f$.
\end{itemize}
Following a perturbative expansion in powers of $(1/\omega)$, we obtain [see Appendix \ref{appendix:effective}]
\begin{align}
&\hat H_{\rm eff}=\hat H_0 + \frac{1}{\omega }  \sum_{j=1}^{\infty} \frac{1}{j } [\hat V^{(j)} , \hat V^{(-j)} ] \label{effective_ham} \\
&+ \frac{1}{2 \omega^2} \sum_{j=1}^{\infty} \frac{1}{j^2} \left ( [[\hat V^{(j)},\hat H_0],\hat V^{(-j)} ] + [[\hat V^{(-j)},\hat H_0],\hat V^{(j)} ] \right ) \!+\! \mathcal{O} (T^3), \notag \\
&\hat K(t)\!=\! \int^t \hat V(\tau) \text{d} \tau + \mathcal{O} (T^2) \!=\!  \sum_{j \ne 0}  \frac{1}{i j \omega }  \hat V^{(j)} e^{i j \omega t} + \mathcal{O} (T^2). \label{F-operator}
\end{align}
In Eq. \eqref{effective_ham}, we have omitted the second-order terms that mix different harmonics, noting that these terms do not contribute in the situations presented in this work; the complete second-order terms contained in $\hat H_{\rm eff}$ and $\hat K(t)$ are presented in the Appendix \ref{appendix:effective}, see Eqs. \eqref{eq:general_mixing}-\eqref{eq:general_mixing_kick}. By construction, and in contrast to the Trotter approach, the expressions \eqref{effective_ham}-\eqref{F-operator} constitute a strong basis to evaluate the relevance of periodic-driving schemes in view of realizing non-trivial and robust effects, such as non-zero effective magnetic fields. 

We conclude this Section by pointing out that effective Hamiltonians can also be obtained through Floquet theory \cite{Maricq}. However, as apparent in  Refs. \cite{Maricq,Grozdanov:1988}, there is  \emph{a priori} no natural constraint within Floquet theory that prevents the ``Floquet" effective Hamiltonians to contain $t_i$-dependent terms. A possible way to get rid of these terms in Floquet theory is to consider an adiabatic launching of the driving \cite{Eckardt:2007,Eckardt:2009}, such that the evolving state is constrained to remain in the same (principal) quasienergy multiplicity at all times (the multiplicity being well separated by the large energy $\hbar \omega$). However, for the sake of generality and clarity, we will follow here the approach based on the partitionment  \eqref{rahav} discussed in this Section, which provides an unambiguous and physically relevant definition for the effective Hamiltonian. Finally, we point out that the convergence of the perturbative expansion in powers of $(1/\omega)$  and leading to Eq. \eqref{effective_ham} is by no means guaranteed, as will be discussed later in Section \ref{section:convergence}.

\subsection{Illustration of the formalism: back to the two-step sequence}\label{back_to_N2}

As a first illustration, let us apply the expressions \eqref{effective_ham}-\eqref{F-operator} to the simple two-step sequence in Eq. \eqref{trivial_sequence}. The Hamiltonian is given by $\hat H(t) = \hat H_0 + f(t) \hat V$, where $f(t)$ is the standard square-wave function. Expanding $f(t)$ into its Fourier components, we obtain a simple expression for the $\hat V^{(j)}$ operators introduced in Eq. \eqref{ham_series}. A direct evaluation of Eqs. \eqref{effective_ham}-\eqref{F-operator} then implies [see Appendix \ref{appendix:two_phase}]
\begin{align}
&\hat H_{\text{eff}}=\hat H_0 + \frac{\pi^2}{24 \omega^2} [[\hat V, \hat H_0],\hat V] + \mathcal{O} (1/\omega^3) , \label{eq:N=2}  \\
& \hat K(t)= - \frac{\pi}{2 \omega} \hat V + \vert t \vert \hat V + \mathcal{O} (1/\omega^2) \text{ , for $t \in \left [- \frac{T}{2} , \frac{T}{2} \right]$ },\label{eq:N=2_kick}
\end{align}
such that the evolution operator is given by [Eq. \eqref{rahav}]
\begin{equation}
\hat U(0 \to t )= e^{-i\hat K(t)}e^{-i t  [ \hat H_0 + \mathcal{O} (1/\omega^2)]} e^{-iT  \hat V/4},
\label{}
\end{equation}
in agreement with Eq. \eqref{decomposition}. Note that the amplitude of the  initial kick $\hat K(t_i)$ is maximal at the initial time $t_i=0$, and that it is zero at time $t_i=T/4$. In contrast with the Trotter analysis, the approach based on the partitionment \eqref{rahav} directly identifies: (a) the absence of first-order term in the effective Hamiltonian $\hat H_{\text{eff}}$, (b) the initial kick produced by the operator $\hat S=\text{exp} (i\hat K(0))=\text{exp} (-iT \hat V/4 )$, and (c) the micro-motion $\text{exp} (-i\hat K(t))$. We note that since the latter operator satisfies $\text{exp} (-i\hat K(NT))\!=\!\text{exp} (-i\hat K(0))\!=\!\text{exp} (iT \hat V/4 )\!=\!\hat S^{\dagger}$, we exactly recover Eq. \eqref{decomposition} for $t=NT$.

The result in Eq. \eqref{eq:N=2}, which is associated with the two-step sequence \eqref{trivial_sequence}, is to be compared with the smooth driving considered in Sections \ref{section-paul_trap}-\ref{section:modulated_lattice},
\be
\hat H (t)=\hat H_0 + \hat V(t)= \hat H_0 + \hat V \cos (\omega t) ,\notag
\ee 
which is readily treated using Eq. \eqref{effective_ham}. Setting $V^{(1)}=V^{(-1)}=\hat V/2$ yields
\begin{align}
&\hat H_{\text{eff}}=\hat H_0 + \frac{1}{4 \omega^2} [[\hat V, \hat H_0],\hat V] + \mathcal{O} (1/\omega^3), \label{eq:N=2_smooth}\\
& \hat K(t)= \left ( \sin (\omega t) / \omega \right ) \hat V + \mathcal{O} (1/\omega^2), \label{eq:N=2_kick_smooth}
\end{align}
which is indeed qualitatively equivalent to Eqs. \eqref{eq:N=2}-\eqref{eq:N=2_kick}. \\

We now apply the formalism to the two examples presented in Sections \ref{section-paul_trap}-\ref{section:modulated_lattice}: \\ 

%\subsection{Back to the Paul trap and modulated optical lattice}

\paragraph{The Paul trap } We readily recover the effective Hamiltonian in Eq. \eqref{paul_result_main_text} by inserting the operators defined in Eq. \eqref{paul_operators} into Eq. \eqref{eq:N=2_smooth}. Furthermore, Eq. \eqref{eq:N=2_kick_smooth} provides an approximate expression for the micro-motion underlying the slow motion in the Paul trap [see also Appendix \ref{appendix:paul_trap} for more details].

%into Eqs. \eqref{eq:N=2_smooth}-\eqref{eq:N=2_kick_smooth} yields
%\begin{align}
%&\hat H_{\text{eff}}=\hat H_0 + \frac{1}{2} m \Omega^2 \hat x^2, \notag \\
%&\hat K (t)= m \omega_0^2 \sin (\omega t) \hat x^2 / 2 \omega ,\notag
%\end{align}
%where we indeed recover the effective Hamiltonian in Eq. \eqref{paul_result_main_text}.

\paragraph{The modulated lattice} Inserting the operators defined in Eq. \eqref{modulated_lattice_operators} into Eqs. \eqref{eq:N=2_smooth}-\eqref{eq:N=2_kick_smooth} yields
\begin{align}
&\hat H_{\text{eff}}= \left ( 1 - \frac{\kappa^2}{4 \omega^2} \right ) \hat H_0 + \mathcal{O} (1/\omega^3), \label{mod_lat_dev} \\
&\hat K (t)= (\kappa / \omega) \hat x \sin (\omega t) + \mathcal{O} (1/\omega^2),\notag
\end{align}
where we indeed recover the first terms of the Bessel function expansion in Eq. \eqref{result_bessel_main_text}. We note that the maximal amplitude of the kick associated with the micro-motion is given by $\exp [-i \hat K(T/4)]= \exp [-i  (\kappa / \omega) \hat x]$, which corresponds to a translation in the Brillouin zone by an amount $\Delta k=\kappa /\omega$. We thus recover the fact that the modification of the hopping rate becomes appreciable when the micro-motion is comparable to the width of the Brillouin zone, $\Delta k \approx \pi$: indeed $\mathcal{J}_0[(\kappa/\omega)=\pi] \approx -1/3$ is very close to the minimal value of the Bessel function, and thus corresponds to a dramatic change in the tunneling rate (the maximal value of the Bessel function is $\mathcal{J}_0(0)=1$, which corresponds to the standard hopping rate in the absence of shaking). 

The result in Eq. \eqref{mod_lat_dev} stems from the perturbative expansion in powers of $(1/ \omega)$. However, the formalism presented in Section \ref{section:formalism} also allows for an \emph{exact} treatment of the modulated-lattice problem. The full derivation is given in Appendix \ref{appendix:bessel}, where we recover the Bessel-renormalized-hopping result of Eq. \eqref{result_bessel_main_text}. Moreover, this derivation provides an exact form for the kick operator, see Eq. \eqref{kick_operator_exact_modulated}. The latter expression indicates that the maximal amplitude of the kick operator is given by $\hat K = \hat x \kappa / \omega $, which is precisely the result discussed above based on the perturbative treatment. 

%\section{Various ways to launch the driving: abrupt launch and two adiabatic protocols}
%\subsection{Abruptly launching the driving}\label{section:abrupt}

\section{Launching the driving: illustration of the initial kick}\label{section:abrupt}

In the last Section, we introduced the partitionment of the evolution operator 
\begin{align}
\hat U(t_i \to t)= e^{-i\hat K(t)}e^{- i (t-t_i) \hat H_{\rm eff}}e^{i\hat K(t_i)},
\label{rahav_encore}
\end{align}
which highlights the fact that the system undergoes an initial kick $\exp [i\hat K(t_i)]$ before evolving according to the (time-independent) effective Hamiltonian. This initial kick depends on the launching time of the sequence $t_i$, through Eq. \eqref{F-operator}, and it can have a great impact on long-time dynamics. It is the aim of this Section to illustrate this effect, based on a basic but enlightening example.

Consider a particle driven by a uniform force that alternates its sign in a pulsed manner: $+F,-F, \dots$ This system obeys the two-step sequence in Eq. \eqref{trivial_sequence} with $\hat H_0=\hat p^2/2m$, $\hat V=-F \hat x$. The effective Hamiltonian and kick operators are readily obtained through Eqs. \eqref{eq:N=2}-\eqref{eq:N=2_kick}, yielding
\begin{align}
&\hat H_{\text{eff}}=\hat p^2/2m + \text{cst}, \label{eq:simple_example}  \\
& \hat K(t)= (F T/4) \hat x  - \vert t \vert F \hat x + \mathcal{O} (1/\omega^2) \text{ , for $t \in \left [- \frac{T}{2} , \frac{T}{2} \right]$ }.\label{eq:N=2_kick_2}
\end{align}
The initial kick operator is $\hat K(t_i)= \hat x FT/4 $ for $t_i\!=\!0$, $\hat K(t_i)= - \hat x FT/4 $ for $t_i\!=\!T/2$, and its effect is thus to modify the initial mean velocity $v(t_i)\rightarrow v(t_i) \pm FT/4m$ before the long-time free evolution. We emphasize that the initial kick operator is zero when starting the sequence at time $t_i\!=\!\pm T/4$.

Further insight is provided by computing the evolution operator \emph{exactly}, at time $t=T$. This reads  
\be
\hat U(T)\!=\!\exp \! \left [ -i T \!\left ( \frac{1}{2m} \left ( \hat p + \mathcal{A}(t_i) \right )^2 \text{+ cst}  \right ) \right ] \!=\! \exp  \left [ -i T \tilde H_{\text{eff}}(t_i)  \right ] ,\notag
\ee
where the gauge potential is given by $\mathcal{A}(t_i)\!=\! \pm FT/4$ whether the starting time is $t_i\!=\!0$ or $t_i\!=\!T/2$, respectively. We note that the effective Hamiltonian $\tilde H_{\text{eff}}(t_i)$ is related to the $t_i$-\emph{independent} effective Hamiltonian $\hat H_{\text{eff}}$ in Eq. \eqref{eq:simple_example} through the unitary (gauge) transformation involving the kick operator $\exp [i \hat K (0)]=\exp [i \mathcal{A} \hat x]$, in agreement with Eq. \eqref{rahav_encore}. 

\begin{figure}[h!]% Figure 1
\centering
\includegraphics[width=8cm]{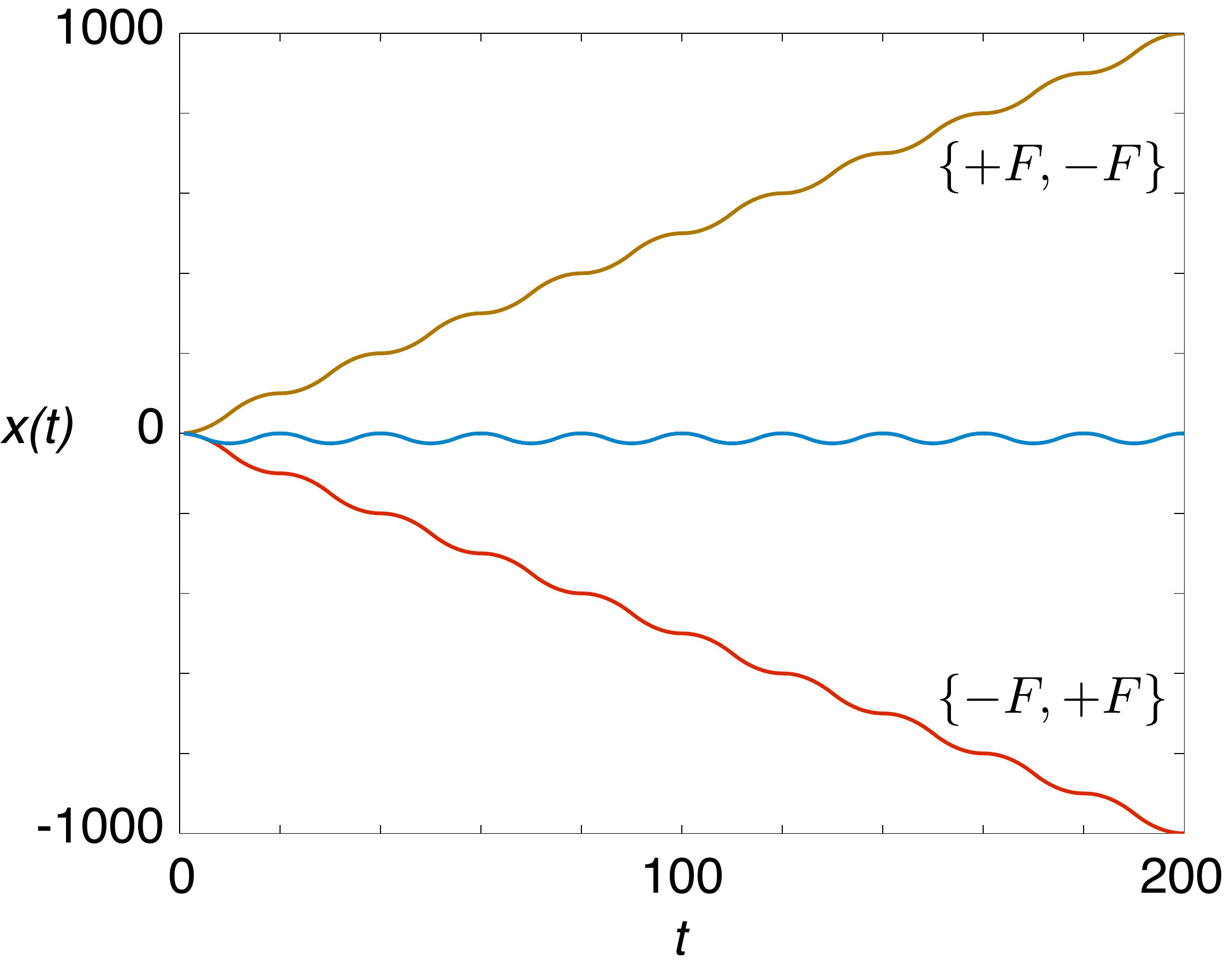}
\caption{\label{fig-1}  Sensitivity to the initial phase of the driving.  (brown) Classical dynamics of a particle driven by a repeating pulse sequence $\{ +F , - F \}$, where $F>0$ is a uniform force; the period is $T=20$ and the particle is initially at rest $v(t_i)=0$ at $x(t_i)=0$. (red) Same but considering the sequence $\{ -F , + F \}$, i.e. shifting the starting time $t_i \rightarrow t_i + T/2$. (blue) Same but starting the sequence with $-F$ during a time $T/4$, and then repeating the sequence $\{ +F , - F \}$, i.e. shifting the starting time $t_i \rightarrow t_i - T/4$. Note that the initial kick $\hat K(t_i)$ is inhibited in the latter case, while it is opposite in the two former cases [Eq. \eqref{eq:N=2_kick_2}]. In all the plots, the particle depicts a small micro-motion,  captured by $\hat K(t)$ in Eq. \eqref{eq:N=2_kick_2}. The brown and red curves  highlight the great sensitivity to the initial phase of the modulation}.
\end{figure}

Treating the basic mechanics problem associated with $\tilde H_{\text{eff}}(t_i)$  semi-classically, we recover that the initial kick modifies the initial velocity $v(t_i) \rightarrow v(t_i) + \mathcal{A}/m$. Hence, for an arbitrary $v(t_i)$, the position of the particle $x(t \gg T)$ will significantly depend on whether the pulse sequence has started with a pulse $+ \hat V$ [$t_i=0$] or $- \hat V$ [$t_i=T/2$]: the dynamics is strongly sensitive to the initial phase of the modulation. In Fig. \ref{fig-1}, we compare these predictions to the real (classical) dynamics of the pulse-driven particle. This figure illustrates the sensitivity to the initial phase of the driving, but also, the effects of micro-motion present at all times for all configurations. \\

 In general, the effects related to the initial phase of the modulation may remain important in more sophisticated driven systems.  The sensitivity to the initial phase will be further illustrated in the context of driven-induced Rashba spin-orbit couplings in Section \ref{section_five} [see Fig. \ref{fig-3}].
   \\

\section{Multistep sequences}
\label{section-three}

\begin{figure}[h!]% Figure 2
\centering
\includegraphics[width=9cm]{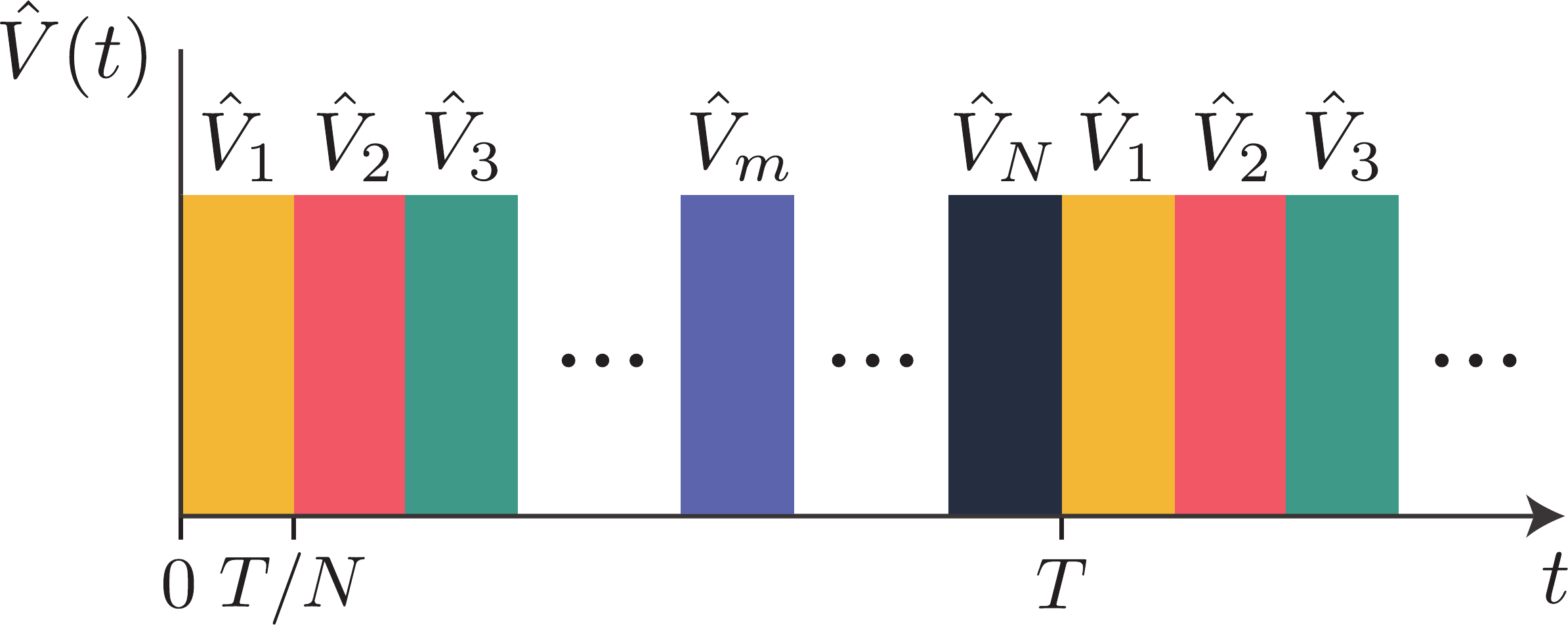}
\caption{\label{fig-gammaN}  Schematic representation of the general $N$-step driving sequence $\gamma_N$ introduced in Eqs. \eqref{general_pulse_sequence}-\eqref{fourier}, where $\hat H(t)=\hat H_0 + \hat V (t)$}.
\end{figure}

Going beyond the two-step driving sequence in Eq. \eqref{trivial_sequence} potentially  increases the possibility to engineer interesting effective Hamiltonians and gauge structures. We now derive the effective Hamiltonian for the general situation where the pulse sequence is characterized by the repeated $N$-step sequence
\be
\gamma_N=\{ \hat H_0 +\hat V_1 , \hat H_0 +\hat V_2, \hat H_0 +\hat V_3, \dots , \hat H_0 +\hat V_N \} , \label{general_pulse_sequence} 
\ee
where the $\hat V_m$'s are arbitrary operators. In the following, we consider that the duration of each step is $\tau=T/N$, where $T$ is the driving period, and we further impose that $\sum_{m=1}^N \hat V_m =0$.  The general driving sequence $\gamma_N$ is illustrated in Fig. \ref{fig-gammaN}. Note that $\gamma_N$ reduces to the simple sequence \eqref{trivial_sequence} of Section \ref{section:Trotter_ambiguity} for $N=2$ and $\hat V_1=-\hat V_2=\hat V$. 

The general Hamiltonian corresponding to the pulse sequence \eqref{general_pulse_sequence} is written as
\begin{align}
&\hat H (t)= \hat H_0 + \sum_{m=1}^{N} f_m(t) \hat V_m, \label{fourier}  \\
&f_m(t)  = 1 \, \text{ for } (m-1)T/N \le t \le mT/N \text{ , \, otherwise zero.} \notag \\
&\quad \quad \, \,= \frac{1}{2 \pi i} \sum_{n \ne 0} \frac{1}{n} e^{-i 2 \pi n m/N} \left (  e^{i n (2 \pi/N)}  -1 \right ) e^{i n \omega  t},  \notag
\end{align}
where the last line of Eq. \eqref{fourier} provides the Fourier series of the square functions $f_m(t)$. In order to apply Eqs. \eqref{effective_ham} and \eqref{F-operator}, we expand the Hamiltonian in terms of the harmonics
\begin{align}
&\hat H (t)= \hat H_0 + \sum_{j \ne 0} \hat V^{(j)} e^{i j \omega  t}, \\
& \hat V^{(j)}= \frac{1}{2 \pi i} \sum_{m=1}^{N} \frac{1}{j} e^{-i 2 \pi j m/N} \left (  e^{i j (2 \pi/N)}  -1 \right ) \hat V_m ,
\end{align}
where we used the Fourier series in Eq. \eqref{fourier}. The effective Hamiltonian and the initial-kick operator $\hat K (0)$ are then given by the general expressions [see Appendix \ref{appendix:Npulse}]
\begin{widetext}
\begin{align}
&\hat H_{\text{eff}}= \hat H_0+ \frac{2 \pi i}{N^3  \omega} \sum_{m<n=2}^{N} \mathcal{C}_{m,n} \, [\hat V_m , \hat V_n]\,  + \frac{\pi^2 (N-1)^2}{6 N^4  \omega^2} \sum_{m=1}^{N} [[\hat V_m , \hat H_0],\hat V_m] \notag \\
&\, \qquad \qquad + \frac{\pi^2}{6 N^4  \omega^2} \sum_{m<n=2}^{N} \mathcal{D}_{m,n} \left ( [[\hat V_m , \hat H_0],\hat V_n] + [[\hat V_n , \hat H_0],\hat V_m] \right )\, + \mathcal{O} (1/\omega^3) , \notag \\
& \hat K (0)= \frac{2 \pi}{N^2  \omega} \sum_{m=1}^{N} \hat V_m \, m + \mathcal{O} (1/\omega^2). 
\label{general_result}
\end{align}
where $\mathcal{C}_{m,n} =   \frac{N}{2} +m -n  $ and $\mathcal{D}_{m,n} =1+ N^2 - 6 N (n-m) +6 (n-m)^2  $.
\end{widetext}
%\begin{widetext}
%\begin{align}
%&\hat H_{\text{eff}}= \hat H_0+ \frac{2 \pi i}{N^3  \omega} \sum_{m<n=2}^{N} [\hat V_m , \hat V_n]\,  \left ( \frac{N}{2} +m -n  \right)    , \notag \\
%& \hat K (0)= \frac{2 \pi}{N^2  \omega} \sum_{m=1}^{N} \hat V_m \, m, 
%\label{general_result}
%\end{align}
%\end{widetext}
We have again omitted the harmonic-mixing terms given in Eq. \eqref{eq:general_mixing}, which do not contribute for the sequences considered here. The result in Eq. \eqref{general_result} clearly highlights the fact that the initial kick $\hat K(0)$ depends on the way the pulse sequence starts, whereas the effective Hamiltonian $\hat H_{\text{eff}}$ is independent of this choice: shifting the pulse sequence, namely redefining the operators $\hat V_m \rightarrow \hat V_{m+p}$, with $p \in \mathbb{Z}$, results in a change in $\hat K(0)$ but leaves $\hat H_{\text{eff}}$ invariant. 

We have applied the formula \eqref{general_result} to general sequences with $N\!=\!3$ and $N\!=\!4$ different steps, and we present the associated results in Eqs. \eqref{ham_eff_3_pulse}-\eqref{ham_eff_4_pulse} in Appendix \ref{appendix:differentN}.  In contrast with the case $N\!=\!2$, we find that sequences with $N\!=\!3$ or $N\!=\!4$ steps can potentially lead to non-trivial effects that are \emph{first order} in $(1/ \omega)$. We point out that the scheme proposed by Kitagawa \emph{et al.} \cite{Kitagawa} to realize the Haldane model using a modulated honeycomb lattice corresponds to the case $N=3$.  Moreover, the model of Refs. \cite{Rudner2013,Reichl:2014}, which features  topological ``Floquet" edge states, corresponds to the case $N=5$. In the following of this work, we will further explore and illustrate the case $N=4$, with a view to creating synthetic magnetic fields and spin-orbit couplings with cold atoms (see also \cite{Sorensen,Ueda,Anderson}).

\section{Sequences with $N=4$ steps}\label{section:4phase}

Motivated by the importance and versatility of four-step sequences to generate non-trivial effective potentials and gauge structures, we now explore two specific examples of such sequences that lead to different effects. The following paragraphs will constitute a useful guide for the applications presented in Sections \ref{section_four} and \ref{section_five}.

\subsection{The class of sequences  $\alpha$}\label{section:alpha}

Let us first consider the following four-step sequence
\begin{equation}
\alpha:\, \{ \hat H_0+ \hat A, \hat H_0+ \hat B, \hat H_0-\hat A,\hat H_0- \hat B \},\label{alpha_sequence}
\end{equation}
which corresponds to $\gamma_4$ in Eq. \eqref{general_pulse_sequence} with $\hat V_1 = -\hat V_3= \hat A$ and $\hat V_2 = -\hat V_4= \hat B$. We now apply the formula  \eqref{general_result} [see also Eq. \eqref{ham_eff_4_pulse}] and we obtain 
\begin{align}
&\hat H_{\text{eff}}\!=\! \hat H_0 \!+\! \frac{i \pi}{8   \omega} [\hat A, \hat B] \!+\! \frac{\pi^2}{48  \omega ^2} \! \left (  [[ \hat A, \hat H_0], \hat A] \!+\!  [[ \hat B, \hat H_0], \hat B] \right ) \!+\! \mathcal{O} (1/ \omega ^3), \notag \\
&\hat K(0) = - \frac{\pi}{4   \omega} (  \hat A + \hat B) + \mathcal{O} (1/ \omega ^2), \label{good_eff_ham}
\end{align}
where the expression for $\hat K(0)$ was obtained for a sequence $\alpha$ starting with the pulse $+ \hat A$ (and we remind that the kick operator $\hat K (0)$ depends on this choice). The result in Eq. \eqref{good_eff_ham} shows that the driving schemes belonging to the class $\alpha$ can generate a combination of \emph{first-order} and \emph{second-order} terms, which might potentially lead to interesting observable effects, see Sections \ref{section_four} and \ref{section_five}. In particular, we note that the scheme of Ref. \cite{Sorensen} to generate synthetic magnetic flux in optical lattices belongs to this class, see Section \ref{section_four}.

Finally, we note that the pulse sequence $\alpha$ can be approximated by the smooth driving $\hat V(t)= \hat A \cos (\omega t) + \hat B \sin (\omega t) $. In this case, a direct evaluation of Eq. \eqref{effective_ham} yields
\begin{align}
&\hat H_{\text{eff}}\!=\! \hat H_0 \!+\! \frac{i}{2   \omega} [\hat A, \hat B] \!+\! \frac{1}{4 \omega ^2}\! \left (  [[ \hat A, \hat H_0], \hat A] \!+\!  [[ \hat B, \hat H_0], \hat B] \right ) \!+\! \mathcal{O} (1/ \omega ^3), \notag
\end{align}
which is indeed approximatively equal to the effective Hamiltonian \eqref{good_eff_ham} associated with the pulsed system.

\subsection{The class of  sequences $\beta$}\label{section_beta}

We now consider an apparently similar four-step sequence
\begin{equation}
\beta:\, \{ \hat H_0+ \hat A, \hat H_0- \hat A, \hat H_0+ \hat B,\hat H_0- \hat B \},
\label{beta_sequence}
\end{equation}
which corresponds to $\gamma_4$ in Eq. \eqref{general_pulse_sequence} with $\hat V_1 = -\hat V_2= \hat A$ and $\hat V_3 = -\hat V_4= \hat B$. Applying the formula  \eqref{general_result} [Eq. \eqref{ham_eff_4_pulse}] yields
\begin{align}
&\hat H_{\text{eff}}= \hat H_0 \!+\! \frac{5 \pi^2}{384 \omega ^2}\! \left (  [[ \hat A, \hat H_0], \hat A] \!+\!  [[ \hat B, \hat H_0], \hat B] \right ) \notag \\
&\quad \quad \quad \quad \, \!-\! \frac{\pi^2}{128 \omega ^2}\! \left (  [[ \hat A, \hat H_0], \hat B] \!+\!  [[ \hat B, \hat H_0], \hat A] \right ) \!+\! \mathcal{O} (1/ \omega ^3) , \notag \\
&\hat K(0) = - \frac{\pi}{8   \omega} (\hat A+ \hat B) + \mathcal{O} (1/ \omega ^2), \label{beta_result} 
\end{align}
where $\hat K(0)$ corresponds to a sequence $\beta$ starting with the pulse $+\hat A$. The result in Eq. \eqref{beta_result} emphasizes two major differences between the $\alpha$ and the $\beta$ classes: (a) the effective Hamiltonian associated with the $\beta$ sequence does not contain any first-order term, and in this sense, this $N\!=\!4$ sequence resembles the case $N\!=\!2$; (b) the $\beta$ sequence generates additional second-order terms that mix the pulsed operators $\hat A$ and $\hat B$, i.e. terms of the form $ [[ \hat A, \hat H_0], \hat B]$, which are not present in the class $\alpha$ [Eq. \eqref{good_eff_ham}].

As illustrated in Section \ref{section_five}, the schemes of Refs. \cite{Anderson,Ueda} to generate spin-orbit couplings in cold gases can be expressed in the form of sequences $\beta$.

\section{Physical illustration: generating synthetic magnetic fields in cold gases}
\label{section_four}

In this Section, we apply the results of Section \ref{section:4phase} to generate synthetic magnetic fields in one-component atomic gases, using a four-step sequence of type $\alpha$, see Eq. \eqref{alpha_sequence}. 
%The corresponding effective Hamiltonian is given by Eq. \eqref{good_eff_ham}.

\subsection{Without a lattice}\label{sorensen_no_lattice}

We first consider atoms moving in two-dimensional free space, such that $\hat H_0=(\hat p_x^2+ \hat p_y^2)/2m$.  Inspired by Ref. \cite{Sorensen}, we drive the system with a pulse sequence $\alpha$, see Eq. \eqref{alpha_sequence}, with the operators $\hat A=(\hat p_x^2- \hat p_y^2)/2m$ and $\hat B= \kappa \hat x \hat y$. We will comment later on the possibility to implement such a scheme practically. Over a period, the evolution is thus given by the  sequence
\be
\left \{   \frac{\hat p^2_x}{m} \, ,  \,  \frac{\hat p^2_x+\hat p^2_y}{2m} + \kappa \hat x\hat y  \, ,  \, \frac{\hat p^2_y}{m}  \, ,  \, \frac{\hat p^2_x+\hat p^2_y}{2m} - \kappa \hat x \hat y  \right \}, \label{bulk_sorensen}
\ee
which consists in repeatedly allowing for the movement in a pulsed and directional manner, while subjecting the cloud to an alternating quadrupolar field. The corresponding effective Hamiltonian is given by Eq. \eqref{good_eff_ham}, which yields, up to second order $(1/\omega^2)$,
\begin{align}
&\hat H_{\text{eff}} = \frac{1}{2m} \left ( \left ( \hat p_x - \mathcal A_x \right)^2 + \left ( \hat p_y - \mathcal A_y  \right) ^2   \right ) + \frac{1}{2} m \omega_h^2 (\hat x^2 + \hat y^2) \notag \\
&\bs{\mathcal{A}} = (-m \Omega \hat y,  m \Omega \hat x) , \quad \Omega=\frac{\pi \kappa}{8 m \omega} , \quad \omega_h=\sqrt{\frac{5}{3}} \Omega,\label{bulk_eff_ham}
\end{align}
which corresponds to the realization of a perpendicular and uniform synthetic magnetic field 
\be
\bs B = 2 m \Omega  \bs 1_z=\frac{\pi \kappa}{4 \omega} \bs 1_z. \label{magnetic_field_bulk}
\ee
We point out that the second-order corrections in Eq. \eqref{good_eff_ham} lead to a harmonic confinement, which dominates over the centrifugal force generated by the first-order term, resulting in an overall trapping potential in Eq. \eqref{bulk_eff_ham}. Defining the cyclotron frequency $\omega_c=B/ m$, we obtain the ratio between the confinement and cyclotron frequencies
\be
\omega_h/\omega_c= (1/2) \sqrt{5/3} \approx 0.64.\label{bulk_cyclotron}
\ee
The induced confinement is a special feature of the driving scheme \eqref{bulk_sorensen}. It has a significant impact on the dynamics, which illustrates the fact that the perturbative expansion in $(1/ \omega)$ should not be limited, in typical applications, to its first-order terms [see Section \ref{section:convergence} for a more detailed discussion].  

\subsection{With optical lattices}\label{sorensen_lattice}

A similar scheme can be applied to cold atoms in optical lattices, where a uniform synthetic magnetic field would provide a platform to simulate the Hofstadter model \cite{Hofstadter:1976,Jaksch:2003,Gerbier:2010,Aidelsburger:2013,Miyake:2013}. Here, we suppose that the atoms evolve within a two-dimensional optical square lattice and that their dynamics is well captured by a single-band tight-binding description. The static Hamiltonian is thus taken in the form
\begin{align}
\hat H_0&= -J \sum_{m,n} \hat a_{m+1,n}^{\dagger} \hat a_{m,n} + \hat a_{m,n+1}^{\dagger} \hat a_{m,n} + \text{h.c.}, \label{TB_ham}\\
&=\bar p_x^2/2m^* + \bar p_y^2/2m^*, \label{TB_ham_bis}
\end{align}
where $J$ is the hopping amplitude, $\hat a_{m,n}^{\dagger}$ creates a particle at lattice site $\bs x = (m a , n a)$, and where $a$ is the lattice spacing. In Eq. \eqref{TB_ham_bis}, we introduced the notation $\bar p_{x,y}^2/2m^*$ to denote hopping along the $(x,y)$ directions, and also, the effective mass $m^*\!=\!1/(2 J a^2)$. In the following of this work, any operator denoted $\bar O$ will be defined on a lattice, with the convention that $\bar O \rightarrow \hat O$ in the continuum limit [see Appendix \ref{appendix:lattice_operators}].

We now apply the pulse sequence in Eq. \eqref{bulk_sorensen} to the lattice system, by substituting $\hat p_{x,y}^2/2m \rightarrow \bar p_{x,y}^2/2m^*$ and $\hat x \hat y \rightarrow \bar x \bar y$. The lattice analogue of the sequence \eqref{bulk_sorensen}, which was originally introduced in Ref. \cite{Sorensen}, now involves pulsed directional hoppings on a lattice, and it can thus be realized using optical-lattice technologies. In this lattice framework, the effective Hamiltonian in Eq. \eqref{good_eff_ham} yields
\begin{align}
\hat H_{\text{eff}}&= -J \sum_{m,n} \left ( 1 - i \pi \Phi n - \frac{4}{3} (\Phi n \pi)^2   \right ) \hat a_{m+1,n}^{\dagger} \hat a_{m,n} , \notag \\
& +  \left ( 1 + i \pi \Phi m - \frac{4}{3} (\Phi m \pi)^2   \right ) \hat a_{m,n+1}^{\dagger} \hat a_{m,n}  + \text{h.c.}  \label{lattice_sorensen_eff} , 
\end{align}
where we introduced the ``flux" $\Phi\!=\! a^2 \kappa / 8   \omega$, and used the commutators presented in Appendix \ref{appendix:1component}. In the small flux regime $ \Phi \!\ll\! 1$, we obtain the Hofstadter Hamiltonian~\cite{Hofstadter:1976}
\begin{align}
\hat H_{\text{eff}}=& -J  \sum_{m,n} e^{-i \pi \Phi n} \hat a_{m+1,n}^{\dagger} \hat a_{m,n} + e^{i \pi \Phi m} \hat a_{m,n+1}^{\dagger} \hat a_{m,n} + \text{h.c.}  \notag \\
& + \frac{1}{2} m^* \omega_h^2 \left (\bar x^2 + \bar y^2 \right)  \label{hofstadter_eff} , 
%&  + \frac{1}{4} m^* \omega_h^2 a^2  \sum_{m,n} n^2 \hat a_{m+1,n}^{\dagger} \hat a_{m,n} + m^2 \hat a_{m,n+1}^{\dagger} \hat a_{m,n} + \text{h.c.}  \label{hofstadter_eff} , 
\end{align}
where the additional term acts as a harmonic confinement in the continuum limit, with frequency $\omega_h\!=\! 2 \pi \Phi J \sqrt{5/3}$. Noting that $\Phi$ is the number of (synthetic) magnetic flux quanta per unit cell \cite{Hofstadter:1976}, and denoting the cyclotron frequency $\omega_c\!=\!B/m^*$, we recover the free-space results \eqref{magnetic_field_bulk}-\eqref{bulk_cyclotron}, namely
\be
\bs B = (2 \pi  \, \Phi / a^2)   \bs 1_z=\frac{\pi \kappa}{4 \omega} \bs 1_z , \qquad   \omega_h/\omega_c= (1/2) \sqrt{5/ 3}   \label{magnetic_field_lattice},
\ee
which  validates the analogy between the free-space and lattice systems. In the present perturbative framework, the flux is limited to $\Phi \ll 1$; however, a partial resummation of the series \eqref{lattice_sorensen_eff}, similar to that of Section \ref{section:exact}, allows one to extend the flux range to $\Phi \sim 1$, see Ref. \cite{Sorensen}.

\begin{figure}% Figure 2
\centering
\includegraphics[width=8.5cm]{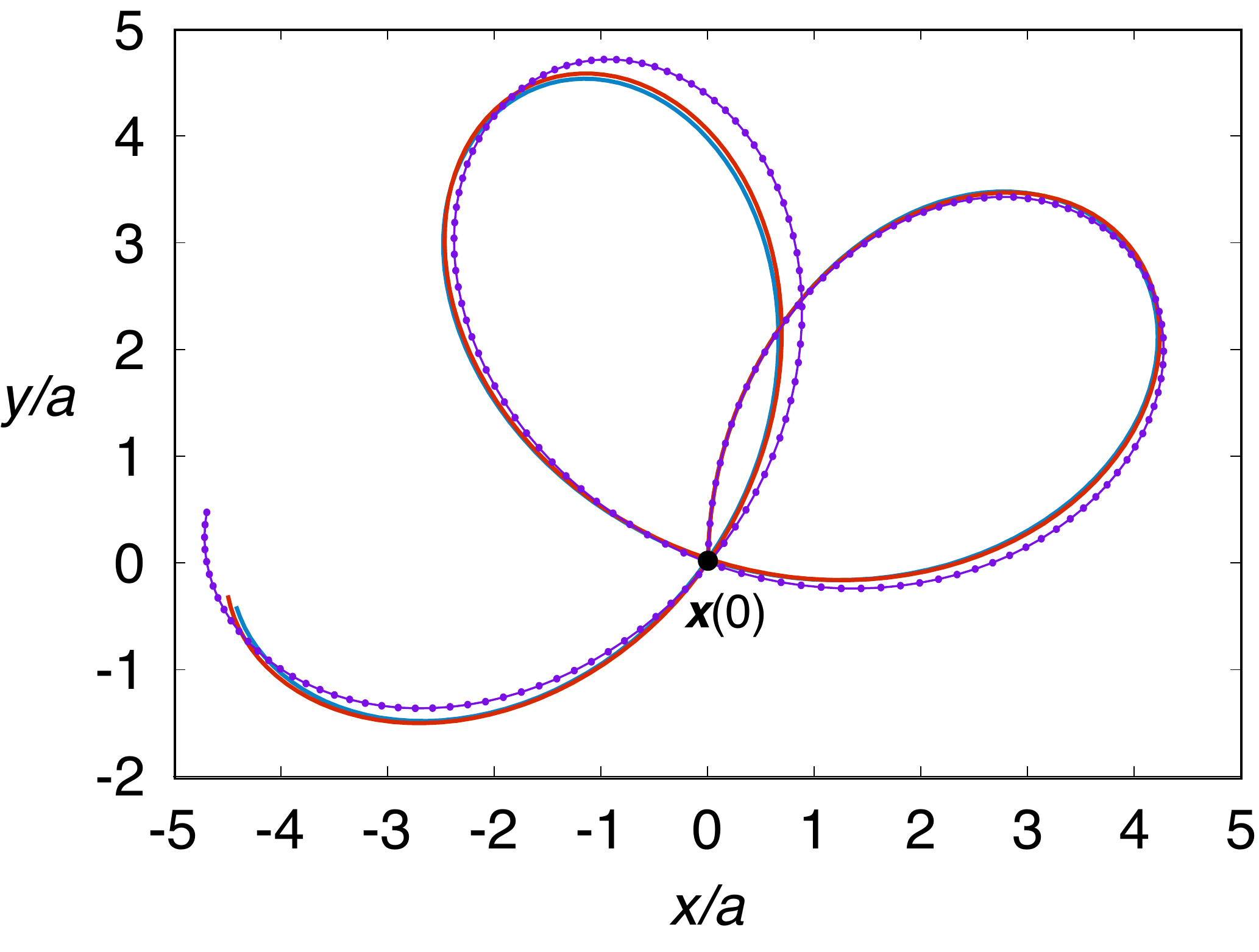}
\caption{\label{fig-2} Comparison between the dynamics of the driven system following the protocol in Eq. \eqref{bulk_sorensen} (red and blue curves), and the dynamics predicted by the effective Hamiltonian \eqref{lattice_sorensen_eff} (purple dotted curve). Shown is the center-of-mass trajectory in the $x-y$ plane for time $t \in [0 , 100] (1/J)$. For all simulations, a gaussian wave packet is initially prepared around $\bs x (0)=0$ with a non-zero group velocity along the $+y$ direction. The blue and red curves correspond to different initial phases of the driving sequence: the blue (resp. red) curve was obtained by starting the sequence \eqref{bulk_sorensen} with $\hat H_0 + \hat A= \bar p^2_x/m^*$ (resp. $\hat H_0 - \hat A=\bar p^2_y/m^*$). We set the values $\kappa=10$ and $T=\pi/80 (1/J)$, such as to fulfill the low-flux condition $\Phi=1/128 \ll 1$. 
}
\end{figure}

The lattice system is convenient for physical implementation, since optical lattices offer a platform to activate and deactivate the hopping terms in a controllable manner, e.g., by simply varying the lattice depths in a directional way. Besides, the lattice configuration described in this Section also allows for direct numerical simulations of the Schrödinger equation (the lattice discretization being physical). We have performed two types of simulations illustrating: (a) the dynamics of a gaussian wave packet subjected to the pulse sequence \eqref{bulk_sorensen}, and (b) the dynamics of the same wave packet evolving according to the effective Hamiltonian \eqref{lattice_sorensen_eff} with $\Phi= a^2 \kappa / 8 \omega$.  In Fig. \ref{fig-2}, we show the center-of-mass dynamics of a wave packet initially prepared around $\bs x (0)=0$ with a non-zero group velocity $\bs v_g (0) = v \bs 1_y$, $v>0$. The non-trivial dynamics associated with the effective Hamiltonian \eqref{lattice_sorensen_eff} is in very good agreement with the real dynamics of the pulsed system. The spirographic and \emph{chiral} motion depicted in Fig. \ref{fig-2}  is well understood in terms of the effective Hofstadter-like Hamiltonian in Eq.\eqref{hofstadter_eff}: it involves the interplay between the cyclotron motion induced by the synthetic magnetic field $\bs B$ and the presence of the harmonic potential with frequency $\omega_h$. Moreover, Fig. \ref{fig-2} compares the trajectories resulting from a different choice of the initial phase of the modulation: the almost identical red and blue trajectories highlight the robustness of this scheme against perturbations in the driving's initial conditions, and incidentally, it shows the negligible role played by the initial kick $\hat K(t_i)$ in this example. Similarly, we also note that the micro-motion is negligible in real space. 

The numerical results presented in Fig. \ref{fig-2} confirm that the driving sequence \eqref{bulk_sorensen}, introduced in Ref. \cite{Sorensen}, produces an effective magnetic field in the optical-lattice setup; moreover, it illustrates the relevant effects associated with second-order $(1/ \omega^2)$ corrections, which will be present even at low flux $\Phi \ll 1$. Moreover, we point out that adding terms in the static Hamiltonian $\hat H_0$, for instance to further control the confinement of the gas or to combine several effects, should be treated with care, as these extra terms will potentially contribute to second-order corrections through the commutators $[[ \hat A, \hat H_0], \hat A] +  [[ \hat B, \hat H_0], \hat B]$, see Eq. \eqref{good_eff_ham}. Finally, we note that the operators $\hat A$ and $\hat B$ entering the sequence in Eq. \eqref{bulk_sorensen} could be slightly modified for the sake of experimental simplicity. 
 
\section{Generating synthetic spin-orbit couplings in cold gases}\label{section_five}

\subsection{Spin-orbit coupling in 2D: general considerations}\label{general_SOC}

In this Section, we investigate the possibility to generate spin-orbit coupling (SOC) terms $\sum_{\mu \nu} \alpha_{\mu \nu} \hat p_{\mu} \hat J_{\nu}$ in a cold-atom gas, where $\hat{\bs J}$ denotes the spin operator associated with the atoms, $\hat{\bs p}$ is the momentum and $\alpha_{\mu \nu}$ are some coefficients. Here, for the sake of simplicity, the focus will be set on the case of two-dimensional and spin-1/2 Rashba SOC, which is given by $\hat H_{\text{R}}= \lambda_{\text{R}} \, \hat{\bs{p}} \cdot \hat{\bs \sigma}$. In the following, we will also encounter another SOC term, $\hat H_{\text{L} \sigma}= \Omega_{\text{SO}} \, \hat L_z \hat \sigma_z$, the so-called ``intrinsic" or ``helical" SOC, which is responsible for the quantum spin Hall effect in topological insulators \cite{Guinea:2010,Kane:2005}. The combination of both terms, e.g. as in the Kane-Mele model \cite{Kane:2005}, will be referred to as the ``helical-Rashba" configuration.

Before presenting different schemes leading to SOC terms, we point out a subtlety associated with Rashba spin-orbit coupled systems, which arises when considering a perturbative treatment in powers of $\lambda_{\text{R}}$. Consider $\hat H_0$ the free Hamiltonian in 2D space, and let us perform the following unitary transformation
\begin{align}
&e^{-i \hat G} \hat H_0 e^{i \hat G}= \hat H_0+ \lambda_{\text{R}} \, \hat{\bs{p}} \cdot \hat{\bs \sigma} + \Omega_{\text{SO}} \hat L_z \hat \sigma_z+ m \lambda_{\text{R}}^2 + \mathcal{O} (\lambda_{\text{R}}^3), \notag \\
& \qquad \qquad \, \, \, \,\,= \hat H_0+ \hat H_{\text{R}} +\hat H_{\text{L} \sigma}+ m \lambda_{\text{R}}^2 + \mathcal{O} (\lambda_{\text{R}}^3), \label{subtle_SOC} \\
&\hat H_0 =\hat p^2/2m, \quad \hat G = m \lambda_{\text{R}} \, \hat{\bs{x}} \cdot \hat{\bs \sigma}, \quad \Omega_{\text{SO}}=m \lambda_{\text{R}}^2, \notag
\end{align}
where we introduced the short notation $\mathcal{O} (\lambda_{\text{R}})\equiv \mathcal{O} (m \lambda_{\text{R}} L)$, and $L$ is the system's length. The result in Eq. \eqref{subtle_SOC} shows that -- up to third order in $\lambda_{\text{R}}$ -- any model combining the Rashba $\hat H_{\text{R}}$ and helical $\hat H_{\text{L} \sigma}$ SOC terms with the weight ratio $\Omega_{\text{SO}}/ \lambda_{\text{R}}=m \lambda_{\text{R}}$, is equivalent to a \emph{trivial} system described by the free Hamiltonian $ \hat H_0$ \cite{Ueda}. This observation highlights the importance of evaluating the SOC terms up to second order in $\lambda_{\text{R}}$, to identify the schemes producing genuinely non-trivial spin-orbit effects. Moreover, in this perturbative framework, Eq. \eqref{subtle_SOC} indicates that the (first-order) Rashba SOC Hamiltonian $\hat H \!=\! \hat H_0  \!+\! \hat H_{\text{R}}$ is \emph{equivalent} to the (second-order) helical SOC Hamiltonian $\hat H \!=\! \hat H_0  \!+\! \hat H_{\text{L} \sigma}$, with $\Omega_{\text{SO}}\!=\!m \lambda_{\text{R}}^2$. 

Finally, we point out that the expansion in powers of $m \lambda_{\text{R}} L$ introduced in Eq. \eqref{subtle_SOC} should be handled with care for two main reasons. First, it should not be mistaken with the $(1/\omega)$ expansion stemming from the effective-Hamiltonian formalism of Section \ref{section:formalism}, which is typically  characterized by the small dimensionless quantity $\Omega_{\text{SO}}/\omega$. Second, the ground-states of the Rashba SOC Hamiltonian $\hat p^2/2m + \lambda_{\text{R}} \, \hat{\bs{p}} \cdot \hat{\bs \sigma}$ are situated along the ``Rashba ring" at $p=p_{\text{R}}=m \lambda_{\text{R}}$ [i.e. the bottom of the mexican hat dispersion \cite{Goldman:2014Review,Zhai:Review}], so that probing this region of the dispersion relation requires to preparing states with $\Delta p \lesssim m \lambda_{\text{R}}$; in this regime the expansion in Eq. \eqref{subtle_SOC} becomes problematic as $m \lambda_{\text{R}} \Delta x \gtrsim 1/2$, which is imposed by Heisenberg inequality.

\subsection{Generating spin-orbit couplings with the $\alpha$ sequence}\label{section_good_SOC}

\subsubsection{The helical-Rashba scheme}\label{section_good_SOC_one}

Inspired by the result presented in Section \ref{sorensen_no_lattice}, we propose a scheme to realize spin-orbit couplings, based on the four-step sequence $\alpha$ in Eq. \eqref{alpha_sequence}. Considering the operators
\be
\hat H_0= \hat p^2/2m , \, \hat A= (\hat p_x^2 - \hat p_y^2)/2m , \, \hat B= \kappa ( \hat x \hat \sigma_x - \hat y \hat \sigma_y),\label{SOC_good_operators}
\ee
the time evolution of the driven system is characterized by the repeated sequence
\be
\left \{ \frac{\hat p_x^2}{m} ,  \hat H_0 + \kappa ( \hat x \hat \sigma_x - \hat y \hat \sigma_y) , \frac{\hat p_y^2}{m} ,  \hat H_0 - \kappa ( \hat x \hat \sigma_x - \hat y \hat \sigma_y) \right \}. \label{good_SOC_sequence}
\ee
The general expression for the effective Hamiltonian \eqref{good_eff_ham} then yields 
\begin{align}
&\hat H_{\text{eff}} \!=\! \hat H_0 + \lambda_{\text{R}} \, \hat{\bs{p}} \cdot \hat{\bs \sigma} + \Omega_{\text{SO}} \hat L_z \hat \sigma_z + \mathcal{O} (1/\omega^3), \notag \\
&\lambda_{\text{R}} = \pi \kappa / 8 m \omega, \quad \Omega_{\text{SO}}= - (8m/3) \lambda_{\text{R}}^2 ,\label{good_soc}
\end{align}
providing a ``helical-Rashba" configuration. In this perturbative approach, we note that one can eliminate the Rashba \emph{or} the helical SOC term in Eq. \eqref{good_soc} via a unitary transformation, but not both since $\Omega_{\text{SO}} \ne m \lambda_{\text{R}}^2$ [see the previous Section \ref{general_SOC}]. The kick operator $\hat K (t)$ is obtained from Eq. \eqref{F-operator}; in particular, the initial kick at $t_i=0$ is given by [Eq. \eqref{good_eff_ham}]
\be
\hat K(t_i=0) \!=\! - \frac{\pi}{4   \omega} \left [  (\hat p_x^2 - \hat p_y^2)/2m + \kappa ( \hat x \hat \sigma_x - \hat y \hat \sigma_y) \right] + \mathcal{O} (1/ \omega ^2). \notag
\ee
In direct analogy with the discussion presented in Section \ref{section:abrupt}, we find that this initial kick can profoundly alter the dynamics of wave packets in the strong SOC regime, where $\kappa /\omega\sim \lambda_{\text{R}}$ is large. In this regime, it is thus desirable to launch the dynamics at a subsequent time $t_i=3T/8$, in which case
\be
\hat K(t_i=3T/8) \!=\!  \frac{\pi}{4   \omega} \left [  (\hat p_x^2 - \hat p_y^2)/2m \right] + \mathcal{O} (1/ \omega ^2), \notag
\ee
no longer depends on the parameter $\kappa / \omega$.

In the same spirit as in Section \ref{sorensen_lattice}, we now consider the lattice analogue of the driven system characterized by the pulse sequence \eqref{good_SOC_sequence}. The corresponding tight-binding operators and commutators are presented in Appendix \ref{appendix:2component}. Similarly as in Section \ref{sorensen_lattice}, this scheme involves a combination of pulsed directional hoppings on the lattice and oscillating quadrupole fields. We obtain that the corresponding effective Hamiltonian $\hat H_{\text{eff}}$ reproduces the helical-Rashba Hamiltonian in Eq. \eqref{good_soc}, after substituting $m$ by the effective mass $m^*=1/(2Ja^2)$ and taking the continuum limit. A specificity of the lattice framework is that the second-order contributions also lead to a  renormalization of the hopping amplitude $J\! \rightarrow \! J(1-\eta^2)$, where  $\eta = (a \pi \kappa / 4 \sqrt{3}\omega)$; this is due to the fact that $[[\bar p_x^2 , \bar x], \bar x] = \bar p_x^2 a^2$ in the lattice formulation [Appendix \ref{appendix:2component}]. 
 
\begin{figure}%[h!]% Figure 4
\centering
\includegraphics[width=9cm]{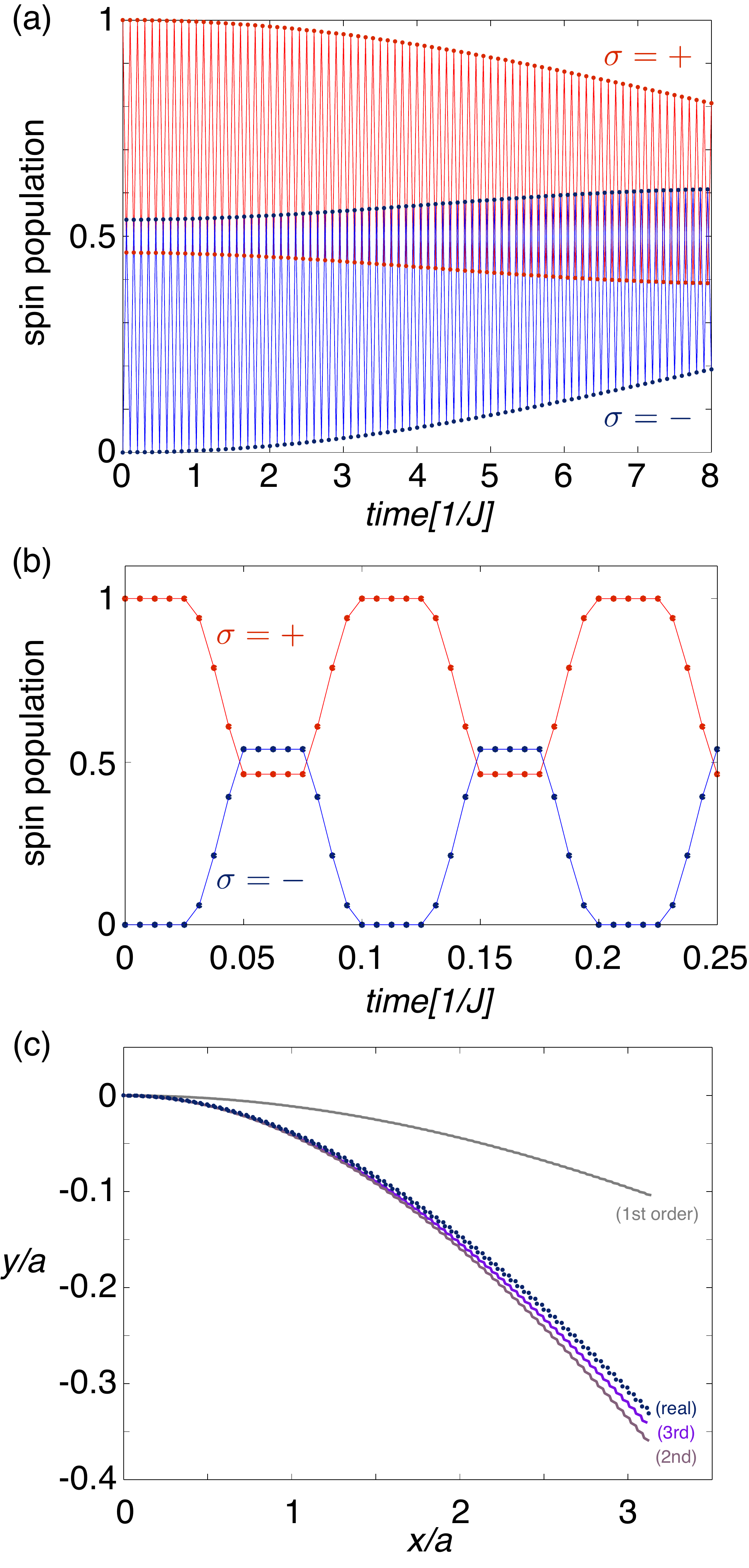}
\caption{\label{fig-4} The helical-Rashba scheme: dynamics of the driven system associated with the driving sequence \eqref{good_SOC_sequence}. Here, $\kappa=10 J/a$ and $T=0.1 /J$. (a) Spin populations as a function of time. The dynamics of the pulsed system [blue and red dots] is compared with the evolution dictated by the evolution operator in Eq. \eqref{rahav} [blue and red lines], where $\hat H_{\text{eff}}$ and $\hat K (t)$ are given by Eqs. \eqref{F-operator}-\eqref{good_soc}-\eqref{third_order_corrections}. The effective and real dynamics are sampled using the time step $\Delta t = T/2$. (b) Same as above, but using a time step $\Delta t = T/16$ to highlight the micro-motion captured by the kick operator $\hat K(t)$ in Eq. \eqref{F-operator}. (c) Center-of-mass displacement in the $x-y$ plane after a time $t=8 /J$.  Here, the real trajectory of the driven system  [blue dots] is compared with the evolution predicted by the effective-Hamiltonian formalism Eq. \eqref{rahav}, considering $\hat H_{\text{eff}}$ at 1st, 2nd and 3rd order in $1/\omega$. The trajectories are sampled using the time step $\Delta t = T/2$. }
\end{figure}

We now demonstrate that the dynamics of the pulse sequence \eqref{good_SOC_sequence} is well captured by the predictions of the effective-Hamiltonian formalism. We consider that the system is initially prepared in a gaussian wave packet with non-zero group velocity along the $+x$ direction and spin component $\sigma=+$. Figure \ref{fig-4} compares the dynamics generated by the pulse sequence \eqref{good_SOC_sequence} with the one associated with the evolution operator in Eq. \eqref{rahav}, with $\hat H_{\text{eff}}$ and $\hat K (t)$ given by Eqs. \eqref{F-operator}-\eqref{good_soc}-\eqref{third_order_corrections}. We show in Figs. \ref{fig-4}(a)-(b) the spin populations as a function of time: using a time step $\Delta t < T$, we observe a wide micro-motion in spin space, which is very well captured by the kick operator $\hat K(t)$ of the effective model. Figure \ref{fig-4}(c) compares the center-of-mass motion of the real and effective evolution operators.  The curved trajectory, together with the evolution of the spin populations, signals the presence of the effective SOC generated by the driving. In agreement with the discussion of Section \ref{general_SOC}, we recover the fact that it is necessary to evaluate the effective Hamiltonian up to (at least) second order in $1/\omega$ to reach a good agreement with the  dynamics of the real pulsed system. Adding third order corrections to the effective Hamiltonian [Eq. \eqref{third_order_corrections} in Appendix \ref{appendix:differentN}] leads to an even better agreement. Finally, similarly as in Fig. \ref{fig-2}, we find that the micro-motion is small in real space.

\subsubsection{The pure Dirac regime}

Interestingly, Eq. \eqref{good_soc} suggests that inhibiting the effect of $\hat H_0$ during the evolution will lead to a \emph{pure Dirac} system
\begin{align}
&\hat H_{\text{eff}}^{\text{D}} =  \lambda_{\text{R}} \, \hat{\bs{p}} \cdot \hat{\bs \sigma} + \mathcal{O} (1/\omega^3), \label{good_soc_Dirac}
\end{align}
noting that the helical SOC term in Eq. \eqref{good_soc} is due to the commutator $[[ \hat B, \hat H_0], \hat B]$ in Eq. \eqref{good_eff_ham}. This corresponds to a pulse sequence
\be
\left \{ (\bar p_x^2 - \bar p_y^2)/2m ,   \kappa ( \bar x \hat \sigma_x - \bar y \hat \sigma_y) , -(\bar p_y^2 - \bar p_x^2)/2m ,  - \kappa ( \bar x \hat \sigma_x - \bar y \hat \sigma_y) \right \}, \notag
\ee
which, in principle, can be realized in an optical-lattice setup. Hence, controlling the static Hamiltonian $\hat H_0$, which is conceivable in a lattice framework, offers the possibility to tune the first-order effective energy spectrum, but also,  to annihilate the second-order contributions to the SOC effects. We note that the pulsed operators $\hat A$ and $\hat B$ in Eq. \eqref{SOC_good_operators} could also be slightly modified to reach other regimes of interest. 
 \\

\subsubsection{Adding terms to the static Hamiltonian $\hat H_0$:  a route towards topological superfluids and topological insulators}

The scheme based on the driving sequence \eqref{SOC_good_operators}-\eqref{good_SOC_sequence} allows for adding a Zeeman term $\hat H_{\text{Z}}=\lambda_{\text{Z}} \hat \sigma_z$ to the effective Hamiltonian \eqref{good_soc}, whose association with the Rashba SOC term could be useful for the quantum simulation of topological superfluids \cite{Sau:2011}.  This could be simply realized by subjecting the system to a static Zeeman field, $\hat H_0 \rightarrow \hat H_0 + \hat H_{\text{Z}}$; this will not perturb the first-order Rashba term $\hat H_{\text{R}}\!\sim\! [\hat A , \hat B]$ in Eq. \eqref{good_soc}, but will add an extra second-order term 
\begin{align}
&\hat H_{\text{eff}} \!=\! \hat H_0 + \hat H_{\text{Z}}+\hat H_{\text{R}} +\hat H_{\text{L} \sigma} + \frac{1}{2} m \Omega_{\text{Z}}^2  (x^2 \!+\!y^2) \hat \sigma_z + \mathcal{O} (1/\omega^3), \notag 
\end{align}
which corresponds to a spin-dependent harmonic potential with frequency $\Omega_{\text{Z}}=2 \sqrt{\lambda_{\text{Z}} \Omega_{\text{SO}}}$. The survival of the topological superfluid phase in the presence of the additional harmonic potential constitutes an interesting open question. In general, we note that any driving scheme aiming to produce Rashba SOC typically presents this potential drawback, namely, the fact that the additional Zeeman term will necessarily generate additional (possibly spoiling) effects. \\

Adding terms to the static Hamiltonian $\hat H_0$ could also be envisaged to generate systems exhibiting the quantum anomalous Hall (AQH) effect \cite{Haldane:1988,Qi:2006}, the so-called Chern insulators. For instance, the AQH model of Ref. \cite{Qi:2006} could be realized by considering the driving sequence \eqref{SOC_good_operators}-\eqref{good_SOC_sequence} on a square lattice, but replacing the static Hamiltonian $\hat H_0 \rightarrow  \hat H_{\text{Z}} + (\bar p^2/2m) \hat \sigma_z$, where the last term  corresponds to a spin-dependent hopping term on the lattice. Another possibility would be to drive lattice systems with more complex geometries (e.g. honeycomb lattice), where the association of Rashba and Zeeman terms directly leads to Chern insulating phases \cite{Beugeling:2012,Qiao:2010}.

\subsubsection{The $xy$ scheme}\label{section:xy}

Finally, we introduce a second scheme based on the $\alpha$ sequence \eqref{alpha_sequence}, which features the standard static Hamiltonian $\hat H_0= \hat p^2/2m$ and the pulsed operators
\be
\hat A= (\hat p_y^2 - \hat p_x^2)/2m - \kappa \hat x \hat \sigma_x, \, \hat B=(\hat p_y^2 - \hat p_x^2)/2m - \kappa \hat y \hat \sigma_y,\label{xy_operators}
\ee
The repeated driving sequence is similar to \eqref{good_SOC_sequence}, reading
\be
\left \{ \frac{\hat p_y^2}{m} - \kappa \hat x \hat \sigma_x,  \frac{\hat p_y^2}{m} - \kappa \hat y \hat \sigma_y , \frac{\hat p_x^2}{m} + \kappa \hat x \hat \sigma_x, \frac{\hat p_x^2}{m} + \kappa \hat y \hat \sigma_y \right \}, \label{xy_SOC_sequence}
\ee
so that this scheme is based on a more regular sequence involving pulsed directional motion and magnetic fields. The effective Hamiltonian \eqref{good_eff_ham} then reads 
\begin{align}
&\hat H_{\text{eff}} \!=\! \hat H_0 + \lambda_{\text{R}} \, \hat{\bs{p}} \cdot \hat{\bs \sigma} - \gamma \, \hat x \hat y \, \hat \sigma_z + \text{cst} +\mathcal{O} (1/\omega^3), \notag \\
&\lambda_{\text{R}} = - \pi \kappa / 8 m \omega, \quad \gamma=  \pi \kappa^2 / 4 \omega .\label{xy_soc}
\end{align}
Here, the second-order terms do not contribute to the effective Hamiltonian, but the first-order terms include a spin-dependent hyperbolic potential. Although the ``$xy$" term is potentially problematic, especially in the strong SOC regime $\kappa/\omega \gg$, we will show in Section \ref{section:exact} that this pulse sequence allows for an almost exact treatment; incidentally, we will see that the ``$xy$" term can be inhibited in the lattice framework through a fine tuning of the driving parameters. 

\subsection{Generating spin-orbit couplings with the $\beta$ sequence: \\ the XA scheme}\label{section_beta_SOC}

The ``XA" scheme introduced by Xu \emph{et al.} and Anderson \emph{et al.} in Refs. \cite{Ueda,Anderson} is based on a repeated pulse sequence, which can be expressed through the evolution operator over one period $T$
\begin{align}
\hat U(T)= & e^{-i (\hat H_0 - \hat B) \tau} e^{-i \hat H_0 \tilde T/2} e^{-i (\hat H_0 + \hat B) \tau} \notag \\ 
& \times e^{-i (\hat H_0 - \hat A) \tau} e^{-i \hat H_0 \tilde T/2}e^{-i (\hat H_0 + \hat A) \tau} ,\label{ueda_sequence}
\end{align}
where $\tau \ll T \approx \tilde T$, and where the operators are explicitly given by
\be
\hat H_0= \hat p^2/2m , \quad \hat A= \kappa \hat x \hat \sigma_x , \quad \hat B= \kappa \hat y \hat \sigma_y.\label{SOC_bad_operators}
\ee
Noting that the sequence \eqref{ueda_sequence} essentially features four non-trivial steps (with pulses $\pm \hat A$ and $\pm \hat B$), we find that it can be qualitatively described by the associated four-step sequence $\beta$ in Eq. \eqref{beta_sequence}. The driving sequence studied in this Section, hereafter referred to as the ``XA" scheme, will thus be taken of the form $\beta$,
\be
\left \{ \hat H_0 + \kappa \hat x \hat \sigma_x ,  \hat H_0 - \kappa \hat x \hat \sigma_x , \hat H_0 + \kappa \hat y \hat \sigma_y , \hat H_0 - \kappa \hat y \hat \sigma_y \right \}. \label{bad_SOC_sequence}
\ee
The corresponding effective Hamiltonian and initial kick operators then read [Eq. \eqref{beta_result}]
\begin{align}
&\hat H_{\text{eff}} \!=\! \hat H_0 - \Omega_{\text{SO}} \hat L_z \hat \sigma_z + \mathcal{O} (1/\omega^3), \quad  \Omega_{\text{SO}}=  m \lambda_{\text{R}}^2, \label{bad_soc}  \\
& \hat K (0)=  - m \lambda_{\text{R}} \, \hat{\bs{x}} \cdot \hat{\bs \sigma} + \mathcal{O} (1/\omega^2), \quad \lambda_{\text{R}} = \pi \kappa / 8 m \omega ,\notag
\end{align}
where the absence of first-order (Rashba) terms is a characteristic of the $\beta$ sequence [Eq. \eqref{beta_result}].  However, using Eq. \eqref{rahav} and comparing Eqs. \eqref{bad_soc}-\eqref{subtle_SOC}, one recovers an effective Rashba SOC term via the unitary transform [$\hat K (T)=\hat K (0)$]
\begin{align}
&\hat U(T)=e^{- i \hat K(T)} \,e^{ -i T \hat H_{\text{eff}}} \, e^{i \hat K(0)}=e^{ -i T \hat H_{\text{eff}}^{\mathcal{T}} } , \notag \\
&\hat H_{\text{eff}}^{\mathcal{T}}= \hat H_0 -  \lambda_{\text{R}} \, \hat{\bs{p}} \cdot \hat{\bs \sigma}  + \mathcal{O} (1/\omega^3) + \mathcal{O} (\lambda_{\text{R}}^3),\label{compatibility_ueda_eff}
\end{align}
where we have introduced the effective Hamiltonian $\hat H_{\text{eff}}^{\mathcal{T}}$. As anticipated at the end of  Section \ref{general_SOC},  Eq. \eqref{compatibility_ueda_eff} implicitly contains two intertwined perturbative expansions: the expansion in powers of $(\Omega_{\text{SO}}/\omega)$ inherent to $\hat H_{\text{eff}}$ and $\hat K (t)$ [Section \ref{section:formalism}], and the expansion in powers of $(m \lambda_{\text{R}} L)$ introduced in Eq. \eqref{subtle_SOC}.

\subsection{Quasi-exact treatments on special cases} \label{section:exact}

In this Section,  we show that specific driving schemes leading to SOC benefit from the fact that they can be treated \emph{almost exactly}. Such an approach is useful, as it allows to go beyond the perturbative treatment of Section \ref{section:formalism}, which has been considered until now to evaluate the evolution operator. Actually, we already encountered such a scheme in our study of the oscillating force in Section \ref{section:abrupt}.

\subsubsection{The XA scheme}\label{section:xa_exact}

Following Refs. \cite{Ueda,Anderson}, the time-evolution operator associated with the sequence \eqref{bad_SOC_sequence} can be conveniently partitioned as $\hat U (T)\!=\! \hat U_y \hat U_x$, where each of the two subsequences
\begin{align}
\hat U_{\mu}= & e^{-i (\hat H_0 - \kappa \hat \mu \hat \sigma_{\mu}) T/4} e^{-i (\hat H_0 + \kappa \hat \mu \hat \sigma_{\mu}) T/4} , \quad \mu=x,y,
\end{align}
can be treated exactly. The calculations presented in Appendix \ref{appendix:ueda} yield the exact result [see also Refs. \cite{Ueda,Anderson}]
\begin{align}
\hat U_{\mu}= e^{-i T \left [\hat H_0/2 - \lambda_{\text{R}}  \hat p_{\mu} \hat \sigma_{\mu} \right]}  , \quad \mu=x,y, \label{exact:ueda_operator}
\end{align} 
where $\lambda_{\text{R}} = \pi \kappa / 8 m \omega$. Finally, the evolution operator after one period is obtained by using the Trotter expansion to the lowest order, $\exp A \exp B \approx \exp (A+B)$, reading 
\begin{align}
&\hat U (T)=\hat U_y \hat U_x = \exp \left (-i \hat H_{\text{eff}}^{\mathcal{T}} T \right ) ,\notag \\
& \hat H_{\text{eff}}^{\mathcal{T}}= \hat H_0 -  \lambda_{\text{R}} \, \hat{\bs{p}} \cdot \hat{\bs \sigma} + \mathcal{O} \left( (\Omega_{\text{SO}}/\omega)^2 \right ). \label{ueda_bulk_result}
\end{align}
We thus recover the result in Eq. \eqref{compatibility_ueda_eff}, with the notable difference that the quasi-exact treatment leads to a partial resummation of the infinite series inherent to Eq.  \eqref{compatibility_ueda_eff}: indeed, the dimensionless parameter $m \lambda_{\text{R}} L$ no longer plays any role in the expression for the evolution operator in Eq. \eqref{ueda_bulk_result}.

Before discussing the result in Eq. \eqref{ueda_bulk_result} any further, we derive its lattice analogue by substituting the operators of the ``XA" sequence \eqref{bad_SOC_sequence} by their lattice counterparts [Appendix \ref{appendix:2component}]. Following the computations presented in Appendix \ref{appendix:ueda},  we obtain the effective Hamiltonian up to first order in $(\Omega_{\text{SO}}/\omega)^2 $,
\begin{align}
&\hat U (T)=\bar U_y \bar U_x = \exp \left [-i \bar H_{\text{eff}}^{\mathcal{T}}  T \right ] , \label{ueda_lattice_result} \\
&\bar H_{\text{eff}}^{\mathcal{T}}\!=\! \frac{\bar p^2}{2 m^*} \left \{ \frac{1}{2} \!+\! \frac{1}{2} \text{sinc} \left ( 4 a m^* \lambda_{\text{R}}  \right )   \right \}  \!+\! \bar{\bs{p}} \cdot \bs{\hat \sigma} \left \{ \frac{\text{cos} \left ( 4a m^* \lambda_{\text{R}} \right ) \!-\!1  }{8 (a m^*)^2 \lambda_{\text{R}}} \right \}, \notag
\end{align}
where $\lambda_{\text{R}} = \pi \kappa / 8 m^* \omega$ and $m^*=1/2Ja^2$ is the effective mass. Hence, we recover the result in Eq. \eqref{ueda_bulk_result} for weak driving $\lambda_{\text{R}} < aJ/2$ and by taking the continuum limit. However, in the lattice framework, the maximum value of the effective Rashba SOC strength is limited: Using Eq. \eqref{ueda_lattice_result}, we find that the ratio Rashba/hopping is maximized for $\lambda_{\text{R}}\!=\!\pi/4 a m^*\!=\!(\pi/2) a J$:
\be
\bar H_{\text{eff}}^{\mathcal{T}}\!=\! \frac{1}{2} \left ( \frac{\bar p^2}{2 m^*} \right ) \!-\! \lambda_{\text{R}}^{*}  \, \bar{\bs{p}} \cdot \bs{\hat \sigma}, \quad \lambda_{\text{R}}^{*}=(2/\pi) a J.\label{lambda_star}
\ee
Note that the appearance of sinc functions in Eq. \eqref{ueda_lattice_result} is a characteristic of lattice systems driven by square-wave modulations \cite{Eckardt:2009}. \\

\begin{figure}% Figure 3
\centering
\includegraphics[width=9.3cm]{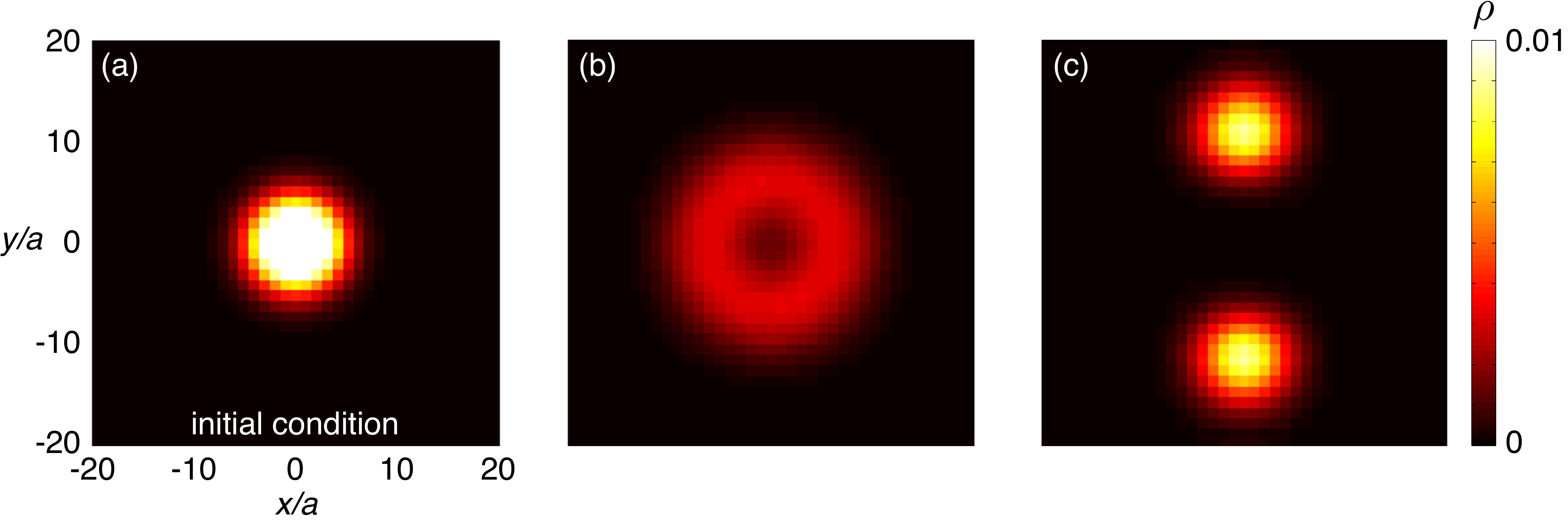}
\caption{\label{fig-3} The dynamics associated with the driving sequence \eqref{bad_SOC_sequence} on a lattice. (a) The gaussian wave packet is initially prepared around $\bs x (0)\!=\!0$ and momentum $\bs k\!=\!0$, so as to probe the Dirac dispersion relation of the effective Rashba Hamiltonian in Eq. \eqref{ueda_lattice_result} [$\Delta k \ll m^* \lambda_{\text{R}}$]. (b)-(c) Shown is the time-evolved particle density $\rho (x,y)$ in the $x-y$ plane at time $t=7.3 (1/J)$, for $\kappa=40 J/a$, and $T=7\pi/100 (1/J)$, i.e. $\lambda_{\text{R}}\approx (\pi/3) J a$. (b) The driving follows the sequence \eqref{bad_SOC_sequence} starting with the pulse $+ \hat A$.  (c) Same but starting with the pulse $- \hat B$, i.e., changing the initial phase of the modulation.}
\end{figure}

The quasi-exact method presented in this Section allows to partially resum the infinite series contained in $\hat H_{\text{eff}}$ and $\hat K (t)$ [Eq. \eqref{compatibility_ueda_eff}]. However, we stress that the evolution operators and the associated effective Hamiltonians in Eqs. \eqref{ueda_bulk_result}-\eqref{ueda_lattice_result} depend on the initial phase of the modulation, in direct analogy with the situation presented in Sections \ref{section:Trotter_ambiguity} and \ref{section:abrupt}: the analysis performed in this Section imposes that the $\beta$ sequence exactly starts with the pulse $+ \hat A$ and ends with the pulse $- \hat B$. Any deviation in the initial phase will alter the evolution operator in Eqs. \eqref{ueda_bulk_result}-\eqref{ueda_lattice_result}, and potentially, the long-time dynamics. Indeed, suppose that the launching time is shifted $t_i\!=\!0\! \rightarrow \!-T/4$, so that the $\beta$ sequence starts with the pulse $- \hat B$ instead of $+ \hat A$: the system will first undergo a kick 
\be
\exp (i \kappa T \hat y \hat \sigma_y/4)\!=\!\exp (i \, \delta p \, \hat y \hat \sigma_y), \quad \delta p = 4 m \lambda_{\text{R}},\label{kick_ueda_minusB}
\ee
 before evolving according to the Rashba Hamiltonian in Eqs. \eqref{ueda_bulk_result}-\eqref{ueda_lattice_result}. Note that $\delta p\!=\!4 p_{\text{R}}$, where $p_{\text{R}}$ is the radius of the Rashba ring along which the ground-states are situated.  Hence, this initial kick, which modifies the group velocity and spin structure of the prepared system, typically has an impact on long-time dynamics [in direct analogy with Fig. \ref{fig-1}]. We illustrate this sensitivity to the initial phase of the driving in Fig. \ref{fig-3}, which shows the time-evolved density of a lattice system driven by the sequence \eqref{bad_SOC_sequence}. Figure \ref{fig-3}(a) shows the initial wave packet in real space, with mean position $\bs x=0$ and momentum $\bs k=0$. The width of the gaussian satisfies $\Delta k \ll m^* \lambda_{\text{R}}$ in k-space, namely, the wave packet is prepared such as to probe the Dirac dispersion relation around $\bs k=0$ [i.e. within the Rashba ring]. Figure \ref{fig-3}(b) shows the time-evolved density at time $t=7.3/J$, in the case where the $\beta$ sequence \eqref{bad_SOC_sequence} is launched with the pulse $+ \hat A$: in agreement with the evolution operator in Eq. \eqref{ueda_lattice_result}, the cloud expands according to the isotropic Dirac dispersion relation associated with the effective Rashba Hamiltonian $\bar H_{\text{eff}}^{\mathcal{T}}$. Figure \ref{fig-3}(c) shows the same time-evolution protocol but for a driving sequence $\beta$ launched with the pulse $- \hat B$ [i.e. shifting the launching time $t_i\!=\!0\! \rightarrow \!-T/4$]. As predicted by Eq. \eqref{kick_ueda_minusB}, the cloud undergoes a sudden kick along the $y$ direction, which ejects the initial momentum distribution out of the Rashba ring, before evolving according to the Rashba Hamiltonian \eqref{ueda_lattice_result}; hence, changing the initial phase of the modulation results in a highly anisotropic dynamics that no longer probes the Dirac dispersion relation proper to the Rashba Hamiltonian \eqref{ueda_lattice_result}. 
 
We also note that the treatment considered in this Section implies that the system is probed stroboscopically ($t=N T$), and in this sense, it does not describe the micro-motion associated with the driving. More importantly, we point out that the exact treatment leading to Eq. \eqref{exact:ueda_operator} cannot be performed when adding Pauli matrices into the static Hamiltonian (e.g. an extra Zeeman term). Finally, we point out that a similar quasi-exact treatment was considered in Ref. \cite{Sorensen}, for the one-component driven lattice discussed in Section \ref{section_four}.\\

\subsubsection{The $xy$ scheme}

The $xy$ scheme introduced in Section \ref{section:xy} can also be treated in an almost-exact manner, which leads to a partial resummation of the series in Eq. \eqref{xy_soc}. The method differs from the one presented in the previous Section \ref{section:xa_exact} for the XA scheme, and importantly, it is only possible under two conditions: (1) the $xy$ sequence in Eq. \eqref{xy_SOC_sequence} should be realized on the lattice, in which case all the operators are replaced by their lattice analogue [Appendix \ref{appendix:2component}]; (2) the driving parameter $\kappa$ and the period $T$ should satisfy the condition $\kappa T = 4 \pi /a$, or equally, using our previous definition, $\lambda_{\text{R}} = \pi \kappa / 8 m^* \omega= (\pi/2) a J$. We note that this particular value was already discussed below Eq. \eqref{ueda_lattice_result}, where it corresponded to the regime where the effective Rashba SOC was maximized on the lattice. In the following, we assume that these two conditions are satisfied.

Here, in contrast with the analysis performed in the previous Section \ref{section:xa_exact}, we  split the evolution operator $\bar U (T)$ associated with the sequence \eqref{xy_SOC_sequence} into its four primitive parts,
\be
\bar U (T)= \bar U_{-B} \bar U_{-A} \bar U_{+B} \bar U_{+A},
\ee 
and we analyze each operator $\bar U_{\pm A,B}$ separately. Using the Zassenhaus formula \cite{Zassenhaus}, we find factorized expressions for the individual operators [see Appendix \ref{appendix:xy} for details]
\begin{align}
&\bar U_{+A}= e^{ -i T \bar p_y^2/4 m^*} e^{i \pi \bar x/a} , \quad \bar U_{-A}= e^{i \pi \bar x/a} e^{i T \lambda_{\text{R}}^{*} \bar p_x \bar \sigma_x},  \notag \\
& \bar U_{+B}= e^{i \pi \bar y/a} e^{ -i T \lambda_{\text{R}}^{*} \bar p_y \bar \sigma_y} , \quad \bar U_{-B}=e^{ -i T \bar p_x^2/4 m^*} e^{i \pi \bar y/a},
\end{align}
where $\lambda_{\text{R}}^{*}$ was defined in Eq. \eqref{lambda_star}. Finally, using Eq. \eqref{useful_SOC_formula}, and applying the Trotter expansion to minimal order, we find
\begin{align}
&\bar U (T)=\bar U_{-B} \bar U_{-A} \bar U_{+B} \bar U_{+A} = \exp \left (-i \bar H_{\text{eff}}^{\mathcal{T}} T \right ) , \notag \\
&\bar H_{\text{eff}}^{\mathcal{T}}\!=\! \frac{1}{2} \left ( \frac{\bar p^2}{2 m^*} \right ) \!+\! \lambda_{\text{R}}^{*}  \, \bar{\bs{p}} \cdot \bs{\hat \sigma} + \mathcal{O} \left( (\Omega_{\text{SO}}/\omega)^2 \right ) , \label{xy_lattice_result} 
\end{align}
which is precisely the Rashba Hamiltonian in Eq. \eqref{lambda_star} up to the sign change $\lambda_{\text{R}}^{*} \rightarrow - \lambda_{\text{R}}^{*}$ (i.e. a gauge transformation). Interestingly, this shows that the $xy$ scheme in Eq. \eqref{xy_SOC_sequence} is equivalent to the XA scheme in Eq. \eqref{bad_SOC_sequence}, in the limit $\lambda_{\text{R}} \rightarrow (\pi/2) a J$ where the Rashba SOC is maximized. Furthermore, Eq. \eqref{xy_lattice_result} shows that the hyperbolic potential, which is predicted by the perturbative treatment [Eq. \eqref{xy_soc}] for the lattice-free case, totally disappears in this special regime; this surprising result is due to the underlying lattice structure  [Appendix \ref{appendix:xy}]
. 

\begin{figure}%[h!]% Figure 5
\centering
\includegraphics[width=9.3cm]{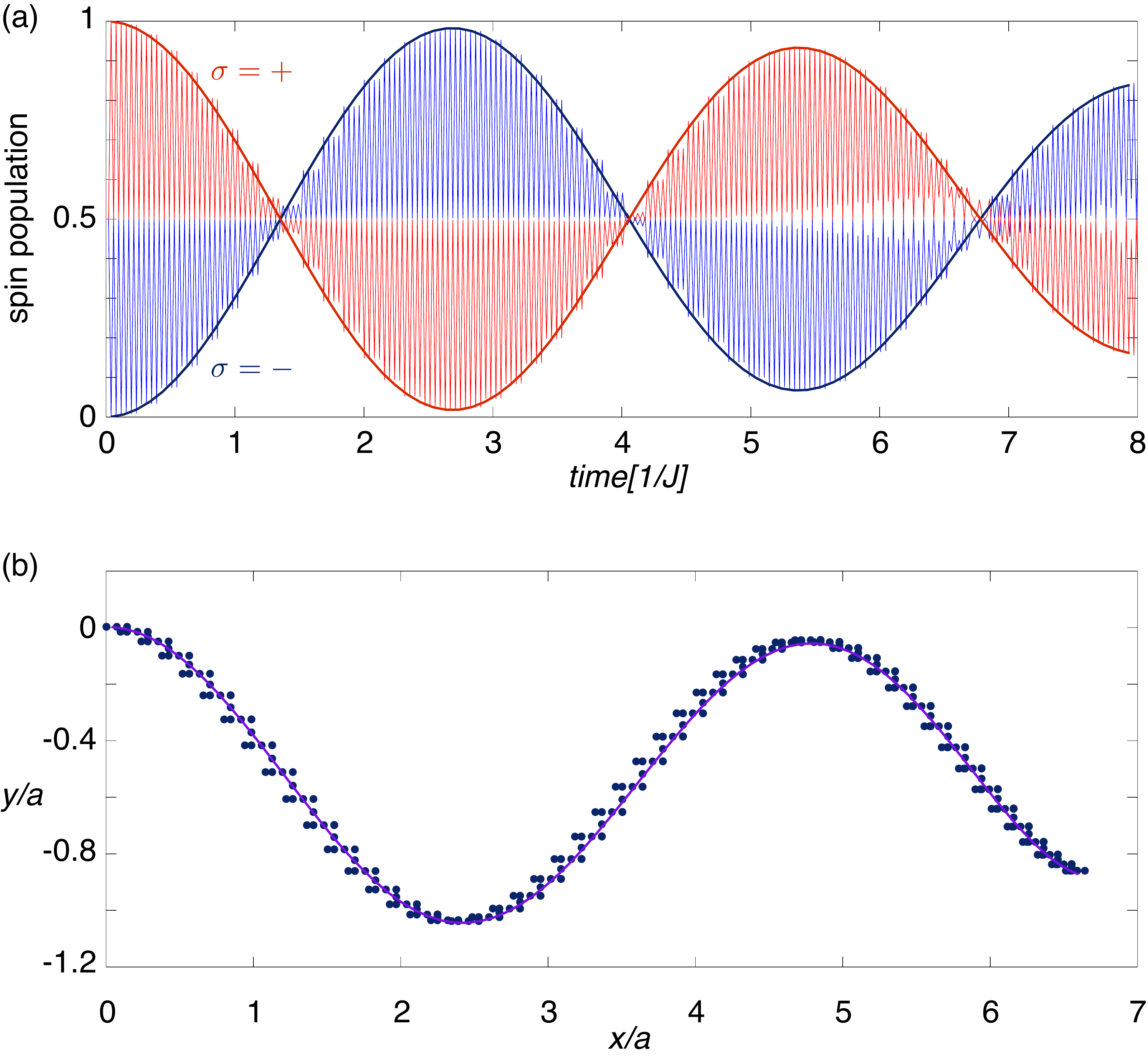}
\caption{\label{fig-5} The $xy$ scheme: dynamics of the driven system associated with the driving sequence \eqref{xy_SOC_sequence}. Here, $\kappa=80 J/a$ and $T=\pi/20  J^{-1}$, which corresponds to the special regime where $\lambda_{\text{R}} \!=\! (\pi/2) a J$. (a) Spin populations as a function of time. Bold [resp. thin] lines correspond to the predictions of the effective Hamiltonian in Eq. \eqref{xy_lattice_result} [resp. to the real dynamics of the pulsed system]. (b) Center-of-mass displacement in the $x-y$ plane after a time $t=8 /J$. The blue dots [resp. purple line] correspond to the real dynamics of the pulsed system [resp. to the effective Hamiltonian in Eq. \eqref{xy_lattice_result}]. In both figures, the real dynamics are sampled using the time step $\Delta t = T/8$ to highlight the micro-motion, which is not captured by the effective Hamiltonian. }
\end{figure}

We show in Fig. \ref{fig-5} the perfect agreement between the dynamics predicted by the effective Hamiltonian \eqref{xy_lattice_result} and the real dynamics of the pulsed lattice system. One should note that this agreement is only valid in the special regime where $\lambda_{\text{R}} \approx (\pi/2) a J$: the spoiling effects associated with the effective hyperbolic potential [Eq. \eqref{xy_soc}] become appreciable as soon as $\Delta \lambda_{\text{R}} \sim 1\%$. Finally, we stress that the analysis leading to Eq. \eqref{xy_lattice_result} cannot be performed when the static Hamiltonian $\hat H_0$ features Pauli matrices (e.g. a Zeeman term).

%%%%%%%%%%%%%%%%%%%%%%%%%%%%%%%%%%

\section{Discussions and Conclusions}\label{section:discussion}

\subsection{Convergence of the $(1/\omega)$ expansion}\label{section:convergence}

As already pointed out in Ref. \cite{Rahav}, one cannot affirm that the perturbative expansion proper to the formalism of Section \ref{section:formalism}  converges in general. Indeed, the small dimensionless parameter associated with the $1/\omega$-expansion that leads to the effective Hamiltonian $\hat H_{\text{eff}}$ in Eq. \eqref{effective_ham} can only be determined \emph{a posteriori}, on a case-by-case basis. 

\subsubsection{Illustration of the convergence issue}\label{section:convergence_illustration}

Let us illustrate this fact based on the driven system discussed in Section \ref{sorensen_no_lattice}. Using Eq. \eqref{effective_ham}, we obtained the second-order effective Hamiltonian in Eq. \eqref{bulk_eff_ham}, which we now decompose in terms of the $1/\omega$ expansion:
\begin{align}
&\hat H_{\text{eff}} = \hat H^{0} + \hat H^{1} + \hat H^{2} + \mathcal{O} (1/\omega^3), \label{sorensen_split}\\
&\hat H^{0}=\frac{\hat p^2_x}{2m} , \, \hat H^{1}= - \Omega \left ( \hat x \hat p_y - \hat y \hat p_x \right ) , \, \hat H^{2}= \frac{1}{2} m \omega_{\text{conf}}^2 (\hat x^2 + \hat y^2),\notag
\end{align}
where $\Omega \sim \kappa/ m \omega$ and $\omega_{\text{conf}}^2 \sim \Omega^2$. At this point, no element justifies the fact that the perturbative expansion should be carried up to second order. In order to simplify the present discussion, we slightly modify the sequence in Eq. \eqref{bulk_sorensen} so that the Hamiltonian in Eq. \eqref{sorensen_split} can be recast in the familiar form 
\be
\hat H_{\text{eff}}=(\hat{\bs{p}} - \bs{\mathcal{A}})^2/2m + \mathcal{O} (1/\omega^3), \quad \bs{\mathcal{A}}=\Omega m (-\hat y , \hat x, 0),  \label{landau_ham}
\ee
which is easily done by changing the prefactor in front of the pulse operator $\hat A \!\sim\! \hat p_x^2 - \hat p_y^2$ entering the sequence in Eq. \eqref{bulk_sorensen}. Having obtained the effective Hamiltonian in Eq. \eqref{landau_ham} up to second-order in $1/\omega$, one can determine its general characteristics: for instance, let us focus on its ground state, which is the standard lowest Landau level (LLL) with cyclotron frequency $\Omega$. This ground-state is characterized by the cyclotron radius $r_0 \!\sim\! 1 /\sqrt{m \Omega}$, such that the typical momentum associated with this state is $p_0 \!\sim\!  \sqrt{m \Omega}\sim \sqrt{\kappa / \omega}$. Hence, as far as the LLL is concerned, we find that all the terms in Eq. \eqref{sorensen_split} are of the same order, $\hat H^{0,1,2} \sim \Omega \sim \kappa/ m \omega$, which indicates that the perturbative expansion leading to Eq. \eqref{sorensen_split} should necessarily be undertaken up to second-order (included). Building on this result, we now evaluate the third-order terms, which have been neglected up to now: including the pulse operators $\hat A \sim \hat p_x^2 - \hat p_y^2$ and $\hat B \sim \hat x \hat y$ into Eqs. \eqref{third_order_alpha_full}-\eqref{third_order_corrections}, we find a cancellation of all the third order terms. We thus need to evaluate the fourth-order terms, which are necessarily of the form
\begin{align}
&\frac{1}{\omega^4}[[[[\hat{\mathcal Q} , \hat B ], \hat B], \hat{\mathcal Q}], \hat{\mathcal Q}] \sim \frac{\kappa^2 p^2}{\omega^4m^3} , \,  \frac{1}{\omega^4}[[[[\hat{\mathcal Q} , \hat B ], \hat B], \hat{\mathcal Q}], \hat B] \sim \frac{\kappa^3 x^2}{\omega^4m^2},\notag\\
&\frac{1}{\omega^4}[[[[\hat{\mathcal Q} , \hat B ], \hat{\mathcal Q}], \hat B], \hat{\mathcal Q}] \sim \frac{\kappa^2 p^2}{\omega^4m^3} , \,  \frac{1}{\omega^4}[[[[\hat{\mathcal Q} , \hat B ], \hat{\mathcal Q}], \hat{B}], \hat B] \sim \frac{\kappa^3 x^2}{\omega^4m^2},\notag
\end{align}
where $\hat{\mathcal Q} \!=\! (\hat H_0, \hat A ) \!\sim\! p^2/m$. In the LLL, these fourth-order terms are all of order $\Omega^2/\omega$ or $\Omega^3/\omega^2$. Hence, we have identified that $\Omega/\omega\sim \kappa / m \omega^2$ constitutes the relevant dimensionless parameter in the problem : the perturbative expansion leading to the effective Hamiltonian in Eq. \eqref{sorensen_split} can indeed be safely limited to the second-order as long as $\Omega \ll \omega$, namely when $\kappa \ll m \omega^2$. \\

Based on this example, we propose a guideline that should be followed in order to validate the convergence of the perturbative expansion: 
\begin{enumerate}
\item Compute the effective Hamiltonian $\hat H_{\text{eff}}$ up to some order $1/\omega^n$, using the formalism presented in Section \ref{section:formalism};
\item Determine the interesting characteristics associated with $\hat H_{\text{eff}}$, e.g. based on its eigenstates or its dispersion relation;
 \item Evaluate the $(n+1)$-order term and identify the condition according to which this term can (possibly) be neglected. This condition defines the small dimensionless parameter of the problem, or equally, a region in the parameter space, out of which higher order terms could come into play.
 \end{enumerate}

\subsubsection{The perturbative approach revisited}

We  emphasize an important aspect related to time-dependent Hamiltonians $\hat H (t) = \hat H_0 + \sum_m f_m( \omega t) \hat V_m $, which is the fact that the operators $\hat H_0, \hat V_m$ are generally not independent with respect to the modulation frequency $\omega$. For instance, in the example discussed above in Section \ref{section:convergence_illustration}, we found that $\hat H_0 \!\sim\! \hat A \!\sim\!  \Omega$ and $\hat B \!=\! \kappa \hat x\hat y \sim \omega$, based on the LLL characteristics [with $\Omega \ll \omega$]. The fact that some driving operators may be \emph{proportional} to the frequency $\hat V_m\!\sim\!\omega$, which can only be determined a posteriori according to the guideline established above, seems  problematic when building a perturbative expansion in powers of $1/\omega$ [Section \ref{section:formalism}]. To treat this seemingly pathological situation, we propose an alternative perturbative approach in Appendix \ref{rahav:type:2}, which is specifically dedicated to the general time-dependent problem 
\be
\hat H (t) = \hat H_0 + \hat{\mathcal{A}} f(t) + \omega \hat{\mathcal{B}} g(t) ,\label{rahavtype2:time-dep}
\ee 
where the functions $f$ and $g$ are assumed to be time-periodic with a zero mean value over one period $T=2 \pi / \omega$. Applying this alternative method to lowest order in $1/\omega$ provides a compact form  for the effective Hamiltonian $\hat H_{\text{eff}}=\hat H_{\text{eff}}^{0}+ \mathcal{O} (1/\omega)$, where
\begin{align}
&\hat H_{\text{eff}}^{0}= \overline{\exp \left ( i G(t) \hat{\mathcal{B}} \right ) \left \{  \hat H_0 + \hat{\mathcal{A}} f(t) \right \} \exp \left ( -i G(t) \hat{\mathcal{B}} \right )}  \notag\\
& \qquad =\hat H_0 +i \overline{G f} [\hat{\mathcal{B}} , \hat{\mathcal{A}}] - \frac{1}{2} \overline{G^2} [\hat{\mathcal{B}} , [\hat{\mathcal{B}}, \hat H_0]] - \frac{1}{2} \overline{G^2 f} [\hat{\mathcal{B}} , [\hat{\mathcal{B}}, \hat{\mathcal{A}}]] \notag \\
&  \qquad - \frac{i}{6} \overline{G^3} [ \hat{\mathcal{B}} [\hat{\mathcal{B}} , [\hat{\mathcal{B}}, \hat H_0]]] - \frac{i}{6} \overline{G^3 f} [ \hat{\mathcal{B}} [\hat{\mathcal{B}} , [\hat{\mathcal{B}}, \hat{\mathcal{A}}]]] +  \dots , \notag\\
&\hat K(t)= G (t) \hat{\mathcal{B}} + \mathcal{O} (1/\omega),\label{rahavtype2}
\end{align}
where $\overline{\zeta (t)}=(1/T) \int_0^T \zeta(t) \text{d}t$ denotes the mean value over one period, and where $G(t)= \omega  \int^t g(\tau) \text{d}\tau$ satisfies $\overline{G(t)}=0$. 

Importantly, the formula in Eq. \eqref{rahavtype2} potentially allows for a partial resummation of the perturbative expansion stemming from the formalism introduced in Section \ref{section:formalism}. To illustrate this point, we again consider the driven system in Section \ref{sorensen_no_lattice}, for which we found that $\hat B \!\sim\! \omega$. Substituting $\hat{\mathcal{A}} \rightarrow \hat A\!=\!(\hat p_x^2 - \hat p_y^2)/2m$ and $\omega \hat{\mathcal{B}} \rightarrow \hat B\!=\! \kappa \hat x \hat y$ into Eq. \eqref{rahavtype2}, and computing the time-averages associated with the $\alpha$ pulse sequence [Appendix \ref{rahav:type:2}], we recover the effective Hamiltonian in Eq. \eqref{bulk_eff_ham}. We find that only a few terms in Eq. \eqref{rahavtype2} have a non-zero contribution, and we stress that, in contrast with the result \eqref{bulk_eff_ham} obtained using the formalism of Section \ref{section:formalism}, the present result based on the formula \eqref{rahavtype2} guarantees the convergence of the perturbative expansion. Indeed, one readily verifies that all the terms that have been identified at this order of the computations [Eq. \eqref{rahavtype2}] are of the same order $ \Omega\!\sim\!\kappa/m\omega$ [see Appendix \ref{rahav:type:2} for more details]. \\

\subsection{Adiabatic launching}

In this work, we assumed that the periodic modulation that drives the system is launched abruptly at some initial time $t_i$. In Section \ref{section:abrupt}, we demonstrated that the long-time dynamics could strongly depend on this choice [Figs. \ref{fig-1} and \ref{fig-3}]. Hence, after a long time $t \gg T$, the system ``\emph{remembers}" the initial phase of the modulation, or equivalently, the very first pulse that was activated [e.g. $+\hat A$ or $-\hat B$ in Fig. \ref{fig-3}]. This sensitivity to the initial phase constitutes an important issue with regards to experimental implementation, where the phase is controlled with a certain uncertainty. One way to ``\emph{erase}" the memory of the system is to ramp up the modulation $\hat H(t)\!=\!=\hat H_0 + \lambda(t) \hat V(t)$, where $\lambda(t)=0 \rightarrow 1$ is turned on very smoothly. This ``adiabatic launching" will effectively annihilate any effect associated with the initial kick $\hat K (t_i)$ in Eq. \eqref{rahav}.   We observe that the time-scales of such memory-eraser ramps depend on the scheme under scrutiny; for example, we find that the ramping time needed to erase the memory in the situation shown in Fig. \ref{fig-3} is of the order of $\sim 1/J$.

Furthermore, the adiabatic launching could be exploited to generate specific target states, although we stress that this strategy is inevitably affected by the micro-motion dictated by the operator $\hat K (t_f)$ in Eq. \eqref{rahav}. Indeed, if one initially prepares the system in the ground state of the static Hamiltonian $\hat H_0$, and then smoothly ramp up the modulation, the system will eventually oscillate around the (target) ground state of the effective Hamiltonian $\hat H_{\text{eff}}$. These micro-motion oscillations may be problematic for specific applications, where the stability of the target state particularly matters; we remind that the micro-motion is inherent to the physics of driven systems, and although it is generally limited in real space, it is typically large in momentum or spin space for the various examples treated in this work [Figs. \ref{fig-4} and \ref{fig-5}]. In view of applications, this also highlights the difference between two possible targets: (1) reaching a specific state [e.g. the ground-state of $\hat H_{\text{eff}}$], or (2) engineering an effective band structure [also associated with $\hat H_{\text{eff}}$]. This aspect of periodically driven systems is left as an open problem for future works. 

Finally, we note that the adiabatic launching is not unique, in the sense that the ramping function $\lambda (t)$ can take arbitrarily many different forms. In particular, one could consider a ``sequence preserving adiabatic" protocol where the function $\lambda (t)$ remains constant during each primitive pulse sequence $\{ \hat H_0 + \hat V_1, \dots ,  \hat H_0 + \hat V_N \}$ [Eq. \eqref{general_pulse_sequence}]. Let us briefly analyze the time-evolution of such a driving scheme, based on the simple 2-step sequence $\gamma_{+}=\{ \hat H_0 + \hat V, \hat H_0 - \hat V \}$, where the $+$ index  indicates that the sequence starts with the pulse $+\hat V$. Suppose that the system is initially prepared in the ground state $\psi_0$ of the static Hamiltonian $\hat H_0$. At the end of the sequence-preserving ramping process, the state has evolved into $\psi_0\rightarrow \psi_{+}$, which is an eigenstate  of the effective Hamiltonian $\hat H_{+}$ associated with the primitive sequence $\gamma_{+}$, namely
\begin{align}
&\hat H_{+} \psi_{+} = E \psi_{+}, \quad e^{-i \hat H_{+} T}=e^{-i (\hat H_0 - \hat V)T/2} e^{-i (\hat H_0 + \hat V)T/2}, \notag \\
&\hat H_{+} = \hat H_0 - i \frac{T}{4} [\hat H_0 , \hat V] + \mathcal{O} (T^2).\notag
\end{align}
To estimate the micro-motion undergone by the target state $\psi_{+}$ after the ramp, we compute its evolution after half a period. We find
\begin{align}
&e^{-i (\hat H_0 + \hat V)T/2} \psi_{+} = \psi_{\text{micro}}, \label{ramp_1}  \\
&\hat H_{-}  \psi_{\text{micro}} = E  \psi_{\text{micro}}, \quad  \hat H_{-} = \hat H_0 + i \frac{T}{4} [\hat H_0 , \hat V] + \mathcal{O} (T^2).\notag
\end{align}
Now suppose that the same driving scheme is performed, but using the alternative sequence $\gamma_{-}=\{ \hat H_0 - \hat V, \hat H_0 + \hat V \}$. In this case, the target state after the ramp $\psi_0\rightarrow \psi_{-}$ satisfies
\begin{align}
&\hat H_{-} \psi_{-} = E \psi_{-}, \quad e^{-i \hat H_{-} T}=e^{-i (\hat H_0 + \hat V)T/2} e^{-i (\hat H_0 - \hat V)T/2}.\label{ramp_2}
\end{align}
If the eigenvalue $E$ is not degenerate, i.e. $\psi_{\text{micro}}\!=\!\psi_{-}$, we find that the target states associated with the two sequences $\gamma_{\pm}$ are sent into each other  $\psi_{+}\!\leftrightarrow\!\psi_{-}$ through  the micro-motion [Eqs. \eqref{ramp_1}-\eqref{ramp_2}]. If the eigenvalue $E$ is degenerate, which is the case for the ground level of spin-orbit-coupled Hamiltonians \cite{Goldman:2014Review,Zhai:Review}, the evolution is non-trivial and should be studied on a case-by-case basis. In both situations, this analysis further highlights the relevance of the micro-motion in modulated systems. 

\subsection{Conclusions}

This work was dedicated to the physics of periodically modulated quantum systems, with a view to realizing  gauge structures in a wide range of physical contexts. 

Our approach was based on the perturbative formalism introduced in Ref. \cite{Rahav}, which clearly highlights three relevant notions associated with the driving: the initial kick captured by the operator $\hat K (t_i)$, the effective Hamiltonian $\hat H_{\text{eff}}$ that rules the long-time dynamics, and the micro-motion described by $\hat K (t_f)$ [Section \ref{section:formalism}]. Based on this perturbative method, we have obtained general formulas and identified diverse driving schemes leading to ``non-trivial" effective Hamiltonians, whose characteristics could be useful for the quantum simulation of gauge structures and topological order. In particular, we have discussed the convergence of the perturbative expansion; building on specific examples,  we have also presented methods allowing for the partial resummation of infinite series. 

This work addresses the general situation where the driving frequency $\omega$ is off-resonant with respect to any energy separation present in the problem. However, the effective-Hamiltonian method presented here can also be generalized to describe schemes based on resonant driving \cite{Nigel}, as recently implemented to generate synthetic magnetic fields in optical lattices \cite{Aidelsburger:2011,Aidelsburger:2013,Miyake:2013}. 

We have also mentioned the possibility to switch on the modulations adiabatically; this minimizes the effects attributed to the initial kick $\hat K (t_i)$, which was shown to be considerable when launching the driving abruptly [Fig. \ref{fig-3}]. Schemes based on adiabatic launching could also be exploited to reach interesting target states.

This works also aimed to highlight the important role played by the micro-motion in periodically driven systems. We have shown that these unavoidable oscillations are typically large in momentum and spin space, for the different examples treated in this work. This is particularly relevant from a detection point of view, noting that various probes are built on (possibly spin-dependent) momentum-distribution imaging. Although these results were obtained in the non-interacting regime, they also suggest that dissipation  due to inter-particle collisions could lead to significant heating in cold-matter systems presenting large micro-motion. This could be particularly problematic in spin systems, where spin-dependent micro-motion could possibly lead to drastic collision processes. The thermodynamics of driven quantum systems has been recently investigated in Refs.  \cite{Lazarides:2014,Langemeyer:2014,Alessio:2013,Choudhury:2014} [see also \cite{Esposito,Gamel:2010}]. We note that interactions could also be modulated in cold-atom systems, using time-dependent magnetic fields in a Feshbach resonance  \cite{Rapp:2012,DiLiberto:2014}.

Finally, we point out that probing interesting effects in cold-matter systems, such as topological order,  generally requires to act on the system with \emph{additional} potentials $\hat V_{\text{probe}}$. For instance, measuring the topologically-invariant Chern number in quantum-Hall atomic systems  \cite{Price:2012,Abanin:2012,Dauphin:2013} could be realized by acting on the cloud with an external force $\hat V_{\text{probe}} \,\sim\, \hat x$ [see also Ref. \cite{Xiao2010}]. These additional potentials will contribute to the static Hamiltonian $\hat H_0 \!\rightarrow\! \hat H_0 \!+\! \hat V_{\text{probe}}$, and hence, they will potentially alter the effective Hamiltonian [Eq. \eqref{effective_ham}] and the corresponding band structure: ``\emph{measuring the topological order associated with an effective Hamiltonian may destroy it}". This phenomenon will be particularly pronounced when $\hat V_{\text{probe}}$ includes Pauli matrices. More generally, adding terms to the static Hamiltonian, either to probe interesting characteristics of the system, or to enrich its topological features, should be handled with care. This issue, which is particularly relevant for the field of quantum simulation, is left as an avenue for future works.

\begin{acknowledgments}
The authors are pleased to acknowledge H. Pichler and F. Gerbier for valuable discussions. They also thank P. Zoller, N. R. Cooper and S. Nascimb\`ene for stimulating inputs,  and G. Juzeli\=unas, M. Ueda and Z. Xu for comments on a preliminary version of this manuscript. We finally thank E. Anisimovas for helping identifying a typo in Eq. \eqref{eq:general_mixing}, see also Ref. \cite{Anisimovas}. N.G.\ is supported by the Universit\'e Libre de Bruxelles (ULB). This research was also funded by IFRAF and ANR (AGAFON), and by the European Research Council Synergy Grant UQUAM. 
\end{acknowledgments}

\newpage

%%%%%%%%%%%%%%%%%%%%%%%%%%%%%%%%%%%%%%%%%%%%%%

\bibliographystyle{apsrev}

\begin{thebibliography}{40}
\bibitem{erratum} The published version of the article contained an erroneous factor of two in the last term of Eq.~\eqref{eq:general_mixing}. This typo, which had no impact on the rest of the paper, has now been corrected.
\end{thebibliography}

\newpage

\appendix

\section{The Paul trap}\label{appendix:paul_trap}

The Paul trap consists in a particle moving in a modulated harmonic trap. The Hamiltonian is taken in the form
\be
\hat H(t)=\hat H_0 + \hat V  \cos (\omega t)= \frac{\hat p^2}{2m} + \frac{1}{2}m \omega_0^2 \hat x^2 \cos (\omega t),\label{paul_definition_appendix}
\ee
where $\omega=2 \pi /T$ [resp. $\omega_0$] denote the modulation [resp. harmonic trap] frequency. It is convenient to write the Schrödinger equation $i \partial_t \psi = \hat H(t) \psi$ in a moving frame by considering the unitary transformation
\be
\psi = \hat R \tilde \psi = \exp \left (\! -i \hat V \int_0^t \cos (\omega \tau) \text{d} \tau \! \right ) \tilde \psi= e^{ -i \sin (\omega t) \hat V /\omega } \tilde \psi ,\label{transf_one}
\ee
so that the transformed state satisfies the Schrödinger equation $i \partial_t \tilde \psi = \tilde H(t) \tilde \psi$, with the modified Hamiltonian
\begin{align}
\tilde H (t)= \hat R^{\dagger} \hat H_0 \hat R = \hat H_0 &- \sin (\omega t) \omega_0^2 (\hat x \hat p + \hat p \hat x)/2 \omega \notag \\
&+ m \omega_0^4 \sin^2 (\omega t) \hat x^2 / 2 \omega^2.\label{transf_two}
\end{align}
Using the Magnus expansion \cite{Maricq} to lowest order, we find that the evolution operator after one period reads
\begin{align}
&\hat U(T) \approx \exp \left (-i \int_0^T \tilde H (\tau) \text{d} \tau \right)= \exp \left (-i T \hat H_{\text{eff}}  \right ), \notag\\
&\hat H_{\text{eff}}=\frac{\hat p^2}{2m} + \frac{1}{2} m \Omega^2 \hat x^2, \quad \Omega=  \omega_0^2 / \sqrt{2} \omega , \label{paul_result}
\end{align}
so that the particle effectively moves in a harmonic trap with frequency $\Omega$. We point out that the non-trivial term in the effective Hamiltonian in Eq. \eqref{paul_result} is \emph{second order} in the period $T$.

To gain further insight on this result, we consider a classical analysis, which consists in seeking  a solution $x(t)$ in the form $x(t)=\bar x(t) + \xi (t)$, where $\bar x(t)$ evolves slowly and where $\xi (t)$ describes the micro-motion. The equations of motion read
\begin{align}
&m \ddot{\bar x}=-m \Omega^2 \bar x (t), \notag \\
& \xi (t) = (\omega_0^2/\omega^2) \cos (\omega t) \bar x (t).
\end{align}
Hence, the effective harmonic potential with frequency $\Omega$ that rules the slow motion $\bar x(t)$ is equal to the average kinetic energy associated with the micro-motion:
\be
\frac{1}{2}m \Omega^2 \bar x^2 = \frac{1}{2}m \langle \dot{\xi}^2 \rangle,
\ee
where $\langle . \rangle$ denotes the average over one period. Furthermore, the momentum dispersion in the ground state of the effective Hamiltonian $\hat H_{\text{eff}}$  [Eq. \eqref{paul_result}] is $\Delta p = \sqrt{m \Omega}$, which is similar to the momentum associated with the average micro-motion $m \dot{\xi}$ over the extension of this ground state. This classical analysis  illustrates the important role played by the micro-motion in modulated systems. \\

We now present the first corrections to the result presented in Eq. \eqref{paul_result}. The Magnus expansion reads \cite{Maricq}
\begin{align}
&\hat U (T)= \exp \left ( - i \,  \mathcal{H} (T)  \right )= \exp \left ( - i \left [ \mathcal{H}^{(0)} (T) + \mathcal{H}^{(1)} (T) + \dots \right ] \right ), \notag \\
& \mathcal{H}^{(0)} (T)= \int_0^T \tilde H (\tau) \text{d} \tau , \notag \\
& \mathcal{H}^{(1)} (T)= - \frac{i}{2}\int_0^T \int_0^{t} [\tilde H (t),\tilde H (\tau)] \text{d} \tau \text{d}t , \label{magnus}
\end{align}
where $\tilde H (t)$ is defined in Eq. \eqref{transf_two}. The zeroth-order term was given in Eq. \eqref{paul_result}, and the first-order corrections read
\be
\mathcal{H}^{(1)} (T)= 4 \gamma T \left (   \hat H_0 - \frac{1}{2} m \Omega^2 \hat x^2    \right ), \quad \gamma= (\Omega/\omega_0)^2.\label{magnus_order1}
\ee
The evolution operator including first-order corrections is finally given by [Eqs. \eqref{magnus}-\eqref{magnus_order1}]
\begin{align}
&\hat U(T) \approx \exp \left (-i T \tilde H_{\text{eff}}  \right ), \notag\\
&\tilde H_{\text{eff}}=\frac{\hat p^2}{2m} (1+4 \gamma) + \frac{1}{2} m \Omega^2 \hat x^2 (1-4 \gamma) , \label{paul_result_2}
\end{align}
where we note that the corrections are small $\gamma \ll 1$ for $\omega_0 \ll \omega$.\\

Finally, we show how the formalism of Appendix \ref{appendix:effective} allows one to recover the result in Eq. \eqref{paul_result_2}. To second order in $(1/\omega)$, the effective Hamiltonian and kick operators read [Eqs. \eqref{eq:general_mixing}-\eqref{eq:general_mixing_kick}]
\begin{align}
&\hat H_{\text{eff}}=\hat H_0 + \frac{1}{4 \omega^2} [[\hat V, \hat H_0], \hat V] + \mathcal{O} (1/\omega^3), \notag \\
&\hat K (0) = \frac{1}{i \omega^2} [\hat V , \hat H_0] + \mathcal{O} (1/\omega^3),\label{paul_comm}
\end{align}
where we used the fact that $V^{(j)}= V^{(-j)} = \hat V /2$ in the single-harmonic case $\hat H(t)=\hat H_0 + \hat V  \cos (\omega t)$. The commutators in Eq. \eqref{paul_comm} are readily computed using the operators defined in Eq. \eqref{paul_definition_appendix}, which yields
\begin{align}
&\hat H_{\text{eff}}=\frac{\hat p^2}{2m} + \frac{1}{2} m \Omega^2 \hat x^2 + \mathcal{O} (1/\omega^3), \notag \\
& \hat K (0) = \gamma \left ( \hat x \hat p + \hat p \hat x   \right ) + \mathcal{O} (1/\omega^3),
\end{align}
where $\Omega$ and $\gamma$ are defined in Eqs. \eqref{paul_result}-\eqref{magnus_order1}. Using Eq. \eqref{appendix_evolution_partition}, we obtain the evolution operator after one period
\begin{align}
&\hat U (T) = e^{-i \hat K(0)} e^{- i T  \hat H_{\text{eff}}} e^{i \hat K(0)} = e^{-i T \hat H_{\text{eff}}^{\mathcal{T}}} \notag \\
& \hat H_{\text{eff}}^{\mathcal{T}} = \frac{\hat p^2}{2m} (1+4 \gamma) + \frac{1}{2} m \Omega^2 \hat x^2 (1-4 \gamma) + \mathcal{O} (1/\omega^4),
\end{align}
such that we recover the result in Eq. \eqref{paul_result_2}.

\section{Renormalization of the hopping \\ in modulated optical lattices} \label{appendix:bessel_one}

In this Appendix, we consider a modulated 1D optical lattice, described by the single-particle Hamiltonian
\be
\hat h (t)= \hat p^2/2m + V_{\text{OL}} \left ( \hat x - x_0 (t) \right ) ,% \quad x_0 (t)= \zeta f(t), 
\ee
where the periodic function $x_0(t)=x_0(t+T)$ is considered to have a zero mean value over one period $T=2 \pi/ \omega$, and where $V_{\text{OL}}(\hat x)$ is the optical lattice potential. In the absence of driving, the static Hamiltonian $\hat H_0$ is written in the form of a second-quantized tight-binding Hamiltonian,
\be
\hat H_0 = -J \left ( \hat T + \hat T^{\dagger} \right ), \quad \hat T= \sum_j \hat a_{j+1}^{\dagger} \hat a_j , \label{ham_tb_1D}
\ee
where the operator $\hat a_j^{\dagger} $ creates a particle at lattice site $x=j a$, $a$ is the lattice spacing and $J$ is the hopping matrix element. The modulated lattice is generally studied in a moving frame, in which case the driving acts through an inertial force,
\be
\hat H (t)= \hat H_0 + \omega \, \xi (t) \hat V=\hat H_0 + \omega \, \xi (t) \sum_j j \hat a_{j}^{\dagger} \hat a_j , \label{inertial_force}
\ee 
where $\xi (t) = (m a/\omega)  \ddot x_0(t)$. In the following, we set 
\be
\xi(t)= \xi_0 \cos (\omega t), 
\ee 
so that the Hamiltonian reads
\be
\hat H (t)=\hat H_0 +  \omega \xi_0 \hat V \cos (\omega t). 
\ee 
Note that the parameter $\kappa$ introduced in Eq. \eqref{modulated_lattice_operators} [main text] is given by $\kappa= \omega \xi_0$. We also introduce the operator
\be
\hat H_1 = i J  \left ( \hat T - \hat T^{\dagger} \right ),\label{ham_one_def}
\ee
which, together with $\hat H_0$ and $\hat V$, form a close set under the action of the commutator
\be
[\hat H_0, \hat V ]=-i \hat H_1 , \quad [\hat H_1, \hat V ]=i \hat H_0, \quad [\hat H_0, \hat H_1 ]=0. \label{cycle_cond}
\ee
The latter relations lead to the useful formula
\be
e^{i \gamma \hat V} \hat H_0 e^{-i \gamma \hat V} = \hat H_0 \cos \gamma - \hat H_1 \sin \gamma, \label{eq:useful}
\ee
which is indeed valid for any triple of operators $\{\hat H_0, \hat H_1 , \hat V \}$ satisfying Eq. \eqref{cycle_cond}, see also Eq. \eqref{expand_eq_C7}. Following the same procedure as in Appendix \ref{appendix:paul_trap}, Eqs. \eqref{transf_one}-\eqref{transf_two}, we consider the unitary transformation
\be
\psi = \hat R \tilde \psi =  e^{ -i  \xi_0 \sin (\omega t) \hat V  } \tilde \psi ,\notag
\ee
and the associated modified Hamiltonian
\be
\tilde H (t)\!=\! \hat R^{\dagger} \hat H_0 \hat R = \hat H_0 \cos \left [ \xi_0 \sin (\omega t)  \right ] -\hat H_1 \sin \left [  \xi_0 \sin (\omega t)  \right ], \notag
\ee
where we used Eq. \eqref{eq:useful}. Finally, the operator evolution after one period is given by
\begin{align}
&\hat U(T)= \exp \left (-i \int_0^T \tilde H (\tau) \text{d} \tau \right)= \exp \left (-i T \hat H_{\text{eff}}  \right ) \notag\\
&\hat H_{\text{eff}} = \mathcal{J}_0 (\xi_0) \hat H_0 , \label{modulated_result}
\end{align}
where we recover the renormalization of the hopping by the Bessel function of the first kind
\be
{\cal J}_0(x)= \frac{1}{\pi} \int_0^{\pi} \cos \left ( x \sin (\tau) \right ) \text{d} \tau.\label{eq:bessel}
\ee
We emphasize that the effective Hamiltonian in Eq. \eqref{modulated_result} is exact.

\section{General expression for the effective Hamiltonian: Derivation of Eqs. \eqref{effective_ham}-\eqref{F-operator}}
\label{appendix:effective}

We start with the Schrödinger equation
\begin{align}
&i  \partial_t \psi(t) = \hat H (t) \psi(t) ,  \notag \\
&\hat H (t)= \hat H_0 + \hat V (t) , \quad \hat V (t+T)=\hat V (t),
\end{align}
and consider the unitary transformation 
\begin{align}
\phi(t) = \hat{\mathcal{U}}(t) \psi(t)= e^{i \hat K(t)} \psi(t). \label{define_unitary}
 \end{align}
The new state $\phi (t)$ satisfies the Schrödinger equation
\begin{align}
&i  \partial_t \phi(t) = \hat H_{\text{eff}} \phi(t) ,\\
& \hat H_{\text{eff}} = e^{i \hat K(t)} \hat H (t) e^{-i \hat K(t)} + i  \left ( \frac{\partial e^{i \hat K(t)}}{\partial t}\right ) e^{-i \hat K(t)},\label{rahav_ham}
\end{align}
where we introduced the effective Hamiltonian $\hat H_{\text{eff}}$. The method of Ref. \cite{Rahav} then consists in constructing the \emph{time-independent} effective Hamiltonian $\hat H_{\text{eff}}$, by transferring all undesired (time-dependent) terms into the operator $\hat K(t)$. The latter has a simple interpretation, which becomes obvious when writing the evolution operator
\be
\hat U (t_i \rightarrow t_f) \psi (t_i)= e^{-i \hat K(t_f)} e^{- i  \hat H_{\text{eff}} (t_f - t_i)} e^{i \hat K(t_i)} \psi (t_i). \label{appendix_evolution_partition}
\ee
This expression indicates that the evolution can be split into three parts: (a) an initial kick associated with the operator $\hat K (t_i)$, (b) the evolution dictated by the \emph{time-independent} effective Hamiltonian $\hat H_{\text{eff}}$, and (c) a final kick associated with the operator $\hat K (t_f)$.

In general, it is not possible to give an analytical expression for the operators $\hat H_{\text{eff}}$ and $\hat K(t)$ in Eq. \eqref{rahav_ham} [see however Appendix \ref{appendix:bessel} for an exactly solvable example]. Thus, it is convenient to build these operators perturbatively, by expanding them in powers of the driving period $T=2 \pi / \omega$, which is assumed to be small in the problem. Following Ref. \cite{Rahav}, we write
\be
\hat H_{\text{eff}} = \sum_{n=0}^{\infty} \frac{1}{\omega^n} \hat H_{\text{eff}}^{(n)} , \quad \hat K = \sum_{n=1}^{\infty} \frac{1}{\omega^n} \hat K^{(n)} ,\label{expand_eq}
\ee
and consider the expansions
\begin{align}
&e^{i \hat K} \hat H  e^{-i \hat K} \!=\! \hat H \!+\! i [\hat K , \hat H] \!-\! \frac{1}{2} [\hat K , [\hat K, \hat H]] \!-\! \frac{i}{6} [\hat K, [\hat K , [\hat K , \hat H]]] \dots \label{expand_eq_C7} \\
&\left ( \frac{\partial e^{i \hat K}}{\partial t}\right ) e^{-i \hat K}= i \frac{\partial \hat K}{\partial t} - \frac{1}{2} [\hat K , \frac{\partial \hat K}{\partial t}] - \frac{i}{6} [\hat K , [\hat K , \frac{\partial \hat K}{\partial t}]] \dots \label{expand_eq_2}
\end{align}
to determine the operators $\hat H_{\text{eff}}$ and $\hat K(t)$ at the desired order $\mathcal{O} (1/\omega ^n)$. Note that we further impose that $\hat K (t)$ should be periodic $\hat K (t)=\hat K (t+T)$, with zero mean value over one period.

We now apply this strategy to the general situation in Eq. \eqref{ham_series}, where the Hamiltonian $\hat H(t)$ of the driven system is given by
\be
\hat H(t)=\hat H_0 + \hat V(t)= \hat H_0 + \sum_{j=1}^{\infty} V^{(j)} e^{i j \omega t} + V^{(-j)} e^{-i j \omega t}.\label{ham_series_bis}
\ee 
Following the expansion procedure \eqref{expand_eq}-\eqref{expand_eq_2} up to second order $\mathcal{O} (1/\omega ^2)$, we obtain the general expressions for the effective Hamiltonian~\cite{erratum}
\begin{widetext}
\begin{align}
\hat H_{\text{eff}}= \hat H_0 + \frac{1}{ \omega} \sum_{j=1}^{\infty} \frac{1}{j} [V^{(j)},V^{(-j)} ] &+ \frac{1}{2  \omega^2} \sum_{j=1}^{\infty} \frac{1}{j^2} \left ( [[V^{(j)},\hat H_0],V^{(-j)} ] + \text{h.c.} \right ) \notag \\
&+ \frac{1}{3  \omega^2} \sum_{j,l=1}^{\infty} \frac{1}{j l} \left ( [V^{(j)},[V^{(l)}, V^{(-j-l)} ]] -  [V^{(j)},[V^{(-l)}, V^{(l-j)} ]] + \text{h.c.} \right ), \label{eq:general_mixing}
\end{align}
and for the kick operator at time $t$
\begin{align}
\hat K (t)&= \frac{1}{i  \omega} \sum_{j=1}^{\infty} \frac{1}{j} \left ( V^{(j)} e^{i j \omega t} - V^{(-j)} e^{-i j \omega t} \right ) + \frac{1}{i  \omega^2} \sum_{j=1}^{\infty} \frac{1}{j^2} \left ( [V^{(j)},\hat H_0] e^{i j \omega t} - \text{h.c.}  \right ) \notag \\
&+  \frac{1}{ 2i  \omega^2} \sum_{j,l=1}^{\infty} \frac{1}{j (j+l)}  \left ( [V^{(j)},V^{(l)}] e^{i (j+l) \omega t} - \text{h.c.}  \right ) +  \frac{1}{ 2i  \omega^2} \sum_{j \ne l=1}^{\infty} \frac{1}{j (j-l)}  \left ( [V^{(j)},V^{(-l)}] e^{i (j-l) \omega t} - \text{h.c.}  \right ) .\label{eq:general_mixing_kick}
\end{align}
\end{widetext}

\section{The two-step sequence: \\ Derivation of Eqs. \eqref{eq:N=2}-\eqref{eq:N=2_kick}}\label{appendix:two_phase}

In the simple case of the two-step sequence $\{ \hat H_0+ \hat  V, \hat H_0- \hat V \}$, the Hamiltonian is given by $\hat H(t) = \hat H_0 + f(t) \hat V$, where $f(t)$ is the standard square-wave function. Expanding $f(t)$ into its Fourier components, we obtain a simple expression for the $\hat V^{(j)}$ operators introduced in Eq. \eqref{ham_series_bis}
\begin{align}
\hat V^{(j)}&=-2i \hat V /\pi j=\left ( \hat V^{(-j)} \right ) ^{\dagger} \text{for $j$ odd,} \notag \\
&=0 \quad \text{otherwise}.
\end{align}
Inserting these operators into Eqs. \eqref{effective_ham}-\eqref{F-operator} yields the result 
\begin{align}
&\hat H_{\text{eff}}=\hat H_0 + \frac{\pi^2}{24 \omega^2} [[\hat V, \hat H_0],\hat V] + \mathcal{O} (1/\omega^3) \label{eq:N=2_app}  \\
& \hat K(t)= - \frac{\pi}{2 \omega} \hat V + \vert t \vert \hat V + \mathcal{O} (1/\omega^2) \text{ , for $t \in \left [- \frac{T}{2} , \frac{T}{2} \right]$ }.\label{eq:N=2_kick_app}
\end{align}
To derive Eq. \eqref{eq:N=2_app}, we used the formula 
\be
\sum_{j=1}^{\infty} 1/(2j-1)^{4}= \pi^4/96.
\ee 
To derive Eq. \eqref{eq:N=2_kick_app}, we used the formula 
\be
\sum_{j=1}^{\infty} \frac{\cos [x (2j-1)] }{ (2j-1)^{2}}= \frac{\pi}{4} \left (\frac{\pi}{2} - \vert x \vert \right),
\ee 
which is valid for $-\pi \le x \le \pi$, see \cite{gradshteyn}.

\section{The modulated optical lattice revisited} \label{appendix:bessel}

In this Appendix, we illustrate how the method described in Appendix \ref{appendix:effective} can be used in a non-perturbative manner, based on the modulated-optical-lattice problem treated in Appendix \ref{appendix:bessel_one}. The present analysis allows to recover the well-known renormalization of the hopping amplitude by a Bessel function, Eq. \eqref{modulated_result}.

As in Appendix \ref{appendix:bessel_one}, we start with the Hamiltonian
\begin{align}
&\hat H (t)= \hat H_0 +m a  \ddot x_0(t) \hat V=\hat H_0 + \omega \, \xi (t) \sum_j j \hat a_{j}^{\dagger} \hat a_j , \label{inertial_force_two}
\end{align} 
where the tight-binding Hamiltonians $\hat H_0$ and $\hat V$ were defined in Eqs. \eqref{ham_tb_1D}- \eqref{inertial_force}. Following the strategy described in Eqs. \eqref{define_unitary}-\eqref{rahav_ham}, we look for a unitary transformation $\hat{\mathcal{U}}(t)$, such that
\begin{align}
& \hat H_{\text{eff}} = \hat{\mathcal{U}}(t) \hat H (t) \hat{\mathcal{U}}^{\dagger}(t) + i  \left ( \frac{ \partial \hat{\mathcal{U}}(t)}{\partial t}\right ) \hat{\mathcal{U}}^{\dagger}(t),\label{rahav_ham_bis}
\end{align}
defines a time-independent effective Hamiltonian. The unknown unitary transformation is taken in the form
\be
\hat{\mathcal{U}} (t)=e^{i \left( \alpha (t) \hat H_0 + \beta (t) \hat H_1 \right )} e^{i \gamma (t) \hat V},\label{unitary_unknown}
\ee
where the operator $\hat H_1$  was introduced in Eq. \eqref{ham_one_def}.

Using the identities \eqref{cycle_cond}-\eqref{eq:useful} together with the ansatz \eqref{unitary_unknown}, we find that the effective Hamiltonian in Eq. \eqref{rahav_ham_bis} takes the simple form
\begin{align}
 \hat H_{\text{eff}} =&  \hat H_0 \left (\cos \gamma -\omega\,\xi\beta + \beta \dot \gamma-\dot \alpha \right ) +\hat V \left( \omega \xi-\dot \gamma \right )
\nonumber\\
&+ \hat H_1 \left ( -\sin \gamma + \omega \xi \alpha -\alpha \dot \gamma -\dot \beta \right ),
\nonumber
\end{align}
where the three time-periodic functions $\alpha, \beta , \gamma$ are still to be determined. Imposing that  $\hat H_{\text{eff}}$ should be time-independent,
\begin{align}
 \hat H_{\text{eff}} =&  c_1 \hat H_0  + c_2 \hat H_1 + c_0 \hat V,
\nonumber
\end{align}
leads to the relations
\begin{align}
&\alpha(t) = \alpha_0+\int_0^t \cos\left[
\omega\int_0^{t'}\xi(t'')\; \text{d} t'' -\gamma_0 \right]\; \text{d} t' -c_1 t , \notag \\
&\beta(t) = \beta_0+\int_0^t \sin\left[ \omega\int_0^{t'}\xi(t'')\;\text{d} t'' -\gamma_0 \right]\; \text{d} t' -c_2 t , \notag \\
&\gamma(t)=\gamma_0 +\omega \int_0^t \xi(t')\; \text{d} t' - c_0 t, \label{alpha_beta}
\end{align}
where we introduced constants of integration. The time-periodicity of $\gamma (t)$ is satisfied by setting $c_0=0$, while imposing the periodicity of the two other functions $\alpha$ and $\beta$ requires further developments. For the sake of clarity, let us consider the driving $x_0(t)= \zeta \cos (\omega t + \phi)$, such that
\be
\xi (t)= \xi_0 \cos \left ( \omega t + \phi \right ), \quad \xi_0=-ma \omega \zeta.
\ee
In this case, the time-periodicity of $\alpha (t)$ and $\beta (t)$ in Eq. \eqref{alpha_beta} is satisfied for
\begin{equation}
c_1={\cal J}_0(\xi_0) \cos\gamma_1,\qquad c_2=-{\cal J}_0(\xi_0) \sin\gamma_1,
\label{}
\end{equation}
where $\gamma_1=\gamma_0 - \xi_0 \sin \phi$, and where ${\cal J}_0(\xi_0)$ is the Bessel function of the first kind in Eq. \eqref{eq:bessel}.
The corresponding effective Hamiltonian reads
\begin{equation}
\hat H_{\text{eff}}= {\cal J}_0(\xi_0) \left( \hat H_0 \cos\gamma_1- \hat H_1 \sin\gamma_1 \right), \label{bessel_ham}
%\label{}
\end{equation}
where we note that the operators $\hat H_{0,1}$ are related by a gauge transformation.

Next, we rewrite the evolution operator in Eq. \eqref{unitary_unknown} in the more usual form [Eq. \eqref{define_unitary}]
\begin{align}
&\hat{\mathcal{U}} (t)=e^{i \hat K (t)}, \label{unitary_unknown_2}\\
&\hat K (t)=\hat H_0 \left \{ \alpha (t) \mathfrak{c}_1 + \beta (t) \mathfrak{c}_2 \right \} + \hat H_1 \left \{ \beta (t) \mathfrak{c}_1 - \alpha (t) \mathfrak{c}_2 \right \} + \gamma (t) \hat V
,\notag
\end{align}
where we used the commutators in Eq. \eqref{cycle_cond}, and where the constants $\mathfrak{c}_{1,2}$ can be deduced from the BCH-Trotter expansion in Eq. \eqref{BCH}. From Eq. \eqref{unitary_unknown_2}, we find that the function $\gamma (t)$ in Eq. \eqref{alpha_beta} should necessarily have a zero mean value over one period to guarantee that the kick operator $\hat K (t)$ satisfies this same constraint. This latter condition sets $\gamma_1=0$, which together with Eq. \eqref{bessel_ham} eventually leads to the familiar renormalization of the hopping amplitude 
\begin{equation}
\hat H_{\text{eff}}= {\cal J}_0(\xi_0) \ \hat H_0 , \quad \xi_0=-ma \omega \zeta,
\label{}
\end{equation} 
which reproduces the result presented in Eq. \eqref{modulated_result}.

The remaining constants $\alpha_0$, $\beta_0$ and $\gamma_0$ define the initial operator $\hat{\mathcal{U}}(0)$, and thus complete the determination of the unitary transformation $\hat{\mathcal{U}}(t)$. The condition $\gamma_1\!=\!0$ implies $\gamma_0 \!=\!  \xi_0 \sin \phi$, such that setting $\alpha_0=\beta_0$ yields
\begin{equation}
\hat U(0)=e^{i \xi_0 \sin \phi \hat V},
\label{kick_operator_exact_modulated}
\end{equation}
which determines the kick operator $\hat K (0)=\xi_0 \sin \phi \hat V$. This latter result illustrates the fact that the initial phase of the lattice modulation $\phi$ has an impact on the initial kick given to the system, but has no consequence on the effective Hamiltonian \cite{Creffield:2011}.

\section{$N$-step sequence: \\ Derivation of the general expression in Eq. \eqref{general_result}} \label{appendix:Npulse}

We derive the general expressions in Eq. \eqref{general_result} for the effective Hamiltonian $\hat H_{\text{eff}}$ and initial-kick operator $\hat K (0)$, in the general case presented in Section \ref{section-three}, where the system is characterized by the repeated pulse sequence
\be
\gamma_N=\{ \hat H_0 +\hat V_1 , \hat H_0 +\hat V_2, \hat H_0 +\hat V_3, \dots , \hat H_0 +\hat V_N \},\label{very_general_sequence}
\ee 
with $N$ an arbitrary integer. We write the Hamiltonian as
\begin{align}
&\hat H (t)= \hat H_0 + \sum_{j \ne 0} V^{(j)} e^{i \omega j t}, \\
& V^{(j)}= \frac{1}{2 \pi i} \sum_{m=1}^{N} \frac{1}{j} e^{-i 2 \pi j m/N} \left (  e^{i j (2 \pi/N)}  -1 \right ) \hat V_m , \label{def_vjs}
\end{align}
where we used the Fourier series in Eq. \eqref{fourier}. Applying Eqs. \eqref{effective_ham} and \eqref{F-operator}, and presently restricting ourselves to the first-order terms, yields
\begin{align}
&\hat H_{\text{eff}}= \hat H_0+ \frac{i}{2 \pi^2  \omega} \sum_{m,n=1}^{N} [\hat V_m , \hat V_n]\,  \mathcal{C}_{m,n} +\mathcal{O}(1/\omega^2), \notag \\
& \hat K (0)= - \frac{2}{\pi  \omega} \sum_{m=1}^{N} \hat V_m \, \tilde{\mathcal{C}}_{m} +\mathcal{O}(1/\omega^2), \notag\\  
& \mathcal{C}_{m,n}=   \sum_{j=1}^{\infty} \frac{1}{j^3} \sin \left ( \frac{2 \pi j}{N} (n-m) \right ) \left ( 1 - \cos \left ( \frac{2 \pi j}{N} \right ) \right ), \notag \\
& \tilde{\mathcal{C}}_{m}= \sum_{j=1}^{\infty} \frac{1}{j^2} \sin \left (\frac{\pi j}{N} (2m-1) \right ) \sin \left (\frac{\pi j}{N} \right ) .  \notag
%\label{general_result}
\end{align}
These expressions can be simplified using the formulas \cite{gradshteyn}
\begin{align}
&\sum_{k=1}^{\infty} \frac{\sin k x}{k^3} = \frac{\pi^2 x}{6} - \frac{\pi x^2}{4} + \frac{\pi^3}{12}, \quad 0 \le x \le 2 \pi \\
&\sum_{k=1}^{\infty} \frac{\cos k x}{k^2} = \frac{\pi^2}{6} - \frac{\pi x}{2} + \frac{\pi^2}{4}, \quad 0 \le x \le 2 \pi
\end{align}
yielding the first-order terms presented in Eqs. \eqref{general_result}.

A similar calculation allows to evaluate the effective Hamiltonian's second-order terms. For the sake of simplicity, we restrict ourselves to the second-order term
\begin{align}
\hat H_{\text{eff}}^{(2)}=\frac{1}{2 \omega^2} \sum_{j=1}^{\infty} \frac{1}{j^2} \left ( [[V^{(j)},\hat H_0],V^{(-j)} ] + \text{h.c.} \right ) \notag ,
\end{align}
noticing that the harmonic-mixing terms (i.e. the second line in Eq. \eqref{eq:general_mixing}) do not contribute for the sequences treated in this work. Using the expansion in Eq. \eqref{def_vjs} for the $V^{(j)}$ operators yields

\begin{align}
&\hat H_{\text{eff}}^{(2)}=\frac{1}{2 \pi^2 \omega^2} \sum_{m<n=2}^{N} \hat C_{m,n}  \mathcal{D}_{m,n} + \frac{1}{4 \pi^2 \omega^2} \sum_{m=1}^{N} \hat C_{m,m}   \mathcal{D}_{m,m}, \notag \\
& \mathcal{D}_{m,n} \!=\!   \sum_{j=1}^{\infty} \frac{1}{j^4} \cos \left ( \frac{2 \pi j}{N} (n-m) \right )\! \left ( 1 - \cos \left ( \frac{2 \pi j}{N} \right ) \right ) \! \label{step_second_order}
   \end{align}
where we introduced the commutators
\be
\hat C_{m,n}=  [[\hat V_m , \hat H_0],\hat V_n] + [[\hat V_n , \hat H_0],\hat V_m].\label{commutators:order2}
\ee
Using the formula \cite{gradshteyn}
\begin{align}
&\sum_{k=1}^{\infty} \frac{\cos k x}{k^4} = \frac{\pi^4}{90} - \frac{\pi^2 x^2}{12} + \frac{\pi x^3}{12} - \frac{x^4}{48}, \quad 0 \le x \le 2 \pi , \notag
\end{align}
one finds simple expressions for the coefficients in Eq. \eqref{step_second_order}
\begin{align}
&\mathcal{D}_{m,n}\!=\! \frac{\pi^4}{3 N^4} \left ( 1+N^2 -6N(n-m) +6 (n-m)^2 \right ), \text{ for } m<n \notag \\
&\mathcal{D}_{m,m}\!=\! \frac{\pi^4}{3 N^4} (N-1)^2, \quad (\text{for } m=n).\notag
\end{align}
Inserting these coefficients into Eq. \eqref{step_second_order} yields the final result presented in Eq. \eqref{general_result}.\\

\section{Effective Hamiltonians for general sequences \\ with $N=3,4$ different steps}\label{appendix:differentN}

In this Appendix, we apply the expression in Eq. \eqref{general_result} to general sequences with $N=3,4$ different steps. The case $N=2$ is fully treated in the main text and in Appendix \ref{appendix:two_phase}.

\subsection{The case $N=3$}

We first consider a general sequence with three different repeated steps: $\gamma_3=\{ \hat H_0 +\hat V_1 , \hat H_0 +\hat V_2, \hat H_0 +\hat V_3 \}$. In this case, the expressions in Eq. \eqref{general_result} yield
\begin{align}
&\hat H_{\text{eff}}=\hat H_0 + \frac{i \pi}{27   \omega} \left ( [\hat V_1 , \hat V_2] + [\hat V_2 , \hat V_3] + [\hat V_3 , \hat V_1] \right ) \label{ham_eff_3_pulse} \\
& \!+\! \frac{\pi^2}{243   \omega^2} \left (\! \hat C_{1,1} \!+\! \hat C_{2,2} \!+\! \hat C_{3,3} \!-\! \hat C_{1,2} \!-\! \hat C_{2,3} \!-\! \hat C_{1,3} \! \right ) \!+\! \mathcal{O} (1/\omega^3) ,  \notag \\
&\hat K(0)=\frac{2 \pi}{9   \omega} \left ( - \hat V_1 + \hat V_3 \right ) \!+\! \mathcal{O} (1/\omega^2),\notag
\end{align}
where the commutators $\hat C_{m,n}$ are defined in Eq. \eqref{commutators:order2}. In contrast with the case $N\!=\!2$, a proper choice of the operators $\hat H_0$ and $\hat V_{1,2,3}$ can potentially lead to non-trivial effects that are \emph{first order} in $(1/ \omega)$. This scenario was considered by Kitagawa \emph{et al.} in Ref. \cite{Kitagawa} to realize the Haldane model \cite{Haldane:1988} in a honeycomb lattice with pulsed hopping terms.

\subsection{The case $N=4$}

We now consider the case of four-step sequences, which are principally explored in this work, and which have also been the basis of several proposals \cite{Anderson,Ueda,Sorensen}. In this case, the expressions in Eq. \eqref{general_result} yield
\begin{widetext}
\begin{align}
&\hat H_{\text{eff}}=\hat H_0 + \frac{i \pi}{32   \omega} \left ( [\hat V_1 , \hat V_2] + [\hat V_2 , \hat V_3] + [\hat V_3 , \hat V_4] + [\hat V_4 , \hat V_1] \right ) \notag \\
&\qquad + \frac{\pi^2}{256   \omega^2} \left [ \frac{3}{4} \left ( \hat C_{1,1} + \hat C_{2,2} + \hat C_{3,3}+ \hat C_{4,4}  \right ) - \frac{1}{6} \left ( \hat C_{1,2} + \hat C_{2,3} + \hat C_{3,4}+ \hat C_{1,4}  \right ) - \frac{7}{6} \left ( \hat C_{1,3} + \hat C_{2,4}  \right ) \right ] + \mathcal{O} (1/\omega^3), \notag \\
&\hat K(0)=\frac{\pi}{16   \omega} \left ( - 3 \hat V_1 - \hat V_2 + \hat V_3+ 3 \hat V_4 \right ) + \mathcal{O} (1/\omega^2),\label{ham_eff_4_pulse}
\end{align}
\end{widetext}
where we note the absence of commutators $[\hat V_1, \hat V_3]$ and $[\hat V_2, \hat V_4]$ in $\hat H_{\text{eff}}$, and where the commutators $\hat C_{m,n}$ are defined in Eq. \eqref{commutators:order2}. As for the case $N\!=\!3$, a proper choice of the operators $\hat H_0$ and $\hat V_{1,2,3,4}$ can potentially lead to non-trivial effects that are \emph{first order} in $(1/ \omega)$. \\

%\subsubsection{Additional results for the $\alpha$ sequence}\label{third_order_alpha}

Let us now focus on the $\alpha$ sequence in Eq. \eqref{alpha_sequence}, introduced in Section \ref{section:alpha}.

First, we note that the third order corrections are easily obtained when the $\alpha$ sequence in Eq. \eqref{alpha_sequence} is approximated by the smooth single-harmonic driving $\hat V(t)= \hat A \cos (\omega t) + \hat B \sin (\omega t) $. Pushing the perturbative expansion of Appendix \ref{appendix:effective} to the next order, we find
\begin{align}
&\hat H_{\text{eff}}\!=\! \hat H_0 \!+\! \frac{i}{2   \omega} [\hat A, \hat B] \!+\! \frac{1}{4 \omega ^2}\! \left (  [[ \hat A, \hat H_0], \hat A] \!+\!  [[ \hat B, \hat H_0], \hat B] \right )\notag \\
& +\frac{i}{4 \omega^3} \left ( [\hat A, [[ \hat B, \hat H_0], \hat H_0]] -  [\hat B, [[ \hat A, \hat H_0], \hat H_0]]   \right ) \notag \\
& +\frac{i}{16 \omega^3} \left ( [[\hat A, [ \hat A, \hat B]], \hat A] + [[\hat B, [ \hat A, \hat B]], \hat B]   \right ) \!+\! \mathcal{O} (1/ \omega ^4). \label{third_order_alpha_full}
\end{align}
 
The third order corrections associated with the original square-wave sequence $\alpha$ [Eq. \eqref{alpha_sequence}]  can also be evaluated through the alternative perturbative method developed in Appendix \ref{rahav:type:2}, reading [Eq. \eqref{third_order_corrections_appendix_type2}]
\begin{align}
\hat H_{\text{eff}}^{(3)}/\omega^3=+\frac{i \pi^3}{768 \omega^3}   [[\hat B, [ \hat A, \hat B]], \hat B] , \label{third_order_corrections}
\end{align}
where we note that $\pi^3/768 \approx 1/16$, in agreement with Eq. \eqref{third_order_alpha_full}. Note that this method disregards the terms 
\be
[\hat A, [[ \hat B, \hat H_0], \hat H_0]], \, \,  [\hat B, [[ \hat A, \hat H_0], \hat H_0]], \, \,  [[\hat A, [ \hat A, \hat B]], \hat A],\label{neglected_terms}
\ee 
at this order of the computation; indeed, the alternative approach implicitly assumes that these terms will contribute to higher orders in $1/ \omega$ [Appendix \ref{rahav:type:2}]. However, we note that the expression in Eq. \eqref{third_order_corrections} applies to the SOC scheme analyzed in Section \ref{section_good_SOC_one}, for which the neglected terms in Eq. \eqref{neglected_terms} all vanish identically. The effects associated with the third order corrections \eqref{third_order_corrections} are illustrated for this specific scheme in Fig. \ref{fig-4}.

\section{Tight-binding operators in 2D lattice systems \\ and useful commutators}\label{appendix:lattice_operators}

In this Appendix, we set a dictionary that translates common operators defined in the bulk ($\hat x, \hat p, \dots$) into their lattice analogues. Here, the lattice framework is treated using a single-band tight-binding approximation. We also provide commutators of these operators, which are useful for the calculations presented in Sections \ref{section_four}-\ref{section_five}. 

In a single-band tight-binding description, the hopping Hamiltonian is taken in the form
\begin{align}
\hat H_0&= \hat T_x + \hat T_y \notag ,\\
&= -J \sum_{m,n} \hat a_{m+1,n}^{\dagger} \hat a_{m,n} + \hat a_{m,n+1}^{\dagger} \hat a_{m,n} + \text{h.c.}, \label{TB_ham}
\end{align}
where the operators $\hat T_{x,y}$ denote hopping along the $(x,y)$ directions
with amplitude $J$, $\hat a_{m,n}^{\dagger}$ creates a particle at lattice site $\bs x \!=\! (m a , n a)$, and where $a$ is the lattice spacing. The indices $(m,n)$ are integers. If we take the continuum limit $a \rightarrow 0$ (or equivalently, if we expand the momentum around the bottom of  the band), we recover the usual kinetic energy term
\be
\hat T_x + \hat T_y \equiv \frac{1}{2m^*} \left (\hat p_x^2 + \hat p_y^2 \right), \quad m^*\!=\!1/(2 J a^2),
\ee
where we introduced the effective mass $m^*$. We exploit this bulk-lattice analogy and use the following notation to denote lattice operators
\be
\frac{\bar p_x^2}{2m^*} \equiv \hat T_x = -J \sum_{m,n} \hat a_{m+1,n}^{\dagger} \hat a_{m,n} + \text{h.c.}
\ee

\subsection{One-component lattices}\label{appendix:1component}

The following Table I defines all the lattice operators used in Section \ref{section_four}.

\begin{table}[h!]
\null\hspace{1cm}
\begin{tabular}{| p{2.5cm} | p{6.5cm} |}
%\begin{tabular}{ | l | l | l | l | l | l | p{5cm} |}
\hline
\emph{{\blue Symbol}} & \emph{{\blue Tight-binding operator}}  \\
\hline
\hline
 $ \bar p_x^2/2m^*$ & $-J \sum_{m,n} \hat a_{m+1,n}^{\dagger} \hat a_{m,n} + \text{h.c.}$\\
 \hline
 $ \bar p_y^2/2m^*$ & $-J \sum_{m,n} \hat a_{m,n+1}^{\dagger} \hat a_{m,n} + \text{h.c.}$\\
 \hline
  $ \bar x  $ & $ a\sum_{m,n} m  \, \hat a_{m,n}^{\dagger} \hat a_{m,n}$\\
 \hline
 $ \bar x \, \bar y $ & $ a^2\sum_{m,n} m \, n \, \hat a_{m,n}^{\dagger} \hat a_{m,n}$\\
 \hline
  $ \bar p_x $ & $ i/2a \sum_{m,n}  \hat a_{m+1,n}^{\dagger} \hat a_{m,n} - \text{h.c.}$\\
 \hline
 $\bar L_z= \bar x \bar{p}_y - \bar y \bar{p}_x$ & $i/2 \sum_{m,n} m\, \hat a_{m,n+1}^{\dagger} \hat a_{m,n} - n \, \hat a_{m+1,n}^{\dagger} \hat a_{m,n} - \text{h.c.}$\\
  \hline
$ \bar x^2 + \bar y^2 $ &  $a^2/2 \sum_{m,n} n^2 \hat a_{m+1,n}^{\dagger} \hat a_{m,n} \!+\! m^2 \hat a_{m,n+1}^{\dagger} \hat a_{m,n} \!+\! \text{h.c.}$\\
  \hline
\end{tabular}
\caption{}
\end{table}
The Table II presents the non-zero commutators required to evaluate $[\hat A, \hat B]$ and $[[\hat H_0, \hat B], \hat B]$ in Section \ref{section_four}, using the lattice operators defined in the previous Table I.\\

\begin{table}[h!]
\null\hspace{1cm}
\begin{tabular}{| p{5cm} | p{4cm} |}
%\begin{tabular}{ | l | l | l | l | l | l | p{5cm} |}
\hline
\emph{{\blue Commutator}} & \emph{{\blue Result}}  \\
\hline
\hline
 $ [ \bar p_x^2/2m^* \!-\!  \bar p_y^2/2m^*, \bar x \, \bar y]$ & $ =2 i  J a^2 \bar L_z=  i \bar L_z/m^* $\\
 \hline
 $ [ \bar p_x^2/2m^* \!+\!  \bar p_y^2/2m^*, \bar x \, \bar y]$ & $ =- 2 i J a^2 (\bar x \bar{p}_y + \bar y \bar{p}_x)$\\
 \hline
  $ i [ \bar x \bar{p}_y + \bar y \bar{p}_x, \bar x \, \bar y]$ & $=\bar x^2 + \bar y^2$\\
 \hline
$ [ \bar p_{\mu}^2/2m^*, \bar \mu] , \quad \mu=x,y$ & $ = - i \bar p_{\mu}/m^* $\\
 \hline
 $ [ \bar p_{\mu}, \bar \mu] / a^2m^*, \quad \mu=x,y$ & $ = i \bar p_{\mu}^2/2m^* $\\
 \hline
\end{tabular}
\caption{}
\end{table}

In particular, setting
\be
\hat H_0=\bar p_{\mu}^2/2m^*, \quad \hat V=\bar \mu/a , \quad \hat H_1=\bar p_{\mu}/a m^*, \quad \mu=x,y , \notag
\ee
we recover the cyclic relations Eqs. \eqref{cycle_cond}, such that Eq. \eqref{eq:useful} yields
\be
e^{i \gamma \bar \mu/a} \left ( \bar p_{\mu}^2/2m^* \right )e^{-i \gamma \bar \mu/a} = \left ( \bar p_{\mu}^2/2m^* \right ) \cos \gamma - \left ( \bar p_{\mu}/a m^* \right ) \sin \gamma. \notag
\ee

\subsection{Two-component lattices}\label{appendix:2component}

In order to set the notations, we write the 2-component tight-binding Hamiltonian in the form
\begin{align}
\hat H_0&= \hat T_x + \hat T_y \notag \\
&=-J \sum_{m,n} \hat \Psi_{m+1,n}^{\dagger} \hat \Psi_{m,n} +\hat \Psi_{m,n+1}^{\dagger} \hat \Psi_{m,n}  + \text{h.c.} ,  
\end{align}
where $\Psi_{m,n}^{\dagger}=(\hat a_{m,n}^{\dagger}, \hat b_{m,n}^{\dagger})$ contains the operators that create a particle at lattice site $\bs x = (m a , n a)$ in state $\sigma=\pm$. The following Table III defines all the lattice operators used in Section \ref{section_five}. 

\begin{table}[h!]
\null\hspace{1cm}
\begin{tabular}{| p{3cm} | p{6.5cm} |}
%\begin{tabular}{ | l | l | l | l | l | l | p{5cm} |}
\hline
\emph{{\blue Symbol}} & \emph{{\blue Tight-binding operator}}  \\
\hline
\hline
 $ \bar p_x^2/2m^*$ & $-J \sum_{m,n} \hat \Psi_{m+1,n}^{\dagger} \hat \Psi_{m,n}+ \text{h.c.}$\\
 \hline
 $ \bar p_y^2/2m^*$ & $-J \sum_{m,n} \hat \Psi_{m,n+1}^{\dagger} \hat \Psi_{m,n}+ \text{h.c.}$\\
 \hline
 $ \bar x \, \hat \sigma_x $ & $a \sum_{m,n} m \, \hat \Psi_{m,n}^{\dagger} \hat \sigma_x \hat \Psi_{m,n}$\\
 \hline
  $ \bar y \, \hat \sigma_y $ & $a \sum_{m,n} n \, \hat \Psi_{m,n}^{\dagger} \hat \sigma_y \hat \Psi_{m,n}$\\
 \hline
  $ \bar p_x \, \hat \sigma_x $ & $i/2a \sum_{m,n} \, \hat \Psi_{m+1,n}^{\dagger} \hat \sigma_x \hat \Psi_{m,n}- \text{h.c.}$\\
 \hline
   $ \bar p_y \, \hat \sigma_y $ & $i/2a \sum_{m,n} \, \hat \Psi_{m,n+1}^{\dagger} \hat \sigma_y \hat \Psi_{m,n}- \text{h.c.}$\\
 \hline
    $\bar L_z \hat \sigma_z\!=\! ( \bar x \bar{p}_y - \bar y \bar{p}_x ) \hat \sigma_z$ & $ i/2\sum_{m,n}  \! \left ( m \! \hat \Psi_{m,n+1}^{\dagger}  -n \hat \Psi_{m+1,n}^{\dagger} \! \right )\hat \sigma_z\! \hat \Psi_{m,n} \! -\! \text{h.c.}$\\
 \hline
  $( \bar x \bar{p}_y + \bar y \bar{p}_x ) \hat \sigma_z$ & $ i/2 \sum_{m,n}  \! \left ( m \! \hat \Psi_{m,n+1}^{\dagger}  +n \hat \Psi_{m+1,n}^{\dagger} \! \right )\hat \sigma_z\! \hat \Psi_{m,n} \! -\! \text{h.c.}$\\
 \hline
\end{tabular}
\caption{}
\end{table}
The Table IV presents the non-zero commutators required to evaluate the effective Hamiltonians in Section \ref{section_five}, using the lattice operators defined in the previous Table III.
\begin{table}[h!]
\null\hspace{1cm}
\begin{tabular}{| p{5cm} | p{4cm} |}
%\begin{tabular}{ | l | l | l | l | l | l | p{5cm} |}
\hline
\emph{{\blue Commutator}} & \emph{{\blue Result}}  \\
\hline
\hline
 $ [ \bar p_x^2/2m^*, \bar x \, \hat \sigma_x]$ & $=(-i/m^*) \bar p_x \hat \sigma_x$\\
 \hline
  $ [ \bar p_x \hat \sigma_x, \bar x \, \hat \sigma_x]$ & $=(i a^2/2) \bar p_x^2$\\
 \hline
   $ \rightarrow [ [ \bar p_x^2/2m^*, \bar x \, \hat \sigma_x], \bar x \, \hat \sigma_x]/a^2$ & $=\bar p_x^2/2m^*$\\
 \hline
  $ [ \bar p_y^2/2m^*, \bar y \, \hat \sigma_y]$ & $=(-i/m^*) \bar p_y \hat \sigma_y$\\
 \hline
  $ [ \bar p_y \hat \sigma_y, \bar y \, \hat \sigma_y]$ & $=(i a^2/2) \hat p_y^2$\\
 \hline
   $ \rightarrow [ [ \bar p_y^2/2m^*, \bar y \, \hat \sigma_y], \bar y \, \hat \sigma_y]/a^2$ & $=\bar p_y^2/2m^*$\\
 \hline
   $ [ \bar p_x \hat \sigma_x, \bar y \, \hat \sigma_y] + [ \bar p_y \hat \sigma_y, \bar x \, \hat \sigma_x]$ & $=-2 i \bar L_z \hat \sigma_z$\\
   \hline
   $ [ \bar p_y \hat \sigma_y, \bar x \, \hat \sigma_x] - [ \bar p_x \hat \sigma_x, \bar y \, \hat \sigma_y]$ & $=-2 i ( \bar x \bar{p}_y + \bar y \bar{p}_x) \hat \sigma_z$\\
 \hline
\end{tabular}
\caption{}
\end{table}

Moreover, it is useful to note the cyclic conditions [see also Eq. \eqref{cycle_cond}]
\begin{align}
&\hat H_0=\bar p_{\mu}^2/2m^*, \quad \hat V=\bar \mu \hat \sigma_{\mu} /a , \quad \hat H_1=\bar p_{\mu} \hat \sigma_{\mu}/a m^*, \quad \mu=x,y , \notag \\
&[\hat H_0, \hat V ]=-i \hat H_1 , \quad [\hat H_1, \hat V ]=i \hat H_0, \quad [\hat H_0, \hat H_1 ]=0. \label{cycle_cond_SOC}
\end{align}
Using Eqs. \eqref{cycle_cond_SOC}-\eqref{eq:useful} yields the useful formulas
\begin{align}
e^{i \gamma \bar \mu \hat \sigma_{\mu}/a} \left( \frac{ \bar p_{\mu}^2}{2m^*} \right )e^{-i \gamma \bar \mu \hat \sigma_{\mu}/a} = \frac{\bar p_{\mu}^2}{2m^*} \cos \gamma - \frac{\bar p_{\mu} \hat \sigma_{\mu} }{a m^* }\sin \gamma , \label{useful_SOC_formula}\\
e^{i \gamma \bar \mu \hat \sigma_{\mu}/a} \left( \frac{ \bar p_{\mu} \hat \sigma_{\mu}}{a m^*} \right )e^{-i \gamma \bar \mu \hat \sigma_{\mu}/a} = \frac{\bar p_{\mu}\hat \sigma_{\mu}}{a m^*} \cos \gamma - \frac{\bar p_{\mu}^2 }{2 m^* }\sin \gamma,\notag
\end{align}
which are used several times in this work.

\section{Exact treatment of the XA scheme}\label{appendix:ueda}

In this Appendix, we derive the effective Hamiltonians in Eqs. \eqref{ueda_bulk_result}-\eqref{ueda_lattice_result}, which are associated with the XA scheme.

\subsection{Without the lattice}

Following Refs. \cite{Ueda,Anderson}, it is convenient to partition the time-evolution operator associated with the ``XA" sequence \eqref{bad_SOC_sequence} as $\hat U (T)\!=\! \hat U_y \hat U_x$, where the two subsequences read
\begin{align}
\hat U_{\mu}= & e^{-i (\hat H_0 - \kappa \hat \mu \hat \sigma_{\mu}) T/4} e^{-i (\hat H_0 + \kappa \hat \mu \hat \sigma_{\mu}) T/4} , \quad \mu=x,y.
\end{align}
We first focus on the subsequence characterized by the evolution operator $\hat U_x$, and perform a unitary transformation in the same spirit as in Eq. \eqref{transf_one},
\be
\psi = \hat R \tilde \psi = \exp \left (\! -i \kappa \hat x \hat \sigma_x \int_0^t f(\tau) \text{d} \tau \! \right ) \tilde \psi , \label{transf_ueda}
\ee
where $f(t)$ denotes the square function
\be
f(t)=1 \quad \text{for t $\in [0 , T/4]$} , \quad f(t)=-1 \quad \text{for t $\in [T/4 , T/2]$}.\notag 
\ee
The transformed state satisfies the Schrödinger equation $i \partial_t \tilde \psi = \tilde H(t) \tilde \psi$, with the modified Hamiltonian
\begin{align}
\tilde H (t)= \hat R^{\dagger} \hat H_0 \hat R = \frac{1}{2m} \left [ \hat p_y^2 + \left ( \hat p_x - \kappa \hat \sigma_x \mathfrak{f}(t)  \right )^2 \right ] , \label{transf_two_ueda}
\end{align}
where $\mathfrak{f}(t) = \int_0^t f(\tau) \text{d} \tau$. Importantly, the result in Eq. \eqref{transf_two_ueda} is exact, and in particular, it is valid for any $\omega$ or $\lambda_{\text{R}} \sim \kappa/\omega$. The evolution operator after half a period thus reads
\begin{align}
&\hat U_{x}= e^{-i \int_0^{T/2} \tilde H (t) \text{d} t} =e^{-i T \left [\hat H_0/2 - (\kappa T/16m)  \hat p_{x} \hat \sigma_{x} \right]} , \label{bulk_ueda_one}
\end{align}
which shows the fact that half of the ``XA" sequence \eqref{bad_SOC_sequence}  has been treated exactly. The next subsequence ($t=T/2 \rightarrow T$) can be treated similarly, yielding the exact result
\begin{align}
&\hat U_{y}= e^{-i T \left [\hat H_0/2 - (\kappa T/16m)  \hat p_{y} \hat \sigma_{y} \right]} . \label{bulk_ueda_two} 
\end{align}
Finally, to lowest order in $T=2\pi / \omega$, the Trotter expansion yields the Rashba SOC Hamiltonian
\begin{align}
&\hat U (T)=\hat U_y \hat U_x = \exp \left ( -i \hat H_{\text{eff}}^{\mathcal{T}} T \right ) , \notag \\
&\hat H_{\text{eff}}^{\mathcal{T}}= \hat H_0 -  \lambda_{\text{R}} \, \hat{\bs{p}} \cdot \hat{\bs \sigma} + \mathcal{O} \left( (\Omega_{\text{SO}}/\omega)^2 \right ), \label{appendix_ueda_bulk_result}
\end{align}
where $\lambda_{\text{R}} = \pi \kappa / 8 m \omega$, and where we introduced the small dimensionless parameter $\Omega_{\text{SO}}/\omega$ with $\Omega_{\text{SO}} \sim m \lambda_{\text{R}}^2 $; here, we supposed that the momentum $p \sim m \lambda_{\text{R}}$ is of the order of the Rashba ring (i.e. the bottom of the mexican hat). 

\subsection{With the lattice}

A similar calculation allows to evaluate the evolution operator for the ``XA" sequence \eqref{bad_SOC_sequence} defined on the lattice, namely, when substituting the operators in \eqref{bad_SOC_sequence} by their lattice analogues [see Appendix \ref{appendix:2component}]. 

As in the lattice-free case, we consider the unitary transformation in Eq. \eqref{transf_ueda}. The transformed state satisfies the Schrödinger equation $i \partial_t \tilde \psi = \tilde H(t) \tilde \psi$, with the modified Hamiltonian
\begin{align}
\tilde H (t)= \frac{1}{2m^*} \left \{ \bar p_y^2 + \bar p_x^2 \cos [a \kappa \frak{f} (t) ] \right \} - \frac{1}{a m^*} \bar p_x \hat \sigma_x \sin [a \kappa \frak{f} (t)]   , \notag
\end{align}
where we used the formula \eqref{useful_SOC_formula}, and where the function $\mathfrak{f}(t) = \int_0^t f(\tau) \text{d} \tau$ was introduced in the last Section. The evolution operator after half a period thus reads
\begin{align}
&\bar U_{x}= e^{-i \int_0^{T/2} \tilde H (t) \text{d} t} =\exp \left ( -i T \bar H_x \right ) , \label{lattice_ueda_one}  \\
&\bar H_x= \frac{\bar p_y^2}{4m^*} + \frac{\bar p_x^2}{a \kappa T m^*}\sin [a \kappa T/4]   - \frac{4}{a^2 \kappa T m^*} \bar p_x \hat \sigma_x \sin^2 [a \kappa T/8] \notag .
\end{align}
A similar calculation can be performed for the next subsequence, yielding
\begin{align}
&\bar U_{y}=\exp \left (-i T \hat H_y \right ) , \label{lattice_ueda_two}  \\
&\bar H_y= \frac{\bar p_x^2}{4m^*} + \frac{\bar p_y^2}{a \kappa T m^*}\sin [a \kappa T/4]   - \frac{4}{a^2 \kappa T m^*} \bar p_y \hat \sigma_y \sin^2 [a \kappa T/8] \notag .
\end{align}
We point out that the expressions Eqs. \eqref{lattice_ueda_one}-\eqref{lattice_ueda_two} are exact, and that they constitute the direct lattice analogues of Eqs. \eqref{bulk_ueda_one}-\eqref{bulk_ueda_two}.  The Trotter expansion then yields the evolution operator after one period
\begin{align}
&\hat U (T)=\bar U_y \bar U_x = \exp \left (-i \hat H_{\text{eff}}^{\mathcal{T}} T \right ) , \label{ueda_lattice_result_appendix} \\
&\bar H_{\text{eff}}^{\mathcal{T}}\!=\! \frac{\bar p^2}{2 m^*} \left \{ \frac{1}{2} \!+\! \frac{1}{2} \text{sinc} \left ( 4 a m^* \lambda_{\text{R}}  \right )   \right \}  \!+\! \bar{\bs{p}} \cdot \bs{\hat \sigma} \left \{ \frac{\text{cos} \left ( 4a m^* \lambda_{\text{R}} \right ) \!-\!1  }{8 (a m^*)^2 \lambda_{\text{R}}} \right \}, \notag
\end{align}
where $\lambda_{\text{R}} = \pi \kappa / 8 m^* \omega$ and $m^*=1/2Ja^2$ is the effective mass. Note that we recover exactly the result in Eq. \eqref{appendix_ueda_bulk_result} for weak driving $\lambda_{\text{R}} < aJ/2$ and by taking the continuum limit. However, in the lattice framework, the maximum value of the effective Rashba SOC strength is limited. In particular, we find that this maximum value is obtained for $\lambda_{\text{R}}=\pi/4 a m^*=a J(\pi/2)$.

\section{Exact treatment of the $xy$ scheme}\label{appendix:xy}

In this Appendix, we derive the effective Hamiltonian in Eq. \eqref{xy_lattice_result}. We consider the $\alpha$-type ``$xy$" sequence [Eq. \eqref{xy_SOC_sequence}],
\be
\left \{ \frac{\bar p_y^2}{m} - \kappa \bar x \bar \sigma_x,  \frac{\bar p_y^2}{m} - \kappa \bar y \bar \sigma_y , \frac{\bar p_x^2}{m} + \kappa \bar x \bar \sigma_x, \frac{\bar p_x^2}{m} + \kappa \bar y \bar \sigma_y \right \}, \label{xy_SOC_sequence_appendix}
\ee 
and consider the special regime where $\kappa T = 4 \pi /a$, that is, $\lambda_{\text{R}} = \pi \kappa / 8 m^* \omega= (\pi/2) a J$. Furthermore, we consider that the system is set on a lattice, such that the operators  in Eq. \eqref{xy_SOC_sequence_appendix} correspond to the lattice operators defined in Appendix \ref{appendix:2component}. 

Here, in contrast with the analysis performed in the previous Appendix \ref{appendix:ueda}, we split the evolution operator $\bar U (T)$ corresponding to the sequence \eqref{xy_SOC_sequence_appendix} into its four primitive parts,
\begin{align}
\bar U (T)&= e^{-\frac{i T}{4 } (\bar H_0-\bar B)} e^{-\frac{i T}{4 } (\bar H_0-\hat A)} e^{-\frac{i T}{4 } (\bar H_0+\bar B)} e^{-\frac{i T}{4 } (\bar H_0+\bar A)}\notag \\
&=\bar U_{-B} \bar U_{-A} \bar U_{+B} \bar U_{+A},\label{evolution_operator_xy}
\end{align} 
and we analyze each operator $\bar U_{\pm A,B}$ separately. 

The operators $\bar U_{+A}$ and $\bar U_{-B}$ are readily simplified: setting $\lambda_{\text{R}} \!=\!  (\pi/2) a J$, we directly obtain the factorized expressions
\be
\bar U_{+A}= e^{ -i T \bar p_y^2/4 m^*} e^{i \pi \bar x/a} , \quad \bar U_{-B}=e^{ -i T \bar p_x^2/4 m^*} e^{i \pi \bar y/a},\label{factozied_1}
\ee
where we used the simplification
\be
\exp (i \pi \bar \mu \bar \sigma_{\mu} / a) = \exp (i \pi \bar \mu / a), \quad \mu=x,y .\label{simplification_lattice}
\ee
The latter is due to the fact that $\exp (i \pi m \hat \sigma_x)=(-1)^{m}$ for $m \in \mathbb{Z}$. The property \eqref{simplification_lattice} is thus specific to the lattice operators $\bar x \bar \sigma_x$, $\bar y \bar \sigma_y$ defined in Appendix \ref{appendix:2component}. 

The two other operators $\bar U_{-A}$ and $\bar U_{+B}$ require more care, as they are both of the form $\exp (X+Y)$ with $[X,Y] \ne 0$. The factorization can be performed through the  Zassenhaus formula \cite{Zassenhaus}
\be
e^{X+Y}= e^{X} e^{Y} e^{-\frac{1}{2}[X,Y]} e^{\frac{1}{3}[Y, [X,Y]]+\frac{1}{6}[X, [X,Y]]} \dots
\ee
We find that the latter formula takes an elegant form when the operators satisfy the following cyclic relations
\be
[X,Y]=Z , \quad [X,Z] = \mathfrak{a} Y , \quad [Y,Z]=0.\label{zassenhaus_cyclic}
\ee
Indeed, under the conditions \eqref{zassenhaus_cyclic}, we find the \emph{exact} result
\be
e^{X+Y}= e^{X} \exp \left \{Y \text{sinh} (\sqrt{\mathfrak{a}})/\sqrt{\mathfrak{a}} - Z \left [-1 + \text{cosh} (\sqrt{\mathfrak{a}}) \right ]/\mathfrak{a}  \right \}.\notag
\ee
Using the latter formula together with the cyclic conditions in Eq. \eqref{cycle_cond_SOC} yields 
\begin{align}
\bar U_{-A}= e^{i \pi \bar x/a} e^{i T \lambda_{\text{R}}^{*} \bar p_x \bar \sigma_x}, \quad 
& \bar U_{+B}= e^{i \pi \bar y/a} e^{ -i T \lambda_{\text{R}}^{*} \bar p_y \bar \sigma_y} ,\label{factozied_2}
\end{align}
where $\lambda_{\text{R}}^{*}=(2/\pi) a J$ was introduced in Eq. \eqref{lambda_star}. 

We now have at our disposal factorized forms \eqref{factozied_1}-\eqref{factozied_2} for the four operators $\bar U_{\pm A,B}$, which can be inserted into the evolution operator in Eq. \eqref{evolution_operator_xy}. Noting that
\be
\exp(-i \pi \bar x/a) \, \left (\bar p_x \, \bar \sigma_x \right ) \, \exp(i \pi \bar x/a)= -\bar p_x \bar \sigma_x,
\ee 
which directly results from Eq. \eqref{useful_SOC_formula}, we obtain the exact result
\be
\bar U (T)= e^{ -i T \bar p_x^2/4 m^*} e^{-i T \lambda_{\text{R}}^{*} \bar p_x \bar \sigma_x}  e^{-i T \lambda_{\text{R}}^{*} \bar p_y \bar \sigma_y} e^{ -i T \bar p_y^2/4 m^*}.
\ee
Applying the Trotter expansion to minimal order, we finally find the effective Rashba Hamiltonian announced in Eq. \eqref{xy_lattice_result},
\begin{align}
&\bar U (T)=\bar U_{-B} \bar U_{-A} \bar U_{+B} \bar U_{+A} = \exp \left (-i \bar H_{\text{eff}}^{\mathcal{T}} T \right ) , \notag \\
&\bar H_{\text{eff}}^{\mathcal{T}}\!=\! \frac{1}{2} \left ( \frac{\bar p^2}{2 m^*} \right ) \!+\! \lambda_{\text{R}}^{*}  \, \bar{\bs{p}} \cdot \bs{\hat \sigma} + \mathcal{O} \left( (\Omega_{\text{SO}}/\omega)^2 \right ) . \notag 
\end{align}

\section{An alternative perturbative approach} \label{rahav:type:2}

In this Appendix, we propose an alternative perturbative approach, which is specifically dedicated to the general time-dependent problem 
\be
\hat H (t) = \hat H_0 +\hat V (t)=\hat H_0 + \hat{\mathcal{A}} f(t) + \omega \hat{\mathcal{B}} g(t) ,\label{rahavtype2:time-dep_appendix}
\ee 
where the functions $f$ and $g$ are time-periodic with a zero mean value over one period $T=2 \pi / \omega$. As motivated in the main text, the following method is particularly suited for problems in which $\hat H_0 \sim \hat{\mathcal{A}} \sim \Omega \ll \omega$ and $ \hat{\mathcal{B}}\sim 1$. Following the general approach of Appendix \ref{appendix:effective}, we write the effective Hamiltonian and kick operators as
\begin{align}
& \hat H_{\text{eff}} = e^{i \hat K(t)} \hat H (t) e^{-i \hat K(t)} + i  \left ( \frac{\partial e^{i \hat K(t)}}{\partial t}\right ) e^{-i \hat K(t)}.\label{rahav_ham_type2}
\end{align}
Since the time-dependent Hamiltonian in Eq. \eqref{rahavtype2:time-dep_appendix} now contains a term that is proportional to the frequency $\omega$, the perturbative expansion in Eq. \eqref{expand_eq} must be slightly modified to include a term $\hat K^{(0)}(t)$,
\be
\hat H_{\text{eff}} = \sum_{n=0}^{\infty} \frac{1}{\omega^n} \hat H_{\text{eff}}^{(n)} , \quad \hat K (t) = \sum_{n=0}^{\infty} \frac{1}{\omega^n} \hat K^{(n)}(t) .\label{expand_eq_type2}
\ee
Using the expansion formulas in Eqs. \eqref{expand_eq_C7}-\eqref{expand_eq_2}, we first identify the terms that are proportional to $\omega$ in Eq. \eqref{rahav_ham_type2}, which simply yields
\be
\hat K^{(0)}(t)= \hat{\mathcal{B}} G (t) , \quad G(t)= \omega  \int^t g(\tau) \text{d}\tau,
\ee
where we impose that $G(t)$ should have a zero mean value over one period, $\overline{G (t)}\!=\!(1/T) \int_0^T G(t) \text{d}t=0$, as required for the kick operator in this formalism. Then, the first-order equation reads
\begin{align}
&\hat H_{\text{eff}}^{(0)}=\!  \hat H_0 + \hat{\mathcal{A}} f \!+\! i [\hat K^{(0)} ,  \hat H_0 + \hat{\mathcal{A}} f] \!-\! \frac{1}{2} [\hat K^{(0)} , [\hat K^{(0)},  \hat H_0 + \hat{\mathcal{A}} f]] \notag \\
&\quad \qquad \!-\! \frac{i}{6} [\hat K^{(0)}, [\hat K^{(0)} , [\hat K^{(0)} ,  \hat H_0 + \hat{\mathcal{A}} f]]] +\dots +\hat{\mathcal{R}}(\hat{\mathcal{B}} , \hat K^{(1)})\notag \\
&= \exp \left ( i G(t) \hat{\mathcal{B}} \right ) \left \{  \hat H_0 + \hat{\mathcal{A}} f(t) \right \} \exp \left ( -i G(t) \hat{\mathcal{B}} \right )  + \hat{\mathcal{R}}(\hat{\mathcal{B}} , \hat K^{(1)}),
\label{rahavtype2_appendix}
\end{align}
where the many terms grouped in the operator $\hat{\mathcal{R}}(\hat{\mathcal{B}} , \hat K^{(1)})$ are used to define the next-order term $\hat K^{(1)}$, which will absorb the time-periodic terms with zero average. Taking the time average on both sides of Eq. \eqref{rahavtype2_appendix} finally yields
\begin{align}
&\hat H_{\text{eff}}= \overline{\exp \left ( i G(t) \hat{\mathcal{B}} \right ) \left \{  \hat H_0 + \hat{\mathcal{A}} f(t) \right \} \exp \left ( -i G(t) \hat{\mathcal{B}} \right )} + \mathcal{O} (1/\omega), \notag\\
& \qquad =\hat H_0 +i \overline{G f} [\hat{\mathcal{B}} , \hat{\mathcal{A}}] - \frac{1}{2} \overline{G^2} [\hat{\mathcal{B}} , [\hat{\mathcal{B}}, \hat H_0]] \notag \\
&\quad \qquad - \frac{1}{2} \overline{G^2 f} [\hat{\mathcal{B}} , [\hat{\mathcal{B}}, \hat{\mathcal{A}}]] - \frac{i}{6} \overline{G^3} [ \hat{\mathcal{B}} [\hat{\mathcal{B}} , [\hat{\mathcal{B}}, \hat H_0]]]\notag \\
&\quad \qquad   - \frac{i}{6} \overline{G^3 f} [ \hat{\mathcal{B}} [\hat{\mathcal{B}} , [\hat{\mathcal{B}}, \hat{\mathcal{A}}]]] +  \dots + \mathcal{O} (1/\omega), \label{appendix_type2_final}
\end{align}
which provides the effective Hamiltonian to lowest order in $1/\omega$.\\

We now apply the expression in Eq. \eqref{appendix_type2_final} to the $\alpha$ pulse sequence introduced in Eq. \eqref{alpha_sequence},
\begin{equation}
\alpha:\, \{ \hat H_0+ \hat A, \hat H_0+ \hat B, \hat H_0-\hat A,\hat H_0- \hat B \}.\label{alpha_sequence_appendix}
\end{equation}
The time-dependent Hamiltonian is of the form
\be
\hat H(t)=\hat H_0+\hat V(t)=\hat H_0+ \hat A f(t) + \hat B g(t), \label{alpha_appendix_t2}
\ee 
where $f(t)$ and $g(t)$ are the square-wave functions associated with the sequence \eqref{alpha_sequence_appendix}. Comparing Eq. \eqref{alpha_appendix_t2} with the notations introduced in Eq. \eqref{rahavtype2:time-dep_appendix}, we note that $\hat{\mathcal{A}}= \hat A$ and $ \hat{\mathcal{B}} = \hat B/\omega$. Using the square-wave functions, we find
\begin{align}
&\overline{Gf}=-\frac{\pi}{8}, \, \overline{G^2}=\frac{\pi^2}{24}, \, \overline{G^2 f}=0, \, \overline{G^3}=0, \, \overline{G^3 f}= - \frac{\pi^3}{128}.\notag
\end{align}
Then substituting $\hat{\mathcal{A}} \rightarrow \hat A$ and $\omega \hat{\mathcal{B}} \rightarrow \hat B$ into the formula \eqref{appendix_type2_final} reads
\begin{align}
\hat H_{\text{eff}}=& \hat H_0 \!+\! \frac{i \pi}{8   \omega} [\hat A, \hat B] \notag \\
&+ \frac{\pi^2}{48 \omega ^2}\!  [[ \hat B, \hat H_0], \hat B] +\frac{i \pi^3}{768 \omega^3}   [[\hat B, [ \hat A, \hat B]], \hat B]  + \dots, \label{third_order_corrections_appendix_type2}
\end{align}
which indeed provides the third-order term announced in Eq. \eqref{third_order_corrections}.\\

Finally, we consider the smooth single-harmonic driving 
\be
\hat H(t)=\hat H_0+\hat V(t)=\hat H_0+ \hat A \cos (\omega t) + \hat B \sin (\omega t), 
\ee 
which approximates the $\alpha$ pulse sequence. In this case $f(t)=\cos (\omega t)$ and $g(t)=\sin (\omega t)$, so that
\begin{align}
&\overline{Gf}=-\frac{1}{2}, \, \overline{G^2}=\frac{1}{2}, \, \overline{G^2 f}=0, \, \overline{G^3}=0, \, \overline{G^3 f}= - \frac{3}{8}.\notag
\end{align}
Then substituting $\hat{\mathcal{A}} \rightarrow \hat A$ and $\omega \hat{\mathcal{B}} \rightarrow \hat B$ into the formula \eqref{appendix_type2_final} reads
\begin{align}
\hat H_{\text{eff}} =& \hat H_0 \!+\! \frac{i}{2   \omega} [\hat A, \hat B] \notag \\
&+ \frac{1}{4 \omega ^2}\! [[ \hat B, \hat H_0], \hat B] +\frac{i}{16 \omega^3}  [[\hat B, [ \hat A, \hat B]], \hat B] +   \dots , \label{third_order_alpha_full_appendix_type2}
\end{align}
so that we recover the result in Eq. \eqref{third_order_alpha_full} partially. Indeed, the present method disregards the terms 
\begin{align}
&[[ \hat A, \hat H_0], \hat A] , \qquad \quad \,\, [\hat A, [[ \hat B, \hat H_0], \hat H_0]], \notag \\
&  [\hat B, [[ \hat A, \hat H_0], \hat H_0]], \quad  [[\hat A, [ \hat A, \hat B]], \hat A], \label{neglected_terms_type2}
\end{align} 
at this order of the computation, as it implicitly attributes different orders to the operators $\hat H_0 \sim \hat A$ and $\hat B\sim \omega$, see Eq. \eqref{rahavtype2:time-dep_appendix}. Let us finally illustrate this last aspect based on the example presented in Section \ref{section:convergence_illustration}, which was based on the $\alpha$ sequence, and for which we found that the operators satisfied
\be
\hat H_0 \sim \hat A \sim \Omega \sim \kappa/m\omega , \quad \hat B \sim \omega, \quad \Omega \ll \omega ,
\ee
according to the lowest Landau level characteristics. In this case, we indeed find that the terms identified by the present perturbative method, and given in Eqs. \eqref{third_order_corrections_appendix_type2}-\eqref{third_order_alpha_full_appendix_type2}, are all of the same order $\sim \Omega$, whereas the neglected terms in Eq. \eqref{neglected_terms_type2} are all of order $\sim \Omega^3/\omega^2 \ll \Omega$. In particular, this illustrates the manner by which the present approach (potentially) allows to partially resum the infinite series inherent to the formalism of Section \ref{section:formalism}, and hence, to guarantee the convergence of the perturbative expansion. %However, we stress that this useful method only applies to time-dependent problems of the form \eqref{rahavtype2:time-dep_appendix}.

 \end{document}